\documentclass[preprint]{aastex631}

\usepackage{amsmath}
\usepackage{rotating}

\shorttitle{Bayesian inference in single-line spectroscopic binaries}
\shortauthors{Videla et al.}
\graphicspath{{./}{figures/}}
\begin{document}

\title{Bayesian inference in single-line spectroscopic binaries with a visual orbit}

\correspondingauthor{Miguel Videla}
\email{miguel.videla@ug.uchile.cl}

\author[0000-0002-7140-0437]{Miguel Videla}
\affiliation{Department of Electrical Engineering\\
Information and Decision Systems Group (IDS) \\
Facultad de Ciencias Físicas y Matemáticas, Universidad de Chile \\
Beauchef 850, Santiago, Chile}

\author[0000-0003-1454-0596]{Rene A. Mendez}
\affiliation{Astronomy Department \\ Universidad de Chile \\ Casilla 36-D, Santiago, Chile}
  
\author{Rub\'en M. Claver\'ia}
\affiliation{Department of Engineering \\ University of Cambridge, UK}

\author{Jorge F. Silva}
\affiliation{Department of Electrical Engineering\\
Information and Decision Systems Group (IDS) \\
Facultad de Ciencias Físicas y Matemáticas, Universidad de Chile \\
Beauchef 850, Santiago, Chile}

\author{Marcos E. Orchard}
\affiliation{Department of Electrical Engineering\\
Information and Decision Systems Group (IDS) \\
Facultad de Ciencias Físicas y Matemáticas, Universidad de Chile \\
Beauchef 850, Santiago, Chile}



\begin{abstract}
We present a Bayesian inference methodology for the estimation of orbital parameters on single-line spectroscopic binaries with astrometric data, based on the No-U-Turn sampler Markov chain Monte Carlo algorithm. Our approach is designed to provide a precise and efficient estimation of the joint posterior distribution of the orbital parameters in the presence of partial and heterogeneous observations. This scheme allows us to directly incorporate prior information about the system - in the form of a trigonometric parallax, and an estimation of the mass of the primary component from its spectral type - to constrain the range of solutions, and to estimate orbital parameters that cannot be usually determined (e.g. the individual component masses), due to the lack of observations or imprecise measurements. Our methodology is tested by analyzing the posterior distributions of well-studied double-line spectroscopic binaries treated as single-line binaries by omitting the radial velocity data of the secondary object. Our results show that the system's mass ratio can be estimated with an uncertainty smaller than 10\% using our approach. As a proof of concept, the proposed methodology is applied to twelve single-line spectroscopic binaries with astrometric data that lacked a joint astrometric-spectroscopic solution, for which we provide full orbital elements. Our sample-based methodology allows us also to study the impact of different posterior distributions in the corresponding observations space. This novel analysis provides a better understanding of the effect of the different sources of information on the shape and uncertainty of the orbit and radial velocity curve.
\end{abstract}



\section{Introduction} \label{sec:intro}
Mass is the most critical parameter which determines the structure and evolution of stars. In binary stellar systems, masses of their individual components can be directly calculated from the orbital parameters through Kepler's laws using astrometric and spectroscopic observations. In some spectroscopic binary systems, the spectral lines of both components are visible (the so-called double-line spectroscopic binaries, or $SB2$ thereinafter). However, in most cases, only one component can be seen in the spectra (single-line spectroscopic binaries, or $SB1$ from now on). In the absence of companion star spectra, the mass ratio of the stellar binary system can not be determined. This lack of information limits the astrophysical study and usefulness of this abundant family of stellar systems. The 9th Catalog of Spectroscopic Binary Orbits (hereafter SB9\footnote{Updated regularly, and available at \url{https://sb9.astro.ulb.ac.be/}.}, \citet{Pouret2004}) is the most comprehensive compilation of $SB1$ and $SB2$ binary systems which contains radial velocity (RV thereafter) amplitudes for all published binary systems for which it has been possible to fit a RV curve. As of January 12 2022, SB9 lists 4014 binary systems with measured radial velocities, of which 2704, or 67\%, are $SB1$ binaries.

The problem of estimating the orbital parameters of binary stellar systems has been studied for decades. The first proposed methods that solved the problem of astrometric orbital fitting on visual binaries used graphical and analytical formulations from physical models. These methods require a set of three complete and highly precise and homogeneous observations of relative position on the apparent orbit, of the form $(t,\rho,\theta)$ (epoch of observation, angular distance between primary (the brightest star), and the secondary, and position angle of the secondary with respect to the primary respectively), and a double areal constant obtained from additional data \citep{thiele1883neue}, an additional incomplete observation of the form $(t,\theta)$ \citep{cid1958necessary}, or an auxiliary angular variable that maps a set of feasible apparent orbits \citep{docobo1985analytic}. All of these methods require highly precise observations because they ignore the different levels of precision and uncertainties of the measurements. To obtain robust solutions to the orbital fitting problem considering multiple observations (with different levels of precision), optimization-based approaches have been then proposed. These family of optimization-based methods minimize a sum of weighted square errors between the physical model estimates and the observations by using, e.g., the Levenberg-Marquardt algorithm \citep{tokovinin1992frequency}, simulated annealing \citep{pourbaix1994trial}, the downhill simplex method \citep{macknight2004calculating}, among others. The strategies mentioned above focus on fitting positional (astrometric) observations of the orbit, ignoring the other important source of evidence obtained through spectroscopic measurements: the radial velocities of each system´s component. The first attempts to estimate the orbital parameters using positional and RV observations were made by fitting each source of information separately. They used the estimates obtained by fitting one of the sources of observations to determine some orbital parameters, where the remaining orbital parameters were estimated by fitting the other source of observations \citep{docobo1992adaptation,hummel1994very}. Unfortunately, this separate fitting strategy yields a sub-optimal determination of the complete set of orbital parameters. Indeed, it has been noted that there is no guarantee that the solutions obtained by both fittings were consistent. To avoid this issue, \cite{morbey1975synthesis} addressed the problem of determining the orbital parameters of a visual-spectroscopic binary system by fitting each source of observations jointly. They used a maximum likelihood estimation and a method based on Lagrange multipliers. More recently, the methods developed for fitting astrometric observations were extended to fit both astrometric and RV observations simultaneously \citep{pourbaix1998simultaneous}. Similar approaches were performed for the estimation of the individual masses in $SB1$ binaries. In this context, a supplementary observation of the system's parallax (using other techniques besides relative astrometry of the binary pair and spectroscopy for RV) was used as a fixed value within the estimation of the orbital parameters \citep{docobo2018visual}, or it was used used as an additional observation to perform the estimation jointly \citep{muterspaugh2010phases}. Overall, one of the major drawbacks of the optimization-based methods is that the obtained solution is entirely deterministic. Therefore, these strategies do not provide a reliable characterization of the estimation uncertainty. In contrast, the Bayesian-based approach is a powerful alternative, because it offers both an estimate (e.g., through the posterior mean, median, MAP, etc.) as well as a robust characterization of the uncertainty about the estimation (using the complete posterior distribution) of all the relevant parameters.

Bayesian methodologies  have been widely used in exoplanet orbit estimation. This approach computes the posterior distribution of the orbital parameters through Markov Chain Monte Carlo (MCMC) sampling. Many variants of the MCMC algorithm have been explored for a characterization of the posterior distributions in exoplanet research, such as the Metropolis-Hastings within Gibbs sampler \citep{ford2005quantifying}, the Parallel Tempering sampler \citep{gregory2005bayesian,gregory2011bayesian}, the Affine Invariant MCMC Ensemble sampler \citep{hou2012affine}, the Differential Evolution Markov Chain sampler \citep{nelson2013run}, and the Hamiltonian Monte Carlo \citep{hajian2007efficient}, among others. One of the most popular MCMC samplers in the statistical community is the No-U-Turn sampler, due to its efficiency on high-dimension problems and its capacity to express complex-correlated scenarios. The No-U-Turn sampler has been explored in the exoplanets context by, e.g., \citet{ji2017investigation} and \citet{shabram2020sensitivity}.

The Bayesian approach was also adapted to the astrometric orbital estimation of visual binaries \citep{burgasser2012discovery,sahlmann2013astrometric,lucy2014mass}. More recently, Bayesian estimation of orbital parameters considering both positional and RV sources of observations jointly has been addressed by \cite{mendez2017orbits, Clavet2019}, and \citet{Mendet2021}. This method uses the Metropolis-Hastings within Gibbs sampler method to provide the posterior distribution of $SB2$ visual-spectroscopic binaries. A similar Bayesian-based approach for the joint orbital parameters estimation on these types of binary system was also developed by \cite{lucy2018binary}, and the Hamiltonian Monte Carlo was used for the determination of the orbital parameters for a binary neutron star \citep{bouffanais2019bayesian}. However, to our knowledge, the Bayesian approach has not been explored with $SB1$ binaries for the task of estimating the individual masses of the system. In this context, the incorporation of suitable priors on specific observable parameters of the system has the potential to characterize the posterior distributions of the individual masses and all the orbital parameters. These informative priors can significantly enrich the analysis of these type of binary systems.

The present work introduces a Bayesian methodology based on the MCMC algorithm No-U-Turn sampler to address the orbital parameters inference problem in $SB1$ binaries, including a determination of the individual component masses. This methodology provides a precise characterization of the uncertainty of the estimates in the form of the joint posterior distribution of the orbital parameters. We address the lack of observations of the RV of the secondary star  by incorporating suitable prior distributions on some critical parameters of the system, such as its trigonometric parallax, and an estimate of the mass of the primary component (for further details, see Section~\ref{sec:massratiosb1}). The methodology is evaluated on several binary systems, providing an exhaustive analysis of the obtained results by comparing the estimated posterior distributions in different information scenarios (incorporating different observational sources and priors). The analysis performed consists not only in the comparison of the estimates and associated errors of the orbital parameters (as is commonly done in binary star research), but also analyzing the complete posterior distributions and the derived uncertainty in the orbit and RV spaces through the projection of the estimated posterior distributions in the observation space. We show that this last analysis (only possible through the Bayesian approach) allows a much richer and complete understanding of the associated uncertainties in the study of binary systems. The software is available on GitHub\footnote{\texttt{BinaryStars} codebase: \url{https://github.com/mvidela31/BinaryStars}.} under a 3-Clause BSD License.

The paper is organized as follows: In Section~\ref{sec:model} we introduce the basics of our Keplerian model and the Bayesian inference with priors. In Section~\ref{sec:benchmarks} we provide an experimental validation of our methodology by comparing our results to a group of benchmark $SB2$ systems treated as $SB1$ binaries. In Section~\ref{sec:application} we provide a proof of concept, by applying our Bayesian inference with priors to a group of $SB1$s for which we compute, for the first time, a combined astrometric-spectroscopic orbit, and an estimate of their mass ratio. Finally, in Section~\ref{sec:conclusions} we provide the main conclusions of our work.

\section{Bayesian inference in single-line spectroscopic binaries} \label{sec:model}

In this section we introduce the proposed Bayesian inference strategy for $SB2$ and $SB1$ binaries with a visual orbit. This section presents a brief explanation of the well-known Keplerian model adopted, the assumptions and re-parametrizations considered for the statistical modeling, and the algorithmic tools used for the inference process. For full details, the reader is referred to the Appendices.

\subsection{Keplerian orbital model} \label{sec:keplermodel}

Neglecting the effects of mass transfer and complex relativistic phenomena as well as the interference of other celestial bodies, the orbit of binary stellar systems is characterized by seven orbital parameters: the \textit{time of periastron passage} $T$, the period $P$, the orbital \textit{eccentricity} $e$, the orbital \textit{semi-major axis} $a$, the \textit{argument of periapsis} $\omega$, the \textit{longitude of the ascending node} $\Omega$, and the orbital \textit{inclination} $i$. The precise definition of these elements is given in Appendix~\ref{Sec1:binary}. The orbit of a binary system, i.e., the position in the plane of the sky $(X(t),Y(t))$ at a given time $t$, can be calculated through the following steps:
\begin{enumerate}
  \item Determination of the so-called eccentric anomaly $E(t)$ at a certain epoch t by numerically solving Kepler´s equation:
  \begin{equation}
      E(t)-e\sin E(t)=2\pi(t-T)/P.
  \label{eccentric_anomaly}
  \end{equation}
  
  \item Calculation of the auxiliary normalized coordinates $(x(t),y(t))$:
  \begin{equation}
    \begin{split}
    x(t)&=\cos{E(t)}-e, \\
    y(t)&=\sqrt{1-e^2}\sin{E(t)}.
\end{split}
  \end{equation}
  
  \item Determination of the Thiele-Innes constants:
  \begin{equation}
    \begin{split}
      A&=a(\cos{\omega}\cos{\Omega}-\sin{\omega}\sin{\Omega}\cos{i}),\\
      B&=a(\cos{\omega}\sin{\Omega}+\sin{\omega}\cos{\Omega}\cos{i}),\\
      F&=a(-\sin{\omega}\cos{\Omega}-\cos{\omega}\sin{\Omega}\cos{i}),\\
      G&=a(-\sin{\omega}\sin{\Omega}+\cos{\omega}\cos{\Omega}\cos{i}).
    \end{split}
    \label{eq:thiele_innes}
  \end{equation}
  
  \item Calculation of position in the apparent orbit $(X(t),Y(t))$:
  \begin{equation}
    \begin{split}
        X(t)&=Ax(t)+Fy(t),\\
        Y(t)&=Bx(t)+Gy(t).
    \end{split}
    \label{eq:pos}
  \end{equation}
\end{enumerate}

To compute the RV of each component of the binary system $(V_{1}(t),V_{2}(t))$ for primary and secondary respectively), it is necessary to incorporate additional parameters: the parallax $\varpi$, the \textit{mass ratio} of the individual components $q=m_2/m_1$ and the \textit{velocity of the center of mass} $V_{0}$. Thereby, the calculation of RV of each component of the system involves the following steps:
\begin{enumerate}
  \item Determination of the true anomaly $\nu(t)$ at a specific time $t$ using the eccentric anomaly $E(t)$ determined in (\ref{eccentric_anomaly}):
  \begin{equation}
    \tan{\dfrac{\nu(t)}{2}}=\sqrt{\dfrac{1+e}{1-e}}\tan{\dfrac{E(t)}{2}}.
  \label{eq:05}
  \end{equation}

  \item Calculation of the RV of the system´s individual components $(V_{1}(t),V_{2}(t))$:
  \begin{equation}
        V_{1}(t)=V_{0}+\dfrac{2\pi a_1 \sin{i}}{P\sqrt{1-e^2}}[\cos(\omega + \nu(t))+e\cos(\omega)],
  \label{eq:V_p}
  \end{equation}
  \begin{equation}
    V_{2}(t)=V_{0}-\dfrac{2\pi a_2 \sin{i}}{P\sqrt{1-e^2}}[\cos(\omega + \nu(t))+e\cos(\omega)],
    \label{eq:V_c}
  \end{equation}
  where $a_1=a''/\varpi\cdot q/(1+q)$, $a_2=a''/\varpi\cdot 1/(1+q)$ and $a''$ the semi-major axis in seconds of arc.
\end{enumerate}
Note that the determination of the true anomaly $\nu(t)$ in (\ref{eq:05}) presents no ambiguity, because this parameter has the same sign as the eccentric anomaly $E(t)$. Furthermore, the expression for the RV (\ref{eq:V_p}) and (\ref{eq:V_c}) contain the orbital parallax $\varpi$ explicitly, with the aim of exploding the full interdependence relations of the orbital parameters, avoiding to condense some of the parameters in those expressions as an independent parameter on the amplitude of the RV curve $K_1=(2\pi a_1\sin i)/(P\sqrt{1-e^2})$ in (\ref{eq:V_p}) and $K_2=(2\pi a_1\sin i)/(P\sqrt{1-e^2})$ in (\ref{eq:V_c}), as discussed in \cite{mendez2017orbits}.

If RV observations of each component ($V_1(t)$ and $V_2(t)$) are available ($SB2$ hereinafter), the combined model that describes the positional and RV observations is characterized by the set of orbital parameters $\vartheta_{SB2}=\{P,T,e,a,\omega,\Omega,i,V_0,\varpi,q\}$. However, if the RV observations of only one component are available ($SB1$ case), the parameters $q$ and $\varpi$ cannot be simultaneously determined.

\subsection{Bayesian model \& inference} \label{sec:bayesmodel}

In this section we introduce the Bayesian model used to perform the inference in $SB2$ and $SB1$ binary systems. These models will be presented as a suitable re-parametrization of the Keplerian orbital model introduced in Section \ref{sec:keplermodel}. We will assume that the positional and RV observations follow a Gaussian distribution, while we consider uniform priors on the model's parameters. These assumptions are design choices of the proposed methodology but not limitations, i.e., any other distributions for the observations and the priors could be assumed instead.

Let $\{t_i,X_i,Y_i\}_{i=1}^{n}$ be a set of $n$ positional observations of the companion star relative to the primary of a binary stellar system in rectangular coordinates and let $\{\bar{t}_i,V_{1i}\}_{i=1}^{n_1},\{\tilde{t}_i,V_{2i}\}_{i=1}^{n_2}$ be a set of $n_1$ and $n_2$ RV observations of the primary and companion stars, respectively. Given a parameter $\theta$ (fixed but unknown), we have that each observation (measurement) distributes as an independent Gaussian distribution centered in the value obtained by the Keplerian model's with a standard deviation equal to the corresponding observational error $\sigma_i$:
\begin{equation}\label{eq_obs_model}
    X_i\sim \mathcal{N}({X}_\theta(t_i),\sigma_i^2),\,\,
    Y_i\sim \mathcal{N}({Y}_\theta(t_i),\sigma_i^2),\,\,
    V_{1i}\sim \mathcal{N}({V}_{1\theta}(\bar{t}_i),\sigma_i^2),\,\,
    V_{2i}\sim \mathcal{N}({V}_{2\theta}(\tilde{t}_i),\sigma_i^2),
\end{equation}
where $({X}_\theta(t_i),{Y}_\theta(t_i))$ denotes the obtained position in the orbit at epoch $t_i$ for the parameter $\theta$, which follows Equation~(\ref{eq:pos}), and $({V}_{1\theta}(\bar{t}_i),{V}_{2\theta}(\tilde{t}_i))$ are the obtained RV of each star at time $\bar{t}_i$ and $\tilde{t}_i$ for the parameter $\theta$, which follows Equations~(\ref{eq:V_p}) and (\ref{eq:V_c}), respectively.

The positional $({X}_\theta(t_i),{Y}_\theta(t_i))$ and RV $({V}_{1\theta}(\bar{t}_i), {V}_{2\theta}(\tilde{t}_i))$ nominal values in (\ref{eq_obs_model}) are determined through the visual-spectroscopic binary system model presented in Section \ref{sec:keplermodel}. Therefore, the set of orbital parameters that characterizes the estimates is $\vartheta_{SB2}=\{P,T,e,a,\omega,\Omega,i,V_0,\varpi,q\}$ for the $SB2$ binaries with a visual orbit, and $\vartheta_{SB1}=\{P,T,e,a,\omega,\Omega,i,V_0,f/\varpi\}$ for the $SB1$ binaries with a visual orbit, where $f=q/(1+q)$ is the so called fractional-mass of the system. For $SB1$ systems, the parameter $f/\varpi$ condenses the pair of parameters $\varpi,q$ in (\ref{eq:V_p}) and (\ref{eq:V_c}), since they are not determinable due to the absence of $\{V_{2i}\}_{i=1}^{n_2}$ observations. The auxiliary parameter $f/\varpi$ has units of parsecs, since it is inversely proportional
to the parallax $\varpi$ which has units of seconds of arc. The range of $f/\varpi$ is $(0, d_{max}/2]$, considering that $q\in(0,1]$ and $\varpi>0$, with $d_{max}$ the maximum distance of observation determined by the measurement instrument.

Denoting the total set of observations as $\mathcal{D}=\{t_i,X_i,Y_i\}_{i=1}^{n}\cup\{\bar{t}_i,V_{1i}\}_{i=1}^{n_1}\cup\{\tilde{t}_i,V_{2i}\}_{i=1}^{n_2}$, the log-likelihood of the observations given the vector of parameters $\theta$ is expressed as:
\begin{equation}
\begin{split}
    \log p(\mathcal{D}|\theta)=&\sum_{i=1}^{n}\log \mathcal{N}(X_i|{X}_\theta(t_i),\sigma_i^2)+\sum_{i=1}^{n}\log \mathcal{N}(Y_i|{Y}_\theta(t_i),\sigma_i^2)\\&+\sum_{i=1}^{n_1}\log \mathcal{N}(V_{1i}|{V}_{1\theta}(\bar{t}_i),\sigma_i^2)+\sum_{i=1}^{n_2}\log \mathcal{N}(V_{2i}|{V}_{2\theta}(\tilde{t}_i),\sigma_i^2).
\end{split}
\label{eq:loglikelihood}
\end{equation}
The prior distribution of each orbital parameter is modeled as independent uniform priors on their valid physical range (defined in Appendix \ref{Sec1:binary}). Therefore, the prior distribution of the complete set of orbital parameters is expressed as:
\begin{equation}
    \log p(\theta)=\sum_{i=1}^{|\theta|}\log U(\min \Theta_i,\max\Theta_i),
\end{equation}
with $\theta_i\in\Theta_i,\forall i \in \{1,...,|\theta|\}$ and $\Theta_i$ the valid physical range of the orbital parameter $\theta_i$, and where $U$ is the density function that generates a uniform distribution of points between $\min \Theta_i$, and $\max\Theta_i$.

According to the Bayes theorem, the posterior distribution is proportional to the likelihood times the prior, i.e., $p(\theta|\mathcal{D})\propto p(\mathcal{D}|\theta)p(\theta)$, and therefore, the complete posterior distribution can be obtained through any sampling technique. For this purpose we use the state-of-art MCMC method No-U-Turn sampler \citep{hoffman2014no}.

The No-U-Turn sampler is an MCMC method that avoids the random-walk behavior and the sensitivity to correlated parameters of other commonly used MCMC algorithms (mentioned in Section \ref{sec:intro}) by incorporating first-order gradient information of the parameters space to guide the sampling steps (such as in the Hamiltonian Monte Carlo \citep{neal2011mcmc}) with an adaptive criterion for determining their lengths. This method has been widely adopted by the statistical community in recent years due to its computational efficiency, effectiveness in high-dimensional problems, and theoretical guarantees, but to the best of our knowledge, it has not been applied in the context of binary stellar systems. A further explanation about the theory behind the No-U-Turn sampler method and the implementation details in the context of binary stellar systems is presented in Appendix \ref{app:hmc&nuts}.

\subsection{Design considerations}

For the inference process, the reparametrization of the time of periastron passage $T$ proposed by \cite{lucy2014mass} is adopted in this work. The author suggests to sample from $T'=(T-t_0)/P$ instead of $T$, since it is beneficial to sample from a well-constrained parameter space, restricting the range of the time of periastron passage to $[0, 1)$. On the other hand, reparametrizations that involve a dimensionality reduction of the parameters space (e.g., \cite{mendez2017orbits}) or transformations of well-constrained parameters (e.g., \cite{ford2005quantifying}) were avoided since it is shown to have a negative impact on the correlation of the obtained  parameters, considerably hindering the exploration of the parameters space through first-order gradient information. This design choice increases the computational cost of the gradient calculation required by the No-U-Turn sampling routine.

Finally, as the parameter space of binary stellar systems (and especially in hierarchical stellar systems, an application which will be presented in a forthcoming paper) is highly correlated, many authors recommend choosing a starting point that lies in areas of high posterior mass (further details are presented in Appendix \ref{app:hmc}) to avoid miss-convergence issues of the sampling process. In this paper, we adopt the quasi-Newton optimization method L-BFGS \citep{liu1989limited} to find a good starting point that alleviates convergence issues of the Bayesian inference, since it allows to perform the optimization using any prior distribution on the parameters. In contrast, other commonly used optimization methods in the astronomical field are restricted to least-squares problems (e.g., the Levenberg-Marquardt algorithm, \citet{more1978levenberg}), limiting considerably the family of prior distributions that can be used.

\subsection{Determining the mass ratio in single-line binaries} \label{sec:massratiosb1}

$SB1$ binaries with a visual orbit are an abundant type of stellar object. Unfortunately, the lack of observations of the RV of the companion (i.e., the $\{V_{2i}\}_{i=1}^{n_2}$ observations in Equation~(\ref{eq:loglikelihood})) does not allow a determination of the mass ratio of the system and hence their individual masses (since the visual orbit provides the mass sum of the system). This limits the use of this type of $SB1$ binaries in astronomical studies and justifies the relevance of the much less abundant $SB2$ binaries. However, as it will be shown below, suitable additional (prior) information about the system can be incorporated to estimate (i.e., resolve with good precision) the individual masses for $SB1$ binaries, such as the system´s trigonometric parallax (e.g., from Gaia), and an estimation of the mass of the primary object, e.g., via its spectral type and luminosity class from low-resolution spectra.

As presented in Section~\ref{sec:bayesmodel}, the Bayesian model for $SB1$ systems is characterized by the set of orbital parameters $\vartheta_{SB1}=\{P,T,e,a,\omega,\Omega,i,V_0,f/\varpi\}$. Here, the parameter $f/\varpi$ replaces the individual parameters $\varpi$ and $q$, since they are non-identifiable in absence of RV observations of the companion object, i.e., there exist different values of the pair $\varpi,q$ that map on the same value of the posterior distribution, preventing a determination of their values individually. The non-identifiability of the mass ratio implies that the individual masses of this type of binary systems can not, in principle, be determined. As the individual masses are relevant for the study of these systems, many authors addresses the non-identifiability problem on $SB1$ systems by incorporating information about the parallax parameter from external measurements (independent from relative astromery and RV observations). This information is usually added in a posteriori manner, i.e., once the initial set of orbital parameters $\vartheta_{SB1}$ is estimated from the observations (e.g., \citet{docobo2018visual}). An important disadvantage of this approach is that it ignores the influence of the external information (in this case a trigonometric parallax) on the estimation, which can lead to a lack of self-consistency. Moreover, if the additional information about parallax is highly uncertain or biased, the estimated mass ratio - given the previously estimated orbital parameters and the adopted parallax - can be out of the valid physical range $q\in(0,1]$. Another common approach in the literature to address the non-identifiability problem is incorporating this external information in a prior manner as an additional observation to fit (e.g., \citet{muterspaugh2010phases}). The main disadvantage of this approach is that the estimation is done deterministically (through optimization-based methods), poorly characterizing the uncertainty of the incorporated information and its impact on the orbital parameters estimation. Accordingly, our approach addresses these disadvantages by providing an all at-once self-consistent solution, while considering the uncertainty of all the information incorporated (in a complete probabilistic manner) and its impact in the orbital parameters estimation.

In order to overcome the non-identifiability problem of the mass ratio $q$ in $SB1$ binaries, two different approaches are proposed in this work: one based on the incorporation of prior information about the trigonometric parallax $p(\varpi)$, and the other based on the incorporation of prior information about the derived parameter (from the set of orbital parameters $\theta$) corresponding to the mass of the primary object $p(m_1|\theta)$. These two sources of information are commonly available for $SB1$s, and are thus natural choices for this exercise.

The first proposed approach makes use of the $SB2$ orbital model described in Section \ref{Sec1:Combined}, characterized by the set of parameters $\vartheta_{SB2}=\{P,T,e,a,\omega,\Omega,i,V_0,\varpi,q\}$, with the incorporation of an informative prior distribution on the parallax $p(\varpi)$. Specifically, $p(\varpi)$ is modeled as a normal distribution with mean and standard deviation determined respectively by the measurement $\bar{\varpi}$ and error $\sigma_{\varpi}$ of the (typically) trigonometric parallax. Nowadays these measurements are precisely determined by {\it Gaia} for most of the observed systems \citep{2000A&AS..143....9W,prusti2016gaia,brown2018gaia}. The addition of the prior $p(\varpi)=\mathcal{N}(\bar{\varpi},\sigma_{\varpi}^2)$ makes the model soft-identifiable, i.e., the indeterminable parameters turn out to be determinable (in a probabilistic sense) through the incorporation of suitable priors. This allows a determination (estimation with good precision) of the mass ratio $q$ of the system.

Alternatively, since only the spectral lines of the primary object of $SB1$ systems are visible, these observations can be used to estimate the mass of the primary object $m_1$ through empirical relations (see e.g., \citet{Abuset2020}), providing additional external information that allow us to alleviate the non-identifiability problem of the mass ratio as well. In fact, the primary object mass $m_1$ can be calculated from the set of orbital parameters $\vartheta_{SB2}$ using the third law of Kepler:
\begin{equation}
    m_1|\theta=\Big(\dfrac{a}{\varpi}\Big)^3\cdot \dfrac{1}{P^2(1+q)}.
    \label{eq::m1}
\end{equation}
Therefore, by assuming that the generated parameter $m_1|\theta$ (random variable) is independent from the other sets of observations (random variables), and that it follows a Gaussian prior distribution with a mean $\bar{m}_1$ and a standard deviation $\sigma_{m_1}$ (from measurement/calibration uncertainty), the distribution $p(m_1|\theta)=\mathcal{N}(\bar{m}_1,\sigma_{m_1}^2)$ can be directly incorporated into the likelihood term in Equation~(\ref{eq:loglikelihood}) to determine the posterior distribution $\log p(\theta | \mathcal{D})$.

Finally, these two approaches can be incorporated simultaneously in the inference routine, i.e., incorporating a prior $p(\varpi)$ and also adding the term $p(m_1|\theta)$ in the likelihood computation. In what follows we present the results of applying this methodology to a set of benchmark systems, to asses the effectiveness of our estimations.

\section{Experimental validation} \label{sec:benchmarks}

We perform the inference of the orbital parameters on eight $SB2$ binaries with a visual orbit, which will be used in subsequent sub-sections as benchmark (full-information) systems. We compare the estimated posterior distributions and their projection on the observations space, which allows us to carry-out an analysis and a quantification of the change in uncertainty due to the absence of RV of the companion object.

Additionally, a comparative study of the estimated posterior distributions of our benchmark systems, treated as $SB1$ binaries, is presented for three cases: incorporating prior information on the parallax alone, on the mass of the primary object alone, and on both the parallax and the mass of the primary object. These results are also compared to the inference from the $SB2$ full-information (benchmark) scenario, with a special focus on the estimation of the mass ratio parameter $q$. Our selected benchmarks, without being exhaustive off all possible orbital configurations, span a wide range in their $q$ values $\in(0.36,0.96)$.

In the case of our tests in the $SB1$ scenario with priors, the adopted trigonometric parallax, primary object's spectral type (both from SIMBAD, \citet{SIMBAD}, unless otherwise noted), and corresponding primary object mass for each of the benchmark systems are presented in Table~\ref{tab:SB2/SB2_plx&m1}. The mass has been derived from the mass versus spectral type and luminosity class calibrations provided by \citet{habets1981empirical}, \citet{1981Ap&SS..80..353S}, \citet{schmidt1982landolt}, \citet{aller1996landolt}, \citet{2006ima..book.....C}, \citet{2008oasp.book.....G}, and \citet{Abuset2020}. The dispersion in mass comes from assuming a spectral type uncertainty of $\pm$ one sub-type which is customary in spectral classification.

\small
\begin{table}[h]
\centering
\caption{Reported trigonometric parallax, primary object's spectral type, and derived primary mass for benchmark $SB2$ systems.}
\begin{tabular}{ccccc}
\hline\hline
HIP~\# & Discovery  &    $\varpi$ & SpTyp & $m_1$\\
       & Designation & [mas] & &   [M$_\odot]$ \\
\hline
677    & MKT11Aa,Ab & $33.62\pm0.35$ &  B8IV &   $3.09\pm0.45$ \\
5531   & HDS155     & $16.6508\pm0.2131$ & G0V &   $1.047\pm0.025$ \\
14157  & HJL1114    & $19.5354\pm0.0557$ & K0V &   $0.851\pm0.040$ \\
20601  & BU1177 & $17.3199\pm0.1297$ & G8V &   $0.912\pm0.020$ \\
89000  & YSC132Aa,Ab & $21.4561\pm0.0937$ &  F6V &   $1.167\pm0.035$ \\
108917 & MCA69Aa,Ab & $32.170\pm0.621$(a) & AV &   $1.740\pm0.075$ \\
111170 & CHR111 & $39.1884\pm0.6249$ & F8V  &   $1.096\pm0.025$ \\
117186 & HJL1116 & $8.2150\pm0.0438$ & F2  &   $1.349\pm0.050$ \\
\hline\hline
(a): From Gaia DR2.
\end{tabular}

\label{tab:SB2/SB2_plx&m1}
\end{table}
\normalsize

\subsection{Initial validation of our methodology and comparisons with previous studies} \label{sec:initvalid}

In this sub-section we compare the results of our code on the benchmark $SB2$s mentioned before, with the equivalent of their $SB1$ counterparts, by omitting altogether the RV observations of the companion object, as well as with previous studies of these systems treated as $SB2$. The obtained estimates and their uncertainties are compared in the parameters space through the visualization of the posterior marginal distributions. The trigonometric parallax and primary object's mass presented in Table~\ref{tab:SB2/SB2_plx&m1} are visualized as error bars ($\pm 2\sigma$) in their corresponding marginal posterior distribution plot for comparison. We emphasize that, in the case of the $SB2$s solutions, the set of parameters to estimate, $\vartheta_{SB2}=\{P,T,e,a,\omega,\Omega,i,V_0,\varpi,q\}$, include the parallax $\varpi$ rather than a value known in advance: This is the so-called orbital parallax, based solely on the orbital motion of the pair, and which is different and independent from the trigonometric parallax, thus putting into practice the somewhat unexplored possibility of utilizing combined data (i.e., astrometry and radial velocity) to estimate parallax-free distances \citep{Pour2000,Mas2015}. The dynamically self-consistent orbital parallaxes may or may not be similar to trigonometric parallaxes, as discussed further below.

We also compute the posterior distribution on the observations space, through the projection of $1000$ randomly selected samples of the posterior distribution on the observation space, drawing trajectories from the time of the first observations $t_0$ to the first completion of the orbit $t_0+P$. The maximum a posteriori MAP estimate (the most probable sample of the posterior distribution) and the $95\%$ high posterior density interval HPDI (defined as the narrowest interval that contains 95\% of the posterior distribution, including the mode) are summarized in Table~\ref{tab:SB2/SB1}. For comparison we also show, in the first line for each object, the values reported for the same parameters by the Sixth Catalog of Orbits of Visual Binary Stars maintained by the US Naval Observatory (\citet{WDSCat2001}, hereafter Orb6\footnote{Updated regularly, and available at \url{https://www.usno.navy.mil/USNO/astrometry/optical-IR-prod/wds/orb6}.}) and by SB9, or from references therein.

The whole inference process is performed through the simulation of $10000$ samples of the respective posterior distributions (discarding the first half for warm-up) on $4$ independent Markov chains using the No-U-Turn sampler algorithm mentioned before, and explained in more detail in Appendix~\ref{app:hmc&nuts}. Each chain starts from an initial point determined by the results of the quasi-Newton optimization method L-BFGS (see previous section).

To avoid redundancy in the analysis, we select three of the eight most representative systems: HIP 89000, HIP 111170, and HIP 117186.

\subsubsection{HIP 89000}

\begin{figure}[!h]
    \centering
    \includegraphics[width=\textwidth]{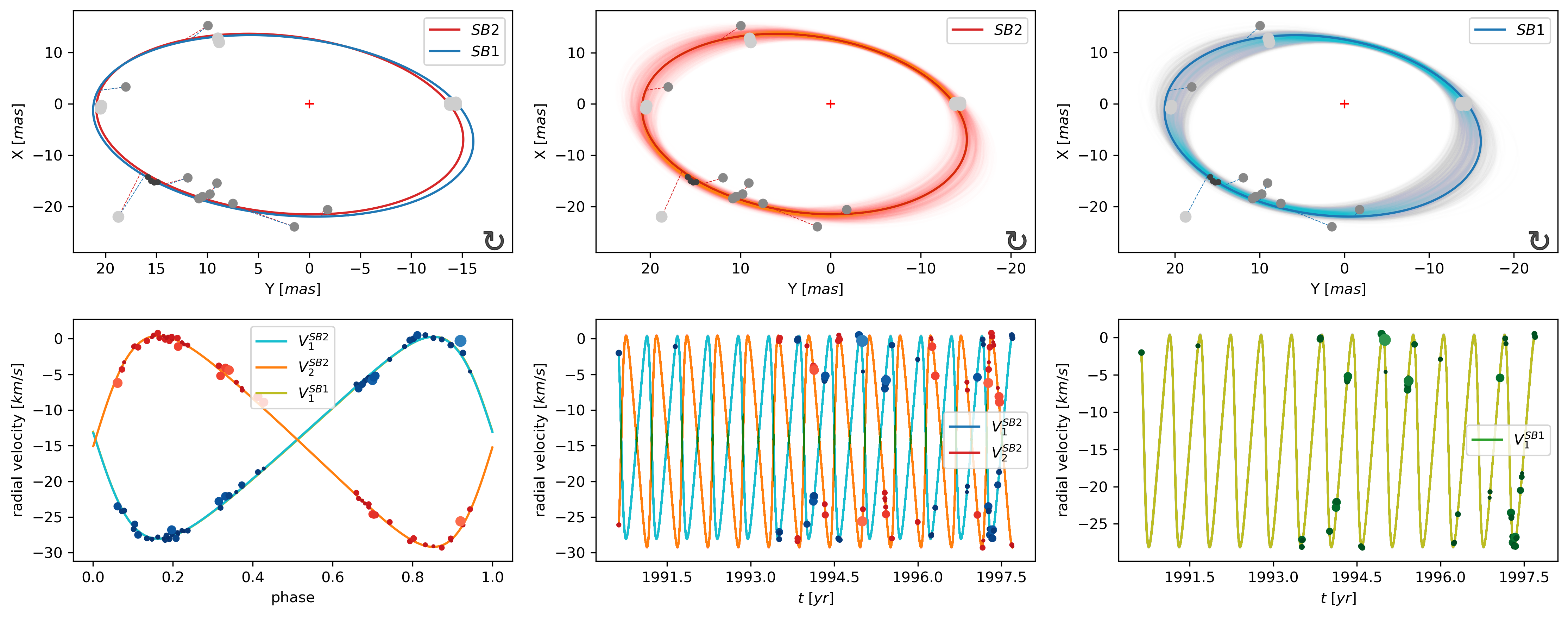}
   \caption{Estimated orbit and RV curves of the HIP 89000 binary system. First column: MAP point estimate projection of the posterior distribution for $SB2$ and $SB1$ cases (note that in the lower left panel, the curves for the $SB2$ and $SB1$ cases overlap). Second and third columns: Projected posterior distribution of the $SB2$ and $SB1$ cases respectively.}
   \label{fig:SB2/YSC132AaAb_obs}
\end{figure}

The system HIP 89000 (discovery designation YSC132AaAb) is a $SB2$ binary presented and solved most recently by \citet{mendez2017orbits}. The available data consists of relatively low precision interferometric observations mostly concentrated around the apoastron passage, but with abundant and precise observations of RV of both components. The observations and their errors are visualized in Figure~\ref{fig:SB2/YSC132AaAb_obs}. This systems has the highest $q$ value of our selected benchmarks, $\sim 0.96$.

\begin{figure}[!h]
    \centering
    \includegraphics[width=\textwidth]{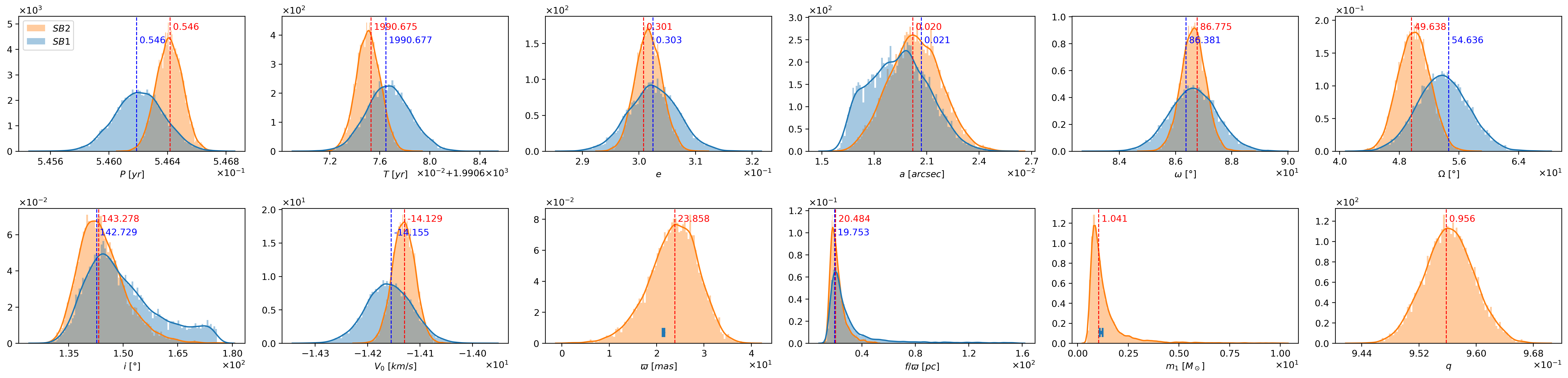}
    \caption{Marginal posterior distribution for the orbital parameters of the HIP 89000 binary system in the $SB2$ and $SB1$ cases. In each panel, we indicate the MAP estimate for the $SB2$ and its $SB1$ analog. Of course, at this stage, we can estimate the orbital parallax, the mass for the primary, and the mass ratio only in the $SB2$ case. The small dot with $2\sigma$ error bars in the PDFs for $\varpi$ and $m_1$ indicate the values shown in Table~\ref{tab:SB2/SB2_plx&m1}. For this particular case, the good correspondence between the orbital and trigonometric parallax is evident, this was documented already by \citet{mendez2017orbits}.}
    \label{fig:SB2/YSC132AaAb_params}
\end{figure}

The estimated posterior distributions of all parameters, presented in Figure~\ref{fig:SB2/YSC132AaAb_params},  show a Gaussian shape with the exception of the parameters $i,f/\varpi$ and $m_1$, whose distributions show a large positive skewness. Slight differences in the means and significant differences in the dispersion of the posterior distribution are observed between the $SB2$ and $SB1$ cases. As expected, the $SB2$ case offers less posterior uncertainty (more concentration) than its $SB1$ analog in all the orbital parameters. This reflects the significant impact in this case of incorporating observations of both RV instead of that for only the primary component.
 
The estimated posterior distribution in Figure~\ref{fig:SB2/YSC132AaAb_params} projected in the observation space is presented in Figure~\ref{fig:SB2/YSC132AaAb_obs}. The first column shows the MAP estimate (curve) in the observation space, and the second and third column show the projection of $1000$ uniformly selected samples (curves in this case) of the posterior distribution for the SB2 and SB1 cases, respectively. A slight difference is observed between MAP orbits of the $SB2$ and $SB1$ cases. In contrast, there is no appreciable difference between the MAP curves in the RV between both cases. The posterior projection in the orbit space of the $SB2$ case shows the lowest uncertainty in the apoastron, which is the zone that has more observations. The zones of the orbit with higher uncertainties are located in between the peri- and apoastron, which coincides with the lack of precise observations there. Notably, the periastron preserves the same amount of projected uncertainty of the apoastron despite the fact that this zone has no observations. This non-intuitive behaviour shows the relevance of analyzing the projected posterior distribution in the observations space, where the obtained uncertainty depends not only on the location of the observations, but also on the parametric configuration of the system itself. The posterior projection in the RV space on the $SB2$ case shows very small uncertainty along all the curves. This is explained by the high number of data points for both components. The projected posterior distribution for the $SB1$ in RV space exhibits no obvious differences with respect to the $SB2$ case. In contrast, the projection of the posterior distribution on the orbit space presents more fuzziness for the $SB1$ in comparison to the $SB2$, which is explained by the differences in their respective estimated posterior distributions in the parameter space (see Figure~\ref{fig:SB2/YSC132AaAb_params}).

\subsubsection{HIP 111170}
\begin{figure}[!h]
    \centering
    \includegraphics[width=\textwidth]{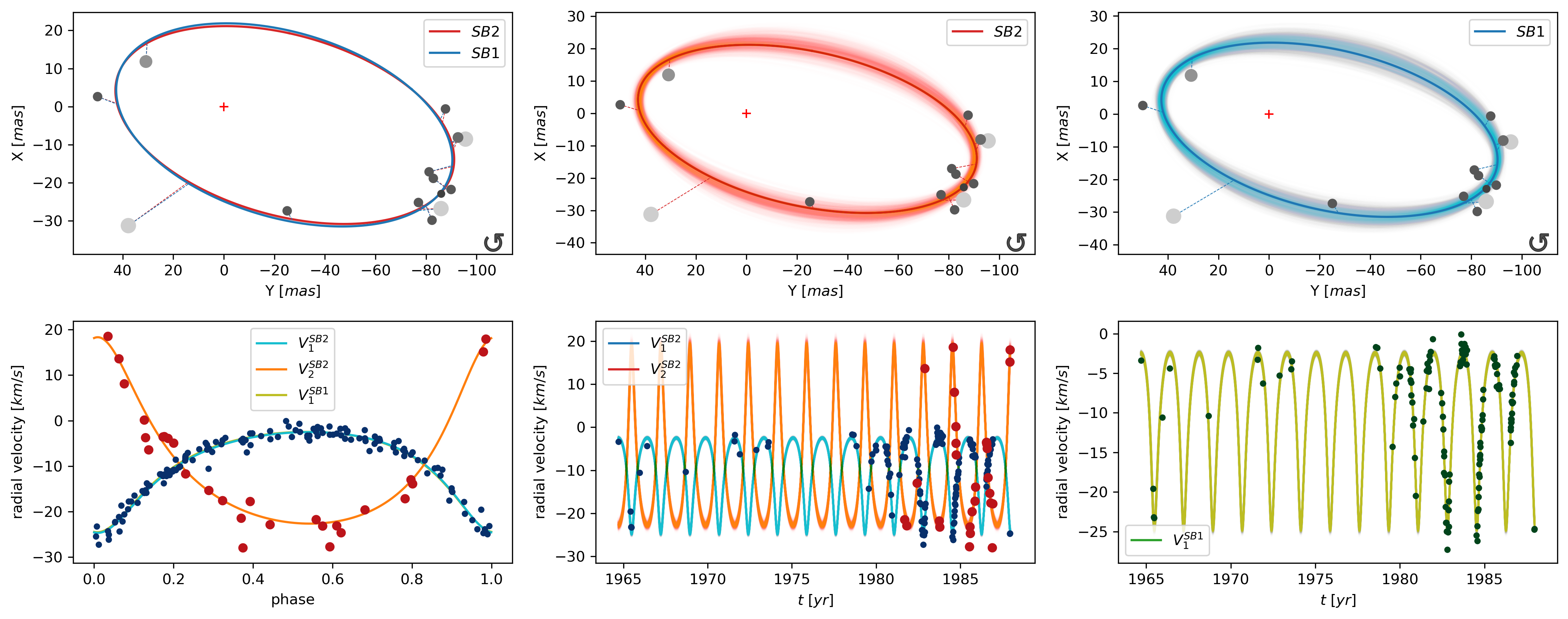}
    \caption{Similar to Figure~\ref{fig:SB2/YSC132AaAb_obs}, but for the HIP 111170 binary system.}
    \label{fig:SB2/CHR111_obs}
\end{figure}

The system HIP 111170 (discovery designation CHR111) is a $SB2$ binary presented and solved most recently by \citet{mendez2017orbits}, with an intermediate value of $q\sim0.54$. The available data consists in astrometric observations mostly concentrated around apoastron passage with a few observations scattered on the rest of the orbit, but with abundant observations of RV of both components. The observations and their errors are visualized in Figure~\ref{fig:SB2/CHR111_obs}.

\begin{figure}[!h]
    \centering
    \includegraphics[width=\textwidth]{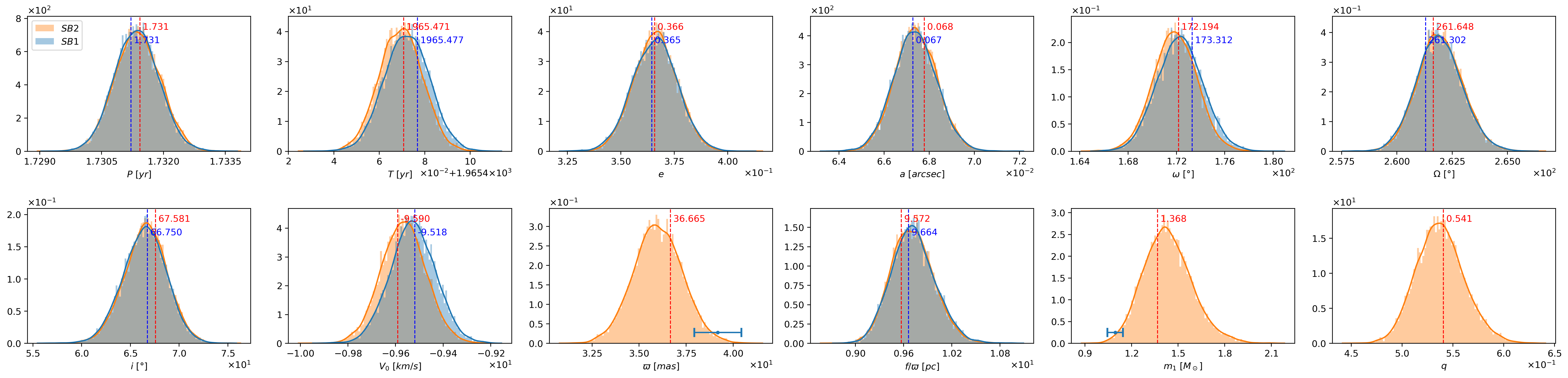}
    \caption{Similar to Figure~\ref{fig:SB2/YSC132AaAb_params}, but for the HIP 111170 binary system. In this case there is a more noticeable difference between the orbital and trigonometric parallaxes, also documented in \citet{mendez2017orbits}, who used the pre-Gaia trigonometric parallax $39.35 \pm 0.70$~[mas].}
    \label{fig:SB2/CHR111_params}
\end{figure}

The estimated posterior distributions shown in Figure~\ref{fig:SB2/CHR111_params} present a Gaussian shape with almost no differences in mean value and dispersion between the $SB2$ and $SB1$ cases. The $SB2$ case is slightly more constrained than its $SB1$ counterpart, but the difference is almost negligible (see also Table~\ref{tab:SB2/SB1}). This reflects the expected precision gain due to the incorporation of both RV observations instead of only the RV for the primary object. However, due to the high precise RV observations and relatively good coverage of the astrometrical observations, the information gain is minimal.

Similarly to the case of HIP 89000, the projection of the estimated posterior distribution in the observation space is presented in Figure~\ref{fig:SB2/CHR111_obs}. In this case, a small (almost negligible) difference is observed between the MAP orbit for the $SB2$ and $SB1$ cases. There is no perceived difference between the MAP posterior projections in the RV space between both cases. However, the posterior projection in the orbit space case shows less uncertainty than in the apoastron, which is the zone that has more observations, like in the case of HIP 89000. Similar uncertainties are noticed in the opposite zone -the periastron- where only two observations are available. The zones of the orbit with a higher uncertainty are located between peri- and apoastron, which coincides with the lack of observations there. The posterior projection in RV space of the $SB2$ case shows very small uncertainty, attributed to the dense phase coverage for both components, save for a slight increase of the uncertainty on the RV curves of both components near their maximum and minimum amplitude. The estimated posterior distributions of the $SB1$ case in both, the orbit and RV spaces, present no appreciable differences with respect to the $SB2$ case. This is consistent with the similarities observed in the parameters space, mentioned in the previous paragraph.
 
\subsubsection{HIP 117186}

\begin{figure}[!h]
    \centering
    \includegraphics[width=\textwidth]{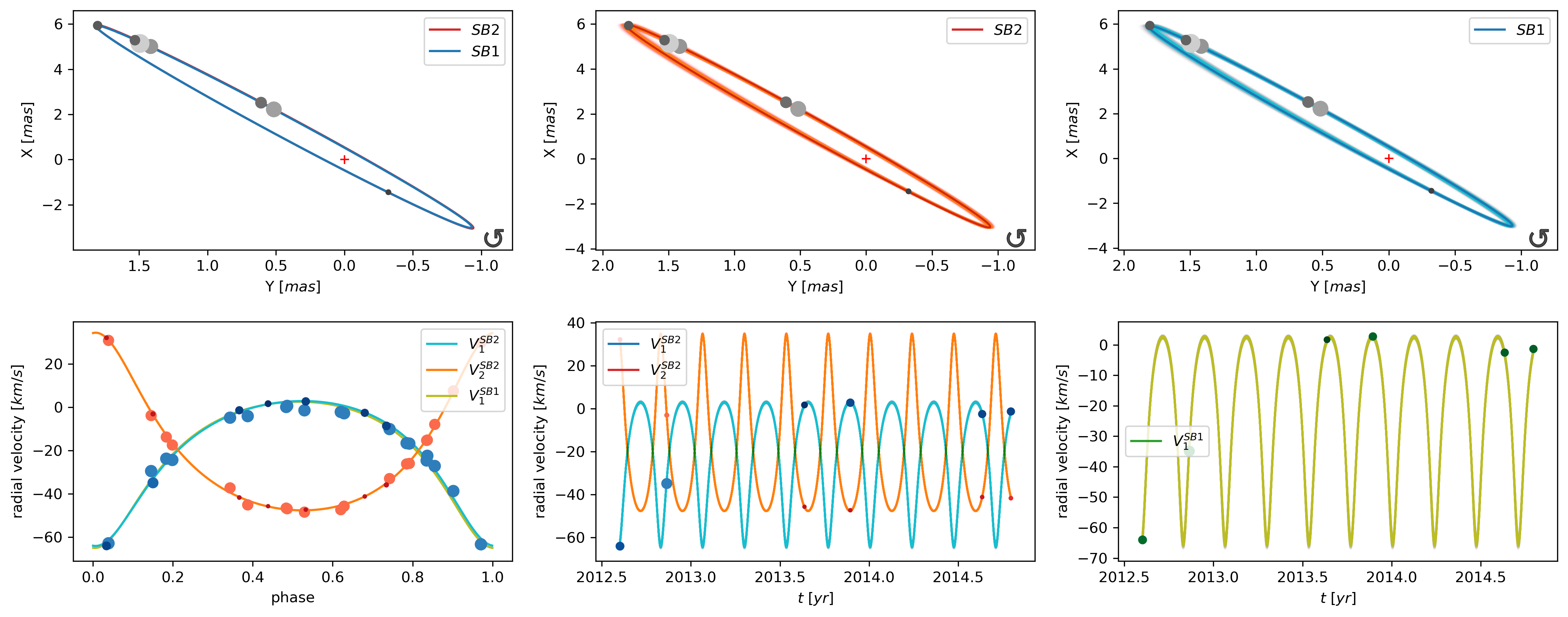}
    \caption{Similar to Figure~\ref{fig:SB2/CHR111_obs}, but for the HIP 117186 binary system.}
    \label{fig:SB2/HIP117186_obs}
\end{figure}

The system HIP 117186 (discovery designation HJL1116) is a $SB2$ binary presented and solved in \cite{halbwachs2016masses}. The available data consists in highly precise astrometric observations dispersed along all the orbit, along with  with abundant and precise observations of RV for both components. The observations and their errors are visualized in Figure~\ref{fig:SB2/HIP117186_obs}.

\begin{figure}[!h]
    \centering
    \includegraphics[width=\textwidth]{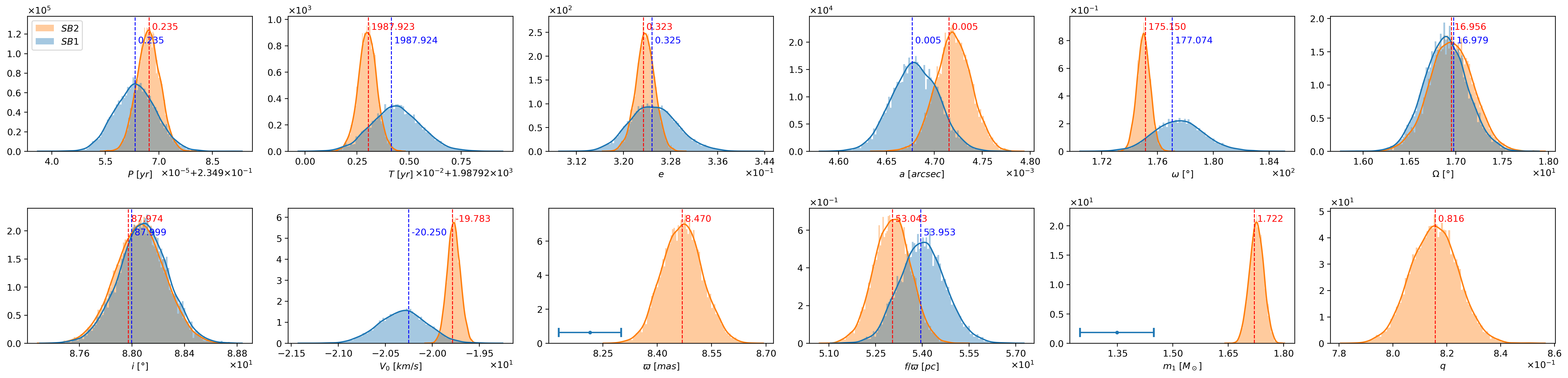}
    \caption{Similar to Figure~\ref{fig:SB2/CHR111_params}, but for the HIP 117186 binary system. In this case we appreciate a large discrepancy between the orbital and trigonometric parallaxes. The impact of this difference is further discussed in Section~\ref{sec:priorstosb1}.}
    \label{fig:SB2/HIP117186_params}
\end{figure}

The estimated posterior distributions are presented in Figure~\ref{fig:SB2/HIP117186_params}. All the posterior marginal distributions exhibit a Gaussian shape. There are noticeable differences in the means and significant differences in the dispersions of the posterior distribution between the SB2 and the SB1 case. As before, the SB2 case offers less posterior uncertainty than the SB1 case in almost all the orbital parameters. The exceptions to this rule are the angular parameters $\Omega$ and $i$ (usually mostly constrained by astrometric observations on visual binaries), where the dispersion between both cases are almost the same. As in the previous two cases, the evident differences on the dispersion of the posterior distribution between the SB2 and SB1 cases reflect the impact of the incorporation of RV observations from both components on the estimated uncertainties.
 
The projection of the estimated posterior distribution in the observation space is presented in Figure~\ref{fig:SB2/HIP117186_obs}. As in the other two cases, a slight difference is observed between MAP projection on the orbit between the SB2 and SB1 cases. There is no difference between the MAP posterior projection in RV space between both cases. The posterior projection in the orbit space for the $SB2$ case shows small uncertainty in the zones with observations and a slight increase of uncertainty in the other zones. Remarkable, the uncertainty throughout the whole orbit is negligible, which is attributed to the extremely high precision and orbital coverage of the positional observations, even considering that only seven observations are available, and that this is our most inclined system (with $i \sim 88^{\circ}$). The posterior projection in RV space for the SB2 case shows also very small uncertainty along the entire curves, which coincides with the large number of observations of RV for both components. Finally, the projected posterior distribution of the $SB1$ counterpart in the observations space shows no appreciable differences with respect to the $SB2$ case.

\small
\begin{sidewaystable}[h]
\centering
\caption{MAP estimates and 95\% HDPIs from the marginal posterior distributions of orbital parameters considering the RV of both components (SB2), and the RV of the primary component only (SB1).}

\resizebox{\textwidth}{!}{\begin{tabular}{ccccccccccccccc}
\hline\hline
     HIP~\# &      Case &               System &                 $P$ &                       $T$ &                     $e$ &              $a$ &                   $\omega$  &                $\Omega$ &                     $i$  &                $V_0$&              $\varpi$ &                $f/\varpi$ &         $m_1$  &                     $q$ \\
 & & & [yr] & [yr] &  & [arcsec] &   [\textdegree] &  [\textdegree] & [\textdegree] & [km/s] & [mas] & [pc] &  [M$_\odot]$ & \\
\hline
   677 & $SB2$ &    (a) &                   0.265 &                        1988.583 &                   0.535 &                   0.024 &                            79.000 &                       104.400 &                       105.600 &                           -11.000 &                     33.620 &                     10.023 &                   3.441 &                   0.508 \\
          & $SB2$ &    (b) & $0.265_{0.265}^{0.265}$ & $1988.847_{1988.842}^{1988.851}$ & $0.535_{0.470}^{0.589}$ & $0.024_{0.022}^{0.025}$ &    $78.339_{73.740}^{81.112}$ & $104.952_{101.653}^{108.823}$ & $106.172_{102.450}^{109.408}$ &  $-12.465_{-14.043}^{-9.567}$ & $32.796_{29.447}^{35.930}$ &  $10.500_{9.113}^{12.139}$ & $3.697_{2.983}^{4.461}$ & $0.525_{0.449}^{0.600}$ \\
          & $SB1$ &    (b) & $0.265_{0.264}^{0.265}$ & $1988.848_{1988.843}^{1988.852}$ & $0.529_{0.461}^{0.599}$ & $0.024_{0.022}^{0.026}$ &    $77.504_{73.201}^{81.426}$ & $104.112_{100.809}^{108.109}$ & $106.075_{102.329}^{109.601}$ &  $-10.194_{-13.237}^{-7.576}$ &                          — &  $10.952_{9.117}^{12.265}$ &                       — &                       — \\
\hline
  5531 & $SB2$ &    (a) &                   7.454 &                        2002.952 &                   0.743 &                   0.083 &                       215.600 &                       151.300 &                        50.900 &                        -9.570 &                     16.400 &                     29.048 &                   1.222 &                   0.910 \\
          & $SB2$ &    (b) & $7.455_{7.450}^{7.458}$ & $1995.494_{1995.490}^{1995.502}$ & $0.744_{0.741}^{0.746}$ & $0.082_{0.082}^{0.083}$ & $215.006_{214.532}^{215.803}$ & $152.001_{151.314}^{152.497}$ &    $50.669_{49.926}^{51.606}$ &    $-9.526_{-9.596}^{-9.477}$ & $16.201_{15.952}^{16.572}$ & $29.379_{28.751}^{29.857}$ & $1.244_{1.176}^{1.294}$ & $0.908_{0.894}^{0.925}$ \\
          & $SB1$ &    (b) & $7.454_{7.450}^{7.458}$ & $1995.495_{1995.488}^{1995.500}$ & $0.743_{0.741}^{0.746}$ & $0.083_{0.082}^{0.083}$ & $214.978_{214.050}^{215.449}$ & $152.032_{151.580}^{152.828}$ &    $50.782_{49.865}^{51.528}$ &    $-9.500_{-9.551}^{-9.417}$ &                          — & $29.345_{28.914}^{30.048}$ &                       — &                       — \\
\hline
 14157 & $SB2$ &    (a) &                   0.119 &                        1999.844 &                   0.759 &                   0.006 &                       174.690 &                        19.141 &                        92.240 &                        30.743 &                     19.557 &                     24.193 &                   0.982 &                   0.898 \\
          & $SB2$ &    (b) & $0.119_{0.119}^{0.119}$ & $1993.795_{1993.795}^{1993.796}$ & $0.761_{0.758}^{0.763}$ & $0.006_{0.006}^{0.006}$ & $175.114_{174.829}^{175.564}$ &    $19.437_{19.092}^{19.924}$ &    $91.532_{90.104}^{92.968}$ &    $30.682_{30.573}^{30.780}$ & $19.436_{19.279}^{19.604}$ & $24.367_{24.145}^{24.616}$ & $1.003_{0.982}^{1.018}$ & $0.900_{0.891}^{0.910}$ \\
          & $SB1$ &    (b) & $0.119_{0.119}^{0.119}$ & $1993.795_{1993.795}^{1993.796}$ & $0.761_{0.758}^{0.765}$ & $0.006_{0.006}^{0.006}$ & $175.044_{174.479}^{175.503}$ &    $19.530_{19.089}^{19.905}$ &    $91.454_{90.131}^{92.966}$ &    $30.695_{30.569}^{30.841}$ &                          — & $24.401_{24.146}^{24.652}$ &                       — &                       — \\
\hline
 20601 & $SB2$ &    (a) &                   0.428 &                        2013.942 &                   0.851 &                   0.011 &                       202.026 &                       340.526 &                       103.138 &                        41.623 &                     16.702 &                     25.486 &                   0.980 &                   0.741 \\
          & $SB2$ &    (b) & $0.428_{0.428}^{0.428}$ & $1971.983_{1971.983}^{1971.983}$ & $0.852_{0.849}^{0.854}$ & $0.011_{0.011}^{0.012}$ & $202.260_{201.535}^{203.109}$ & $340.812_{339.649}^{341.814}$ & $103.636_{101.114}^{105.151}$ &    $42.097_{41.940}^{42.251}$ & $16.710_{16.429}^{16.991}$ & $25.677_{25.178}^{26.099}$ & $0.984_{0.947}^{1.013}$ & $0.752_{0.742}^{0.759}$ \\
          & $SB1$ &    (b) & $0.428_{0.428}^{0.428}$ & $1971.983_{1971.983}^{1971.983}$ & $0.852_{0.849}^{0.855}$ & $0.011_{0.011}^{0.012}$ & $202.428_{201.639}^{203.518}$ & $340.506_{339.570}^{341.959}$ & $102.627_{100.916}^{105.761}$ &    $42.225_{41.922}^{42.433}$ &                          — & $25.623_{25.163}^{26.184}$ &                       — &                       — \\
\hline
 89000 & $SB2$ &    (a) &                   0.546 &                        1990.675 &                   0.302 &                   0.019 &                        86.650 &                        51.189 &                       146.167 &                       -14.131 &                     21.308 &                     22.938 &                   1.214 &                   0.956 \\
          & $SB2$ &    (b) & $0.546_{0.546}^{0.547}$ & $1990.675_{1990.673}^{1990.677}$ & $0.301_{0.297}^{0.307}$ & $0.020_{0.017}^{0.023}$ &    $86.775_{85.787}^{87.484}$ &    $49.638_{45.879}^{53.981}$ & $143.278_{133.337}^{157.681}$ & $-14.129_{-14.173}^{-14.090}$ & $23.858_{13.194}^{33.457}$ & $20.484_{13.426}^{34.668}$ & $1.041_{0.495}^{3.494}$ & $0.956_{0.949}^{0.963}$ \\
          & $SB1$ &    (b) & $0.546_{0.546}^{0.547}$ & $1990.677_{1990.674}^{1990.680}$ & $0.303_{0.294}^{0.311}$ & $0.021_{0.016}^{0.022}$ &    $86.381_{84.949}^{88.248}$ &    $54.636_{47.403}^{60.572}$ & $142.729_{135.808}^{173.393}$ & $-14.155_{-14.247}^{-14.077}$ &                          — & $19.753_{13.071}^{89.385}$ &                       — &                       — \\
\hline
108917 & $SB2$ &    (a) &                   2.241 &                        1970.992 &                   0.496 &                   0.080 &                        92.870 &                       268.341 &                        74.479 &                       -10.743 &                     32.170 &                      8.246 &                   2.246 &                   0.361 \\
          & $SB2$ &    (b) & $2.245_{2.245}^{2.246}$ & $1968.750_{1968.744}^{1968.756}$ & $0.463_{0.452}^{0.469}$ & $0.072_{0.071}^{0.072}$ &    $90.343_{89.919}^{90.644}$ & $273.135_{272.559}^{273.573}$ &    $67.582_{67.044}^{67.983}$ & $-10.796_{-11.394}^{-10.035}$ & $37.243_{34.622}^{40.090}$ &    $7.239_{6.073}^{8.181}$ & $1.043_{0.807}^{1.266}$ & $0.369_{0.300}^{0.425}$ \\
          & $SB1$ &    (b) & $2.245_{2.245}^{2.246}$ & $1968.751_{1968.745}^{1968.759}$ & $0.461_{0.452}^{0.468}$ & $0.072_{0.071}^{0.072}$ &    $90.306_{89.940}^{90.695}$ & $273.266_{272.696}^{273.690}$ &    $67.505_{67.005}^{67.956}$ & $-10.791_{-11.745}^{-10.193}$ &                          — &    $7.130_{6.002}^{8.091}$ &                       — &                       — \\
\hline
111170 & $SB2$ &    (a) &                   1.731 &                        1965.475 &                   0.367 &                   0.066 &                       172.100 &                       261.393 &                        67.141 &                        -9.573 &                     35.542 &                      9.842 &                   1.389 &                   0.538 \\
          & $SB2$ &    (b) & $1.731_{1.730}^{1.732}$ & $1965.471_{1965.452}^{1965.488}$ & $0.366_{0.348}^{0.386}$ & $0.068_{0.066}^{0.069}$ & $172.194_{168.423}^{175.407}$ & $261.648_{260.027}^{263.951}$ &    $67.581_{62.243}^{70.742}$ &    $-9.590_{-9.742}^{-9.377}$ & $36.665_{33.559}^{38.620}$ &   $9.572_{9.214}^{10.242}$ & $1.368_{1.132}^{1.739}$ & $0.541_{0.493}^{0.584}$ \\
          & $SB1$ &    (b) & $1.731_{1.730}^{1.732}$ & $1965.477_{1965.453}^{1965.492}$ & $0.365_{0.346}^{0.387}$ & $0.067_{0.066}^{0.069}$ & $173.312_{168.692}^{176.002}$ & $261.302_{259.930}^{263.951}$ &    $66.750_{62.154}^{70.758}$ &    $-9.518_{-9.711}^{-9.341}$ &                          — &   $9.664_{9.213}^{10.262}$ &                       — &                       — \\
\hline
117186 & $SB2$ &    (a) &                   0.235 &                        2013.301 &                   0.327 &                   0.005 &                       176.070 &                        16.928 &                        88.054 &                       -19.890 &                      8.445 &                     53.509 &                   1.686 &                   0.824 \\
          & $SB2$ &    (b) & $0.235_{0.235}^{0.235}$ & $1987.923_{1987.922}^{1987.924}$ & $0.323_{0.320}^{0.327}$ & $0.005_{0.005}^{0.005}$ & $175.150_{174.106}^{175.971}$ &    $16.956_{16.500}^{17.415}$ &    $87.974_{87.683}^{88.420}$ & $-19.783_{-19.903}^{-19.632}$ &    $8.470_{8.366}^{8.580}$ & $53.043_{51.939}^{54.212}$ & $1.722_{1.690}^{1.765}$ & $0.816_{0.799}^{0.833}$ \\
          & $SB1$ &    (b) & $0.235_{0.235}^{0.235}$ & $1987.924_{1987.922}^{1987.927}$ & $0.325_{0.317}^{0.333}$ & $0.005_{0.005}^{0.005}$ & $177.074_{174.193}^{181.159}$ &    $16.979_{16.446}^{17.354}$ &    $87.999_{87.717}^{88.459}$ & $-20.250_{-20.830}^{-19.843}$ &                          — & $53.953_{52.613}^{55.459}$ &                       — &                       — \\
\hline
\hline
\end{tabular}
}
\label{tab:SB2/SB1}
\raggedright
\footnotesize{\small (a) Results reported by Orb6, SB9 and references therein, see text, (b) Results obtained in this work.}
\end{sidewaystable}
\normalsize

\subsubsection{Concluding remarks}

The experiments presented in this section show an uncertainty reduction of the estimated posterior distributions when RV observations of both components are available instead of only one component, as well as a slight shift on the MAP value of the posterior distributions in some orbital parameters. The orientation and magnitude of the shift between the posterior distribution of the $SB1$ and $SB2$ cases, as well as the magnitude of the uncertainty reduction does not follows an evident pattern along the dimensions of the posterior distribution, neither in between the different systems studied. The magnitude of the shift and the dispersion differences among the posterior distribution between the $SB2$ and $SB1$ cases depends on the system itself, as well as on the quality and quantity of the observations available, so it is difficult to draw general conclusions. However, we can say that all the studied systems exhibit an almost negligible difference in the MAP estimations and the dispersion of the posterior distribution on both cases. Finally, the MAP estimates on the $SB2$ and $SB1$ cases for all the systems studied are very similar compared to the values reported by other authors, using different methodologies (see Table~\ref{tab:SB2/SB1}). This is an remarkable finding that validates our general approach. We note that our estimated orbits and RV curves for all the other benchmark systems, as well as their respective marginal posterior distribution for the orbital parameters, in a format similar to those of Figures~\ref{fig:SB2/YSC132AaAb_obs} and ~\ref{fig:SB2/YSC132AaAb_params} can be found in this site \url{http://www.das.uchile.cl/~rmendez/B_Research/MV_RAM_SB1/SB2/}.

We find that the projected uncertainty in the observations space is lower in the zones where observations are available, and higher in the zones without observations, as intuitively expected. The only exception to this rule was observed in the system HIP 89000, where the observations in the zone of the orbit populated with observations (the apoastron) allow to reduce the uncertainty in the opposite zone of the orbit (the periastron) even considering that this zone has no observations. This result shows the relevance of analyzing the uncertainty on the observation space (through the projection of the posterior distribution), avoiding to waste resources on observing zones of the orbit that are apparently unresolved (to the complete absence of observations there), but that are actually accurately resolved due to the parametric configuration of the system itself. The joint estimation of the orbit and RV curves allows to share the knowledge provided by both sources of information, reducing the uncertainty of the estimates in the observations space significantly, even if one source of information is highly noisy. This is observed specially in the system HIP 108917, incidentally our lowest $q\sim0.36$ system (not discussed here explicitly, but available on our web page), where the projected RV curves exhibit low uncertainty despite the fact that the respective observations are very noisy. The projected orbits and RV curves of the $SB2$ and $SB1$ cases show almost no differences in all the studied systems, as well as the MAP estimate projections (curve) obtained from the posterior distributions. The only appreciable difference between the uncertainty estimated by projections on the observations space was in the orbit of the HIP 89000 system, were the $SB1$ case was slightly more uncertain than its $SB2$ counterpart. We attribute this to the higher uncertainty of its positional observations compared to the other studied systems.

Finally, as all the studied systems are well determined through abundant and good quality observations, the differences observed between the posterior distributions on the parameter and observations spaces were in general negligible. This means that the information provided by the RV observations of the companion object is somewhat redundant in these cases, being less relevant in the inference process, and hence, in the orbital parameters estimation. However, we anticipate that in regimes where the observations are not abundant or precise enough (as in the HIP 89000 system), the use of RV observations for both components might clearly reduce the posterior uncertainty compared to the use of only one of them.

\subsection{Incorporation of priors for estimating the mass ratio in $SB1$ systems} \label{sec:priorstosb1}

In this section, the inference for the eight benchmark well-studied $SB2$ binaries with a visual orbit (described in Section~\ref{sec:initvalid}) is compared with its counterpart omitting the RV observations of the companion object. For this comparison, we use three different approaches to determine the mass ratio: the incorporation of a prior on the parallax $p(\varpi)$, denoted as $SB1+p(\varpi)$ hereinafter, the incorporation of a prior on the primary object mass $p(m_1|\theta)$, denoted as $SB1+p(m_1|\theta)$ hereinafter, and incorporating both priors, denoted as $SB1+p(\varpi)+p(m_1|\theta)$ hereinafter. The adopted parallax and primary object's mass for the priors, as presented in Table~\ref{tab:SB2/SB2_plx&m1}, are visualized as error bars ($\pm 2\sigma$) in their corresponding marginal posterior distribution plot. Note that, in this case, the inferred parallaxes can not be properly called orbital parallaxes, since, while they are derived self-consistently from the model and data, they are only resolvable by the incorporation of the priors.

The estimates and their uncertainties are compared, again, in the parameters space through visualization of the posterior marginal distributions, as well as in the observations space, through the projection of $1000$ randomly selected samples of the posterior distribution on the observation space. For the last analysis, we draw trajectories from the time of the first observations $t_0$ to the first completion of the orbit $t_0+P$. The maximum a posteriori estimation (MAP) and the $95\%$ confidence interval around the MAP solution are summarized in Table~\ref{tab:SB2/SB1+}. The MAP estimation error, high densities intervals lengths, and estimated Kullback-Leibler divergence (KLD thereafter\footnote{The KLD is a measure of similarity between probability distributions and, in this work, it has been estimated through the k-nearest neighbor method \citep{wang2009divergence}. The KLD between two identical probability distributions is $0$, while the greater the discrepancy between them, the higher its corresponding KLD value.}) between the marginal posterior distributions for the mass ratio $q$ between the full-information $SB2$ case and the $SB1$ cases with priors $SB1+p(\varpi)$, $SB1+p(m_1|\theta)$ and $SB1+p(\varpi)+p(m_1|\theta)$ are presented in Table~\ref{tab:SB2/q_tab}.

Just as done in Section~\ref{sec:initvalid}, the inference process is performed through the simulation of $10000$ samples of the respective posterior distributions (discarding the first half for warm-up) on $4$ independent Markov chains using the No-U-Turn sampler algorithm as presented in Section \ref{sec:bayesmodel}.

While the analysis is done over the same eight benchmark objects introduced on Section~\ref{sec:initvalid}, for brevity the analysis is focused on the same three systems discussed in detail previously, namely, HIP 89000, HIP 111170, and HIP 117186.

\subsubsection{HIP 89000}

\begin{figure}[!h]
    \centering
    \includegraphics[width=\textwidth]{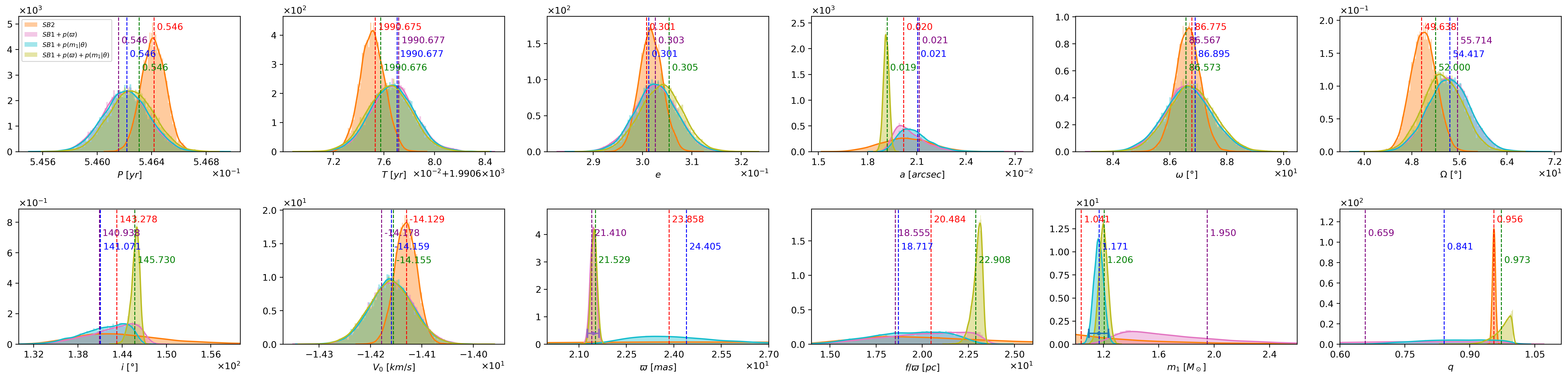}
    \caption{Marginal posterior distribution and MAP estimates of orbital parameters for the HIP 89000 binary system as an $SB2$, and in the $SB1+p(\varpi)$, $SB1+p(m_1|\theta)$ and $SB1+p(\varpi)+p(m_1|\theta)$ scenarios.}
    \label{fig:SB2/YSC132AaAb_params+}
\end{figure}

Figure~\ref{fig:SB2/YSC132AaAb_params+} shows that the posterior distributions of the $SB1$ cases with priors are almost equal except for the parameters $\varpi$, $m_1$ and $q$, which are identified through the incorporation of the priors $p(\varpi)$ or $p(m_1|\theta)$ (and in this particular case, also the orbital parameters $a$ and $i$). The posterior distribution of the other orbital parameters are equal to the posterior distributions of the $SB1$ case presented in the previous section. Naturally, for the $SB1+p(\varpi)$ case, the posterior distribution of $\varpi$ is equal to the prior $p(\varpi)$ (represented with the purple error bar), while for the $SB1+p(m_1|\theta)$ case, the uncertainty of the posterior distribution of $m_1$ is equal to the prior $p(m_1|\theta)$ (represented with the blue error bar). Interestingly, all the cases with priors present a significant reduction on the uncertainty of the posterior distribution of $\varpi,m_1,a,i$ with respect to the full-information scenario $SB2$, attributed to the narrow uncertainty on the priors. In contrast, they show an increase of the uncertainty of the posterior distribution of $q$. This is an interesting results that differs with that observed in the system HIP 111170 (see below), showing that narrower priors (on $\varpi$ or $m_1$) do not necessarily lead into narrower marginal distributions on the mass ratio $q$, depending on the priors, the observations, and the geometry of the system itself, and which highlights the fact that $SB2$s are still the best way to determine individual masses. The mixed priors case $SB1+p(\varpi)+p(m_1|\theta)$ presents the lowest uncertainty on $\varpi,m_1,q$, followed by the $SB1+p(m_1|\theta)$ and $SB1+p(\varpi)$ cases. The posterior distribution of the angular parameters $a,i$ of the $SB1+p(\varpi)$ and $SB1+p(m_1|\theta)$ cases are pretty similar between each other, but significantly different to the mixed priors case $SB1+p(\varpi)+p(m_1|\theta)$. Finally, the posterior distribution of $q$ in the $SB1+p(\varpi)+p(m_1|\theta)$ scenario is significantly different to all other cases. This last result is particularly interesting, since it shows that the mixed prior case can fit both priors individually at the same time, but deriving into different estimates of the mass ratio's posterior distribution.

\begin{figure}[!h]
    \centering
    \includegraphics[width=\textwidth]{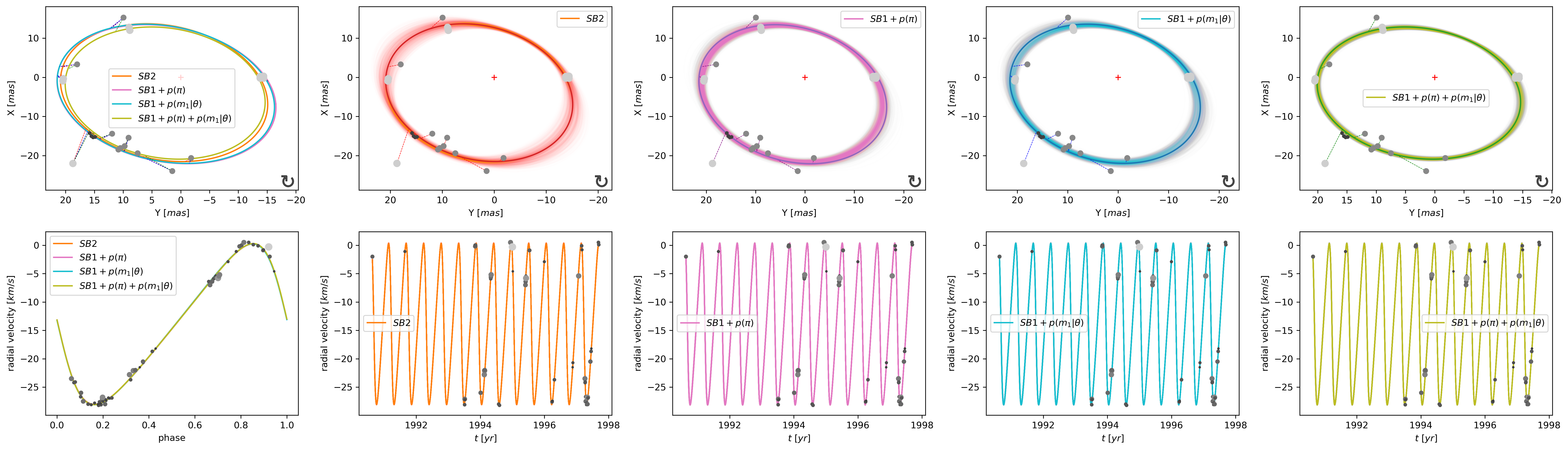}
    \caption{Estimated orbit and RV curves for the HIP 89000 binary system. First column: MAP point estimate projection of the posterior distribution for the $SB2$, $SB1+p(\varpi)$, $SB1+p(m_1|\theta)$ and $SB1+p(\varpi)+p(m_1|\theta)$ cases. Second to fifth columns: Projected posterior distribution for the $SB2$, 
    $SB1+p(\varpi)$, $SB1+p(m_1|\theta)$ and $SB1+p(\varpi)+p(m_1|\theta)$ cases respectively.}
    \label{fig:SB2/YSC132AaAb_obs+}
\end{figure}

The projection of the posterior distributions on the observation space are presented in Figure~\ref{fig:SB2/YSC132AaAb_obs+}. The trajectories of the MAP estimates (i.e., the most likely curves in the orbit space) present some differences between all the cases. The $SB1+p(\varpi)$ and $SB1+p(m_1|\theta)$ cases presents a slight reduction of the projected uncertainty in the orbit space with respect to the $SB2$ case, while the mixed priors case $SB1+p(\varpi)+p(m_1|\theta)$ presents a significant uncertainty reduction. The orbital posterior distribution of the mixed case presents also a different orbital shape with respect to all other cases, with a worst fitting on the most precise observations (in rectangular coordinates $X\sim-15$~[mas],$Y\sim+15$~[mas]). This exemplifies how the narrow priors incorporated play a major role on the inference procedure, to the detriment on the fitting of some positional observations. Finally, no significant differences are observed for the MAP and uncertainty projections in the RV space between all the four cases.

\subsubsection{HIP 111170}
\begin{figure}[!h]
    \centering
    \includegraphics[width=\textwidth]{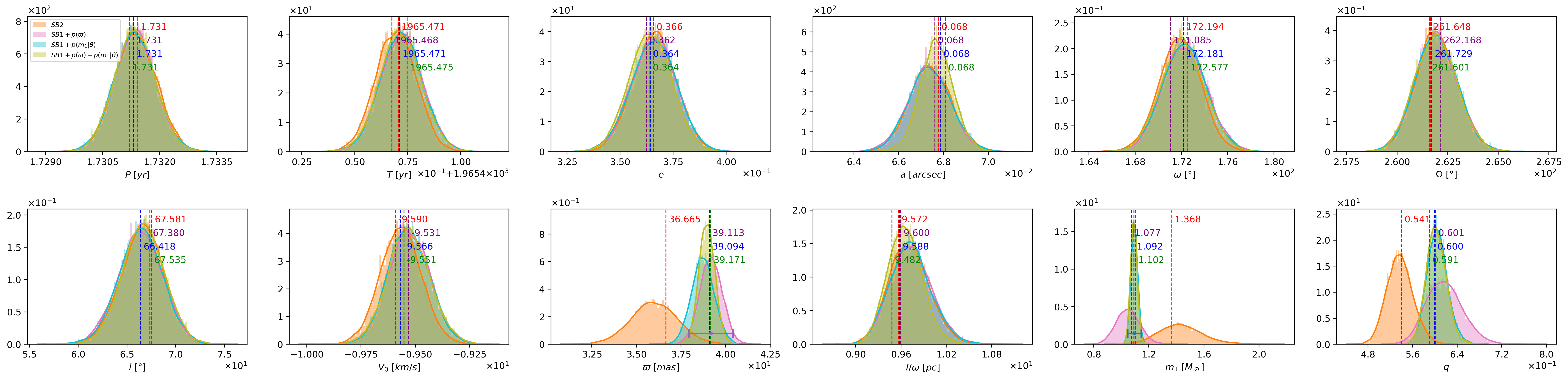}
    \caption{Similar to Figure~\ref{fig:SB2/YSC132AaAb_params+} but for the HIP 111170 binary system.}
    \label{fig:SB2/CHR111_params+}
\end{figure}

Figure~\ref{fig:SB2/CHR111_params+} shows that the posterior distribution are almost equal for all the orbital parameters, except for the parameters $\varpi$, $m_1$ and $q$, which are identified through the incorporation of the priors in the $SB1$ case. It is noted that, for these last trio of parameters, their distributions are shifted with respect to the distributions of the $SB2$ case, i.e., they show a slight discrepancy with respect to the full-information scenario $SB2$. Again, as was the case for HIP 89000, we see that the posterior distribution of $\varpi$ for the $SB1+p(\varpi)$ case is equal to the prior $p(\varpi)$ (represented with the purple error bar), while for the $SB1+p(m_1|\theta)$ case, the uncertainty of the posterior distribution of $m_1$ is equal to the prior $p(m_1|\theta)$ (represented with the blue error bar), which follows the soft-identifiability of both models on the corresponding parameters. Here, too, all the cases with priors offer a significant reduction on the uncertainty of the posterior distribution of $\varpi,m_1,q$ relative to the full-information scenario $SB2$. This apparently non-intuitive behavior is explained to the fact that the sources of information of the $SB2$ case (astrometric+RV1+RV2) are different than those in the $SB1$ cases with priors (astrometric+RV1+prior), and therefore, the inference exercise renders different results too. Hence, very constrained priors could derive into more constrained distributions than the $SB2$ case. The mean values of the posterior distribution of $\varpi,m_1,q$ is almost the same for the $SB1$ cases with priors, but are slightly biased with respect to the full-information scenario $SB2$, denoting a slight bias of the trigonometrical parallax and/or the spectral primary object's mass with respect to their orbital counterparts, as was already mentioned in Figure~\ref{fig:SB2/CHR111_params}. The mixed priors case $SB1+p(\varpi)+p(m_1|\theta)$ presents the lowest uncertainty on $\varpi,m_1,q$, followed by the $SB1+p(m_1|\theta)$ and $SB1+p(\varpi)$ cases, denoting the information gain of incorporating both priors simultaneously, instead of only one. The $SB1+p(\varpi)+p(m_1|\theta)$ scenario is the only case that presents a variation on mean and variance of the posterior distribution of the semi-major axis $a$, showing that very narrow priors can also affect the inference of the orbital parameters that are already identifiable from the astrometric and RV1 observations. This is an important point that is further discussed in the context of the inference of the mass ratio $q$ in Section~\ref{sec:conclremarkspriors}. Finally, it can be observed that in the mixed priors case the posterior distribution of $\varpi$ is in between the posterior distributions of the $SB1+p(\varpi)$ and $SB1+p(m_1|\theta)$ cases, and the posterior distributions of $m_1$ and $q$ are almost equal to the posterior distribution of the $SB1+p(m_1|\theta)$ and different to the $SB1+p(\varpi)$ case.

\begin{figure}[!h]
    \centering
    \includegraphics[width=\textwidth]{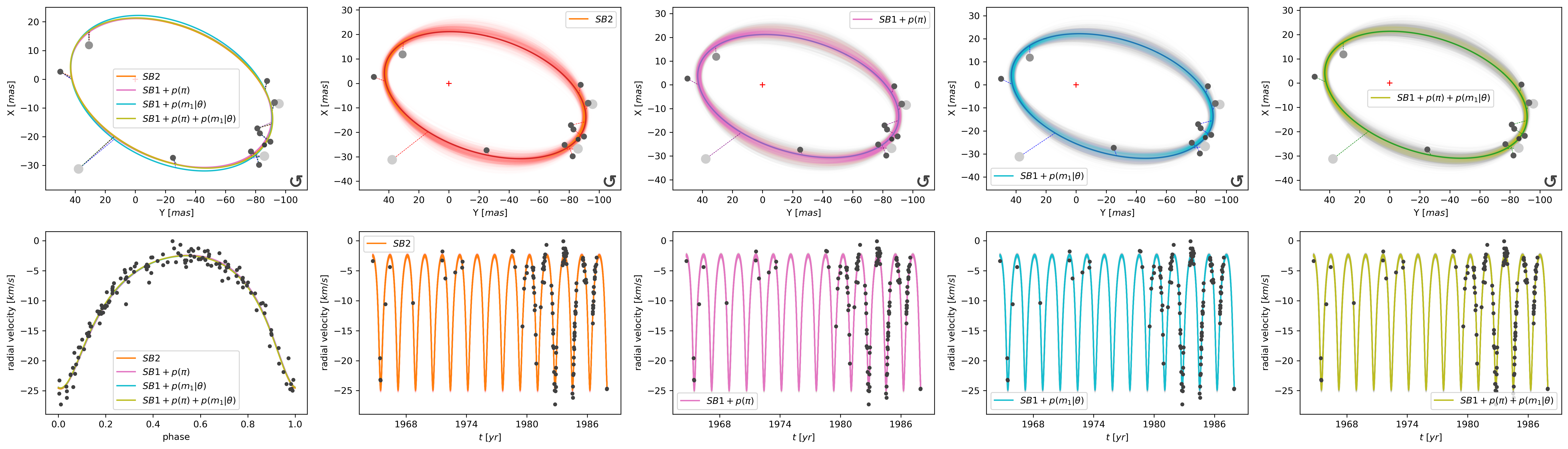}
    \caption{Similar to Figure~\ref{fig:SB2/YSC132AaAb_obs+} but for the HIP 111170 binary system.}
    \label{fig:SB2/CHR111_obs+}
\end{figure}

Moving to the estimated (by projection) posterior distributions in the observation space, these distributions are presented in Figure~\ref{fig:SB2/CHR111_obs+}. No significance differences in the distribution are observed in each of the four cases, which translates into no significant difference in the MAP curves and uncertainty projections between all the cases considered.

\subsubsection{HIP 117186}

\begin{figure}[!h]
    \centering
    \includegraphics[width=\textwidth]{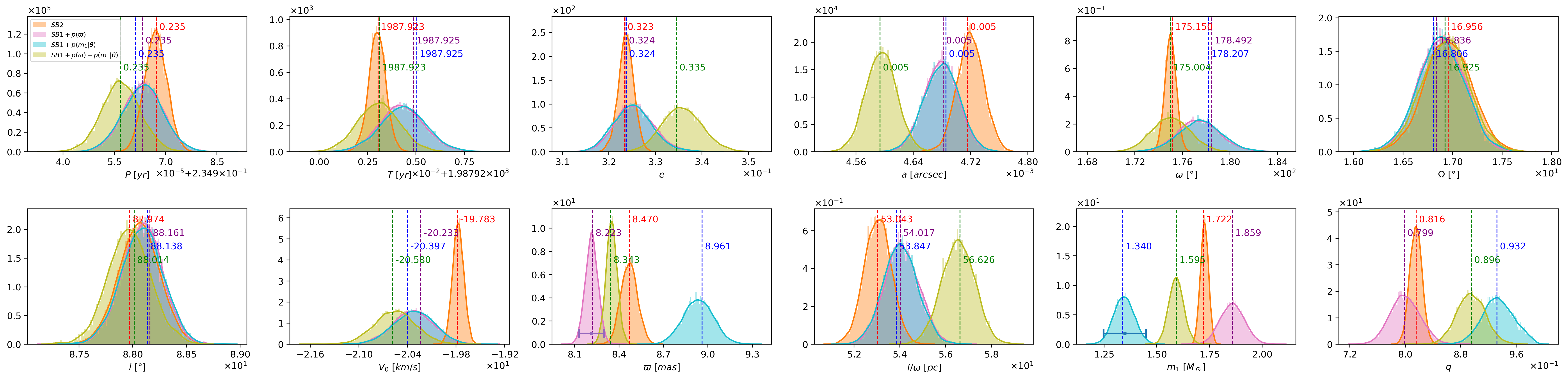}
    \caption{Similar to Figure~\ref{fig:SB2/CHR111_params+} but for the HIP 117186 binary system.}
    \label{fig:SB2/HIP117186_params+}
\end{figure}

To conclude this analysis, Figure~\ref{fig:SB2/HIP117186_params+} shows that the posterior distributions of the $SB1+p(\varpi)$ and $SB1+p(m_1|\theta)$ cases are almost equal except in the parameters $\varpi$, $m_1$ and $q$, which are identified through the incorporation of the priors $p(\varpi)$ or $p(m_1|\theta)$. The posterior distribution of the other orbital parameters are equal to the posterior distributions of the $SB1$ case presented in the previous section. For the $SB1+p(\varpi)$ and $SB1+p(m_1|\theta)$ cases, the posterior distribution of $\varpi$ is equal to the prior $p(\varpi)$ and the posterior distribution of $m_1$ is equal to the prior $p(m_1|\theta)$ respectively, as in the previous systems, and as intuitively expected. All the studied cases with priors presents a significant increase on the uncertainty of the posterior distribution of all the orbital parameters (with the exception of the angular parameters $\Omega$ and $i$) with respect to the full-information scenario $SB2$. The mixed priors case $SB1+p(\varpi)+p(m_1|\theta)$ presents the lowest uncertainty on $\varpi,m_1,q$, closely followed by the $SB1+p(m_1|\theta)$ and $SB1+p(\varpi)$ cases. The posterior distribution of the mixed case $SB1+p(\varpi)+p(m_1|\theta)$ presents a slight bias in all the orbital parameters except the angular parameters $\Omega$ and $i$, with respect to the $SB1$ case, and therefore, also with respect to the $SB1+p(\varpi)$ and $SB1+p(m_1|\theta)$ cases. It can be observed that, in the mixed priors case, the posterior distribution of $\varpi$ is in between the $SB1+p(\varpi)$ and $SB1+p(m_1|\theta)$ but nearest to the first one. On the other hand, the posterior distribution of $m_1$ is in between the $SB1+p(\varpi)$ and $SB1+p(m_1|\theta)$ distributions with a similar distance between them, and the posterior distribution of $q$ is in between the $SB1+p(\varpi)$ and $SB1+p(m_1|\theta)$ but nearest to the second one.

\begin{figure}[!h]
    \centering
    \includegraphics[width=\textwidth]{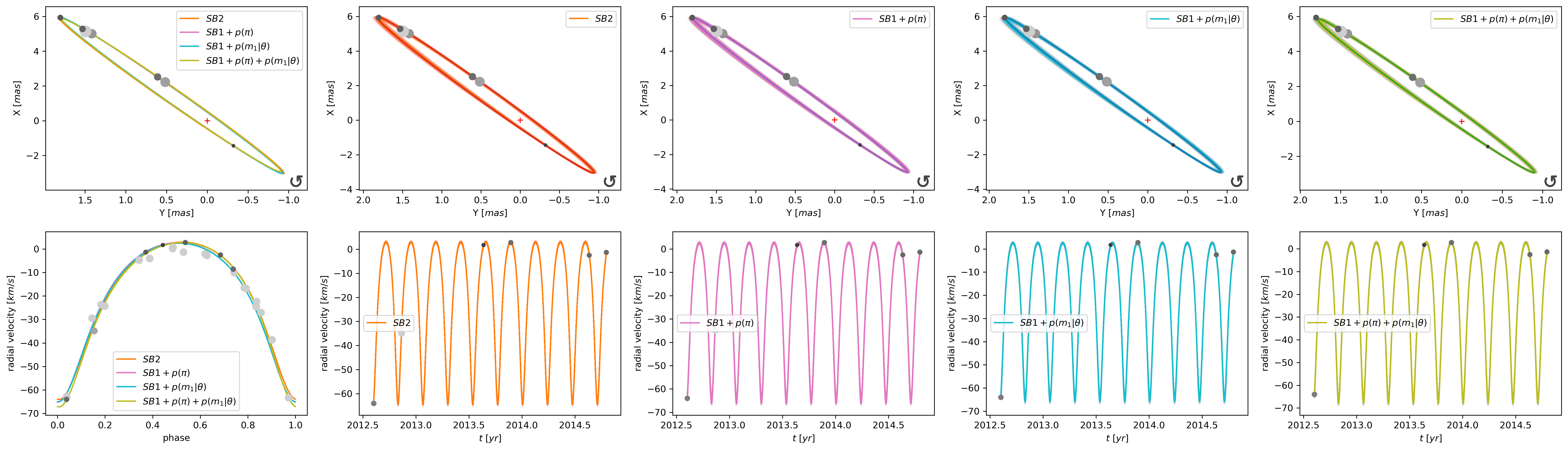}
    \caption{Similar to Figure~\ref{fig:SB2/CHR111_obs+} but for the HIP 117186 binary system.}
    \label{fig:SB2/HIP117186_obs+}
\end{figure}

Finally, the estimated (by projection) posterior distributions in the observation space are presented in Figure~\ref{fig:SB2/HIP117186_obs+}, where again no significant difference are observed in the MAP curves and uncertainties between all the four cases.

\small
\begin{sidewaystable}[h]
\centering
\caption{MAP estimates and 95\% HDPIs from the marginal posterior distributions of orbital parameters incorporating priors on the mass of the primary ($m_1$) and the trigonometric parallax ($\varpi$).}
\resizebox{\textwidth}{!}{\begin{tabular}{cccccccccccccc}
\hline\hline
     HIP~\# &                     Case &                 $P$ &                       $T$ &                     $e$ &              $a$ &                   $\omega$  &                $\Omega$ &                     $i$  &                $V_0$&              $\varpi$ &                $f/\varpi$ &         $m_1$  &                     $q$ \\
 & & [yr] & [yr] &  & [arcsec] &   [\textdegree] &  [\textdegree] & [\textdegree] & [km/s] & [mas] & [pc] &  [M$_\odot]$ & \\
\hline
   677 &                      $SB2$ & $0.265_{0.265}^{0.265}$ & $1988.847_{1988.842}^{1988.851}$ & $0.535_{0.470}^{0.589}$ & $0.024_{0.022}^{0.025}$ &    $78.339_{73.740}^{81.112}$ & $104.952_{101.653}^{108.823}$ & $106.172_{102.450}^{109.408}$ &  $-12.465_{-14.043}^{-9.567}$ & $32.796_{29.447}^{35.930}$ &  $10.500_{9.113}^{12.139}$ & $3.697_{2.983}^{4.461}$ & $0.525_{0.449}^{0.600}$ \\
          &               $SB1+p(\varpi)$ & $0.265_{0.264}^{0.265}$ & $1988.847_{1988.843}^{1988.852}$ & $0.520_{0.459}^{0.594}$ & $0.023_{0.022}^{0.025}$ &    $77.123_{73.047}^{81.221}$ & $105.402_{100.932}^{108.246}$ & $106.471_{102.514}^{109.817}$ &  $-10.481_{-13.177}^{-7.438}$ & $33.732_{32.956}^{34.333}$ &  $10.922_{9.367}^{12.515}$ & $3.031_{2.408}^{4.170}$ & $0.583_{0.453}^{0.718}$ \\
          &        $SB1+p(m_1|\theta)$ & $0.265_{0.264}^{0.265}$ & $1988.848_{1988.843}^{1988.852}$ & $0.523_{0.462}^{0.596}$ & $0.024_{0.022}^{0.026}$ &    $78.617_{72.973}^{81.262}$ & $105.264_{100.708}^{108.133}$ & $106.524_{102.460}^{109.760}$ &  $-10.154_{-13.267}^{-7.524}$ & $34.209_{30.839}^{38.681}$ &  $10.857_{9.255}^{12.434}$ & $3.063_{2.032}^{3.879}$ & $0.591_{0.477}^{0.734}$ \\
          & $SB1+p(\varpi)+p(m_1|\theta)$ & $0.265_{0.264}^{0.265}$ & $1988.848_{1988.843}^{1988.851}$ & $0.530_{0.461}^{0.582}$ & $0.024_{0.022}^{0.025}$ &    $77.903_{73.038}^{80.879}$ & $103.764_{100.862}^{108.126}$ & $105.992_{102.974}^{109.822}$ &   $-9.832_{-13.250}^{-7.476}$ & $33.724_{32.968}^{34.315}$ &  $10.897_{9.693}^{12.377}$ & $3.223_{2.548}^{3.737}$ & $0.581_{0.478}^{0.704}$ \\
\hline
  5531 &                      $SB2$ & $7.455_{7.450}^{7.458}$ & $1995.494_{1995.490}^{1995.502}$ & $0.744_{0.741}^{0.746}$ & $0.082_{0.082}^{0.083}$ & $215.006_{214.532}^{215.803}$ & $152.001_{151.314}^{152.497}$ &    $50.669_{49.926}^{51.606}$ &    $-9.526_{-9.596}^{-9.477}$ & $16.201_{15.952}^{16.572}$ & $29.379_{28.751}^{29.857}$ & $1.244_{1.176}^{1.294}$ & $0.908_{0.894}^{0.925}$ \\
          &               $SB1+p(\varpi)$ & $7.454_{7.450}^{7.458}$ & $1995.495_{1995.488}^{1995.500}$ & $0.744_{0.741}^{0.746}$ & $0.083_{0.082}^{0.083}$ & $214.783_{214.054}^{215.452}$ & $152.208_{151.574}^{152.808}$ &    $50.819_{49.962}^{51.529}$ &    $-9.481_{-9.550}^{-9.418}$ & $16.435_{16.247}^{16.987}$ & $29.390_{28.909}^{29.962}$ & $1.179_{1.029}^{1.223}$ & $0.934_{0.915}^{1.000}$ \\
          &        $SB1+p(m_1|\theta)$ & $7.455_{7.450}^{7.458}$ & $1995.494_{1995.488}^{1995.501}$ & $0.744_{0.741}^{0.746}$ & $0.083_{0.082}^{0.083}$ & $214.764_{214.090}^{215.469}$ & $152.228_{151.622}^{152.848}$ &    $51.032_{50.093}^{51.629}$ &    $-9.474_{-9.553}^{-9.418}$ & $16.910_{16.636}^{17.138}$ & $29.275_{28.853}^{29.855}$ & $1.057_{1.015}^{1.098}$ & $0.980_{0.965}^{1.000}$ \\
          & $SB1+p(\varpi)+p(m_1|\theta)$ & $7.454_{7.450}^{7.458}$ & $1995.493_{1995.489}^{1995.501}$ & $0.744_{0.741}^{0.746}$ & $0.082_{0.082}^{0.083}$ & $214.597_{214.045}^{215.429}$ & $152.306_{151.589}^{152.832}$ &    $50.849_{50.006}^{51.444}$ &    $-9.490_{-9.553}^{-9.419}$ & $16.806_{16.619}^{17.041}$ & $29.429_{28.983}^{29.899}$ & $1.074_{1.027}^{1.102}$ & $0.979_{0.963}^{1.000}$ \\
\hline
 14157 &                      $SB2$ & $0.119_{0.119}^{0.119}$ & $1993.795_{1993.795}^{1993.796}$ & $0.761_{0.758}^{0.763}$ & $0.006_{0.006}^{0.006}$ & $175.114_{174.829}^{175.564}$ &    $19.437_{19.092}^{19.924}$ &    $91.532_{90.104}^{92.968}$ &    $30.682_{30.573}^{30.780}$ & $19.436_{19.279}^{19.604}$ & $24.367_{24.145}^{24.616}$ & $1.003_{0.982}^{1.018}$ & $0.900_{0.891}^{0.910}$ \\
          &               $SB1+p(\varpi)$ & $0.119_{0.119}^{0.119}$ & $1993.795_{1993.795}^{1993.796}$ & $0.762_{0.758}^{0.765}$ & $0.006_{0.006}^{0.006}$ & $174.912_{174.469}^{175.445}$ &    $19.532_{19.092}^{19.937}$ &    $91.432_{89.917}^{92.890}$ &    $30.692_{30.581}^{30.844}$ & $19.524_{19.425}^{19.640}$ & $24.333_{24.138}^{24.645}$ & $0.985_{0.947}^{1.009}$ & $0.905_{0.891}^{0.932}$ \\
          &        $SB1+p(m_1|\theta)$ & $0.119_{0.119}^{0.119}$ & $1993.795_{1993.795}^{1993.796}$ & $0.761_{0.758}^{0.765}$ & $0.006_{0.006}^{0.006}$ & $174.952_{174.476}^{175.429}$ &    $19.445_{19.093}^{19.925}$ &    $91.272_{90.098}^{92.977}$ &    $30.705_{30.573}^{30.837}$ & $20.064_{19.798}^{20.543}$ & $24.379_{24.146}^{24.639}$ & $0.882_{0.810}^{0.924}$ & $0.957_{0.937}^{0.998}$ \\
          & $SB1+p(\varpi)+p(m_1|\theta)$ & $0.119_{0.119}^{0.119}$ & $1993.795_{1993.795}^{1993.796}$ & $0.762_{0.758}^{0.765}$ & $0.006_{0.006}^{0.006}$ & $175.002_{174.532}^{175.496}$ &    $19.527_{19.080}^{19.940}$ &    $91.349_{89.958}^{92.911}$ &    $30.703_{30.579}^{30.841}$ & $19.610_{19.470}^{19.677}$ & $24.505_{24.233}^{24.722}$ & $0.954_{0.932}^{0.990}$ & $0.925_{0.900}^{0.939}$ \\
\hline
 20601 &                      $SB2$ & $0.428_{0.428}^{0.428}$ & $1971.983_{1971.983}^{1971.983}$ & $0.852_{0.849}^{0.854}$ & $0.011_{0.011}^{0.012}$ & $202.260_{201.535}^{203.109}$ & $340.812_{339.649}^{341.814}$ & $103.636_{101.114}^{105.151}$ &    $42.097_{41.940}^{42.251}$ & $16.710_{16.429}^{16.991}$ & $25.677_{25.178}^{26.099}$ & $0.984_{0.947}^{1.013}$ & $0.752_{0.742}^{0.759}$ \\
          &               $SB1+p(\varpi)$ & $0.428_{0.428}^{0.428}$ & $1971.983_{1971.983}^{1971.983}$ & $0.852_{0.849}^{0.855}$ & $0.011_{0.011}^{0.012}$ & $202.630_{201.691}^{203.725}$ & $340.619_{339.625}^{341.903}$ & $102.796_{100.728}^{105.558}$ &    $42.311_{41.920}^{42.486}$ & $17.347_{17.072}^{17.577}$ & $25.712_{25.184}^{26.188}$ & $0.849_{0.797}^{0.924}$ & $0.805_{0.768}^{0.836}$ \\
          &        $SB1+p(m_1|\theta)$ & $0.428_{0.428}^{0.428}$ & $1971.983_{1971.983}^{1971.983}$ & $0.852_{0.849}^{0.855}$ & $0.011_{0.011}^{0.012}$ & $202.553_{201.665}^{203.542}$ & $340.930_{339.542}^{341.811}$ & $103.282_{100.685}^{105.232}$ &    $42.262_{41.913}^{42.476}$ & $17.044_{16.746}^{17.346}$ & $25.667_{25.171}^{26.113}$ & $0.916_{0.873}^{0.954}$ & $0.778_{0.755}^{0.797}$ \\
          & $SB1+p(\varpi)+p(m_1|\theta)$ & $0.428_{0.428}^{0.428}$ & $1971.983_{1971.983}^{1971.983}$ & $0.852_{0.849}^{0.855}$ & $0.011_{0.011}^{0.012}$ & $203.013_{201.789}^{203.748}$ & $340.443_{339.572}^{341.980}$ & $102.590_{100.697}^{105.291}$ &    $42.255_{41.930}^{42.498}$ & $17.186_{17.031}^{17.421}$ & $25.450_{25.069}^{25.976}$ & $0.905_{0.865}^{0.934}$ & $0.777_{0.761}^{0.806}$ \\
\hline
 89000 &                      $SB2$ & $0.546_{0.546}^{0.547}$ & $1990.675_{1990.673}^{1990.677}$ & $0.301_{0.297}^{0.307}$ & $0.020_{0.017}^{0.023}$ &    $86.775_{85.787}^{87.484}$ &    $49.638_{45.879}^{53.981}$ & $143.278_{133.337}^{157.681}$ & $-14.129_{-14.173}^{-14.090}$ & $23.858_{13.194}^{33.457}$ & $20.484_{13.426}^{34.668}$ & $1.041_{0.495}^{3.494}$ & $0.956_{0.949}^{0.963}$ \\
          &               $SB1+p(\varpi)$ & $0.546_{0.546}^{0.547}$ & $1990.677_{1990.673}^{1990.680}$ & $0.303_{0.294}^{0.311}$ & $0.021_{0.019}^{0.022}$ &    $86.567_{85.092}^{88.254}$ &    $55.714_{47.554}^{61.296}$ & $140.938_{136.198}^{147.662}$ & $-14.178_{-14.241}^{-14.075}$ & $21.410_{21.271}^{21.636}$ & $18.555_{15.978}^{23.353}$ & $1.950_{1.182}^{2.524}$ & $0.659_{0.524}^{1.000}$ \\
          &        $SB1+p(m_1|\theta)$ & $0.546_{0.546}^{0.547}$ & $1990.677_{1990.673}^{1990.680}$ & $0.301_{0.294}^{0.311}$ & $0.021_{0.019}^{0.023}$ &    $86.895_{84.988}^{88.268}$ &    $54.417_{47.335}^{61.069}$ & $141.071_{135.668}^{146.270}$ & $-14.159_{-14.240}^{-14.077}$ & $24.405_{21.731}^{27.054}$ & $18.717_{15.412}^{22.656}$ & $1.171_{1.100}^{1.236}$ & $0.841_{0.727}^{1.000}$ \\
          & $SB1+p(\varpi)+p(m_1|\theta)$ & $0.546_{0.546}^{0.547}$ & $1990.676_{1990.673}^{1990.680}$ & $0.305_{0.296}^{0.312}$ & $0.019_{0.019}^{0.019}$ &    $86.573_{85.116}^{88.296}$ &    $52.000_{46.786}^{59.727}$ & $145.730_{144.563}^{146.826}$ & $-14.155_{-14.243}^{-14.078}$ & $21.529_{21.296}^{21.660}$ & $22.908_{22.432}^{23.417}$ & $1.206_{1.141}^{1.260}$ & $0.973_{0.935}^{1.000}$ \\
\hline
108917    &                      $SB2$ & $2.245_{2.245}^{2.246}$ & $1968.750_{1968.744}^{1968.756}$ & $0.463_{0.452}^{0.469}$ & $0.072_{0.071}^{0.072}$ & $90.343_{89.919}^{90.644}$ & $273.135_{272.559}^{273.573}$ &    $67.582_{67.044}^{67.983}$ & $-10.796_{-11.394}^{-10.035}$ & $37.243_{34.622}^{40.090}$ &   $7.239_{6.073}^{8.181}$ & $1.043_{0.807}^{1.266}$ & $0.369_{0.300}^{0.425}$ \\
          &               $SB1+p(\pi)$ & $2.245_{2.245}^{2.246}$ & $1968.752_{1968.745}^{1968.758}$ & $0.459_{0.452}^{0.468}$ & $0.072_{0.071}^{0.072}$ & $90.308_{89.925}^{90.659}$ & $273.234_{272.691}^{273.706}$ &    $67.377_{67.005}^{67.944}$ & $-11.090_{-11.747}^{-10.181}$ & $31.858_{30.990}^{33.415}$ &   $7.116_{6.027}^{8.186}$ & $1.749_{1.473}^{1.930}$ & $0.293_{0.241}^{0.361}$ \\
          &        $SB1+p(m_1|\theta)$ & $2.245_{2.245}^{2.246}$ & $1968.753_{1968.745}^{1968.758}$ & $0.459_{0.452}^{0.468}$ & $0.072_{0.071}^{0.072}$ & $90.297_{89.944}^{90.665}$ & $273.152_{272.676}^{273.707}$ &    $67.455_{66.985}^{67.941}$ & $-10.889_{-11.743}^{-10.158}$ & $32.096_{30.969}^{32.948}$ &   $6.936_{6.107}^{8.256}$ & $1.722_{1.593}^{1.893}$ & $0.286_{0.242}^{0.348}$ \\
          & $SB1+p(\pi)+p(m_1|\theta)$ & $2.245_{2.245}^{2.246}$ & $1968.751_{1968.746}^{1968.758}$ & $0.460_{0.452}^{0.469}$ & $0.072_{0.071}^{0.072}$ & $90.354_{89.936}^{90.669}$ & $272.995_{272.694}^{273.697}$ &    $67.447_{67.004}^{67.955}$ & $-10.901_{-11.760}^{-10.195}$ & $31.982_{31.284}^{32.814}$ &   $7.217_{6.003}^{8.099}$ & $1.721_{1.601}^{1.855}$ & $0.300_{0.240}^{0.346}$ \\
\hline
111170 &                      $SB2$ & $1.731_{1.730}^{1.732}$ & $1965.471_{1965.452}^{1965.488}$ & $0.366_{0.348}^{0.386}$ & $0.068_{0.066}^{0.069}$ & $172.194_{168.423}^{175.407}$ & $261.648_{260.027}^{263.951}$ &    $67.581_{62.243}^{70.742}$ &    $-9.590_{-9.742}^{-9.377}$ & $36.665_{33.559}^{38.620}$ &   $9.572_{9.214}^{10.242}$ & $1.368_{1.132}^{1.739}$ & $0.541_{0.493}^{0.584}$ \\
          &               $SB1+p(\varpi)$ & $1.731_{1.730}^{1.732}$ & $1965.468_{1965.454}^{1965.492}$ & $0.362_{0.346}^{0.386}$ & $0.068_{0.066}^{0.069}$ & $171.085_{168.731}^{176.150}$ & $262.168_{259.960}^{263.976}$ &    $67.380_{61.975}^{70.685}$ &    $-9.531_{-9.713}^{-9.336}$ & $39.113_{37.957}^{40.408}$ &   $9.600_{9.207}^{10.270}$ & $1.077_{0.884}^{1.211}$ & $0.601_{0.556}^{0.681}$ \\
          &        $SB1+p(m_1|\theta)$ & $1.731_{1.730}^{1.732}$ & $1965.471_{1965.452}^{1965.491}$ & $0.364_{0.345}^{0.386}$ & $0.068_{0.066}^{0.069}$ & $172.181_{168.497}^{175.902}$ & $261.729_{260.019}^{263.988}$ &    $66.418_{62.114}^{70.835}$ &    $-9.566_{-9.713}^{-9.349}$ & $39.094_{37.470}^{39.923}$ &   $9.588_{9.203}^{10.243}$ & $1.092_{1.046}^{1.143}$ & $0.600_{0.569}^{0.641}$ \\
          & $SB1+p(\varpi)+p(m_1|\theta)$ & $1.731_{1.730}^{1.732}$ & $1965.475_{1965.453}^{1965.491}$ & $0.364_{0.344}^{0.383}$ & $0.068_{0.066}^{0.069}$ & $172.577_{168.539}^{175.718}$ & $261.601_{259.927}^{263.912}$ &    $67.535_{62.458}^{70.829}$ &    $-9.551_{-9.723}^{-9.345}$ & $39.171_{38.068}^{39.817}$ &   $9.482_{9.229}^{10.106}$ & $1.102_{1.046}^{1.140}$ & $0.591_{0.568}^{0.637}$ \\
\hline
117186 &                      $SB2$ & $0.235_{0.235}^{0.235}$ & $1987.923_{1987.922}^{1987.924}$ & $0.323_{0.320}^{0.327}$ & $0.005_{0.005}^{0.005}$ & $175.150_{174.106}^{175.971}$ &    $16.956_{16.500}^{17.415}$ &    $87.974_{87.683}^{88.420}$ & $-19.783_{-19.903}^{-19.632}$ &    $8.470_{8.366}^{8.580}$ & $53.043_{51.939}^{54.212}$ & $1.722_{1.690}^{1.765}$ & $0.816_{0.799}^{0.833}$ \\
          &               $SB1+p(\varpi)$ & $0.235_{0.235}^{0.235}$ & $1987.925_{1987.922}^{1987.926}$ & $0.324_{0.318}^{0.334}$ & $0.005_{0.005}^{0.005}$ & $178.492_{174.029}^{180.950}$ &    $16.836_{16.451}^{17.332}$ &    $88.161_{87.722}^{88.457}$ & $-20.233_{-20.812}^{-19.847}$ &    $8.223_{8.131}^{8.299}$ & $54.017_{52.632}^{55.526}$ & $1.859_{1.747}^{1.968}$ & $0.799_{0.759}^{0.841}$ \\
          &        $SB1+p(m_1|\theta)$ & $0.235_{0.235}^{0.235}$ & $1987.925_{1987.922}^{1987.927}$ & $0.324_{0.317}^{0.334}$ & $0.005_{0.005}^{0.005}$ & $178.207_{173.818}^{180.990}$ &    $16.806_{16.406}^{17.320}$ &    $88.138_{87.693}^{88.451}$ & $-20.397_{-20.797}^{-19.828}$ &    $8.961_{8.735}^{9.127}$ & $53.847_{52.571}^{55.490}$ & $1.340_{1.251}^{1.439}$ & $0.932_{0.892}^{0.980}$ \\
          & $SB1+p(\varpi)+p(m_1|\theta)$ & $0.235_{0.235}^{0.235}$ & $1987.923_{1987.921}^{1987.925}$ & $0.335_{0.328}^{0.344}$ & $0.005_{0.005}^{0.005}$ & $175.004_{172.090}^{178.279}$ &    $16.925_{16.460}^{17.383}$ &    $88.014_{87.595}^{88.392}$ & $-20.580_{-21.030}^{-20.051}$ &    $8.343_{8.271}^{8.421}$ & $56.626_{55.163}^{57.942}$ & $1.595_{1.530}^{1.669}$ & $0.896_{0.854}^{0.933}$ \\
\hline
\hline
\end{tabular}
}
\label{tab:SB2/SB1+}
\end{sidewaystable}
\normalsize

\small
\begin{sidewaystable}[h]
\centering
\caption{Comparison of the difference in the marginal posterior distribution for the mass ratio $q$ between the $SB1$ cases with priors, and the full-information (benchmark) $SB2$.}
\resizebox{\textwidth}{!}{\begin{tabular}{cccccccccccccc}
\hline\hline   & \multicolumn{3}{c}{MAP estimate error [\%]} &  \multicolumn{3}{c}{HPDI length} & \multicolumn{3}{c}{KLD}\\
       HIP~\# & $SB1+p(\varpi)$ & $SB1+p(m_1|\theta)$ & $SB1+p(\varpi)+p(m_1|\theta)$ & $SB1+p(\varpi)$ & $SB1+p(m_1|\theta)$ & $SB1+p(\varpi)+p(m_1|\theta)$ & $SB1+p(\varpi)$ & $SB1+p(m_1|\theta)$ & $SB1+p(\varpi)+p(m_1|\theta)$ \\
\hline
677 &        -5.81 &               -6.57 &                      -5.58 &        0.265 &               0.257 &                      0.226 &        0.485 &               0.834 &                      0.728 \\
5531 &         -2.60 &               -7.21 &                      -7.03 &        0.085 &               0.035 &                      0.037 &        2.507 &               8.844 &                      8.821 \\
14157 &        -0.54 &               -5.78 &                      -2.55 &        0.041 &               0.061 &                      0.039 &        0.826 &               7.038 &                      2.346 \\
20601 &        -5.36 &               -2.62 &                      -2.58 &        0.068 &               0.042 &                      0.045 &        4.723 &               3.471 &                       4.76 \\
89000 &        +29.67 &               +11.49 &                      -1.73 &        0.476 &               0.273 &                      0.065 &        3.533 &               3.014 &                      2.146 \\
108917 &          +7.60 &                +8.28 &                       +6.91 &         0.12 &               0.106 &                      0.106 &        2.064 &               2.856 &                      2.932 \\
111170 &        -6.05 &               -5.88 &                      -5.02 &        0.125 &               0.072 &                      0.069 &        3.804 &               5.778 &                      5.647 \\
117186 &         +1.66 &              -11.67 &                      -7.98 &        0.082 &               0.088 &                      0.079 &        0.772 &               9.201 &                      7.285 \\
\hline
absmean &         7.41 &                7.44 &                       4.92 &        0.158 &               0.117 &                      0.083 &        2.339 &               5.129 &                      4.333 \\
std &         11.20 &                7.56 &                       4.35 &        0.136 &               0.088 &                      0.058 &        1.481 &               2.857 &                      2.609 \\
\hline\hline
\end{tabular}
}
\label{tab:SB2/q_tab}
\end{sidewaystable}
\normalsize

\subsubsection{Concluding remarks}\label{sec:conclremarkspriors}

The experiments described in the previous subsections demonstrate a relevant uncertainty reduction of the estimated posterior distributions with respect to the $SB1$ case when priors on the parallax or mass of the primary object are incorporated. In general, we observe that the more information is available, the less is the uncertainty obtained in the estimates, as one would expect. Consequently, the case that incorporates both priors, i.e., $SB1+p(\varpi)+p(m_1|\theta)$ presents the lowest uncertainty. Due to the non-identifiability of the parameters $\varpi$, $m_1$ and $q$ in the $SB1$ system with a visual orbit model, the prior $p(\varpi)$ results equal to the marginal posterior distribution of $\varpi$ and the prior $p(m_1|\theta)$ results equal to the marginal posterior distribution of $m_1$. The major differences in the posterior distributions of the cases with priors are observed in the trio of orbital parameters $\varpi,m_1,q$. Here, the marginal posterior distribution of these parameters on the mixed case $SB1+p(\varpi)+p(m_1|\theta)$ lie in between the posterior distributions of the $SB1+p(\varpi)$ and $SB1+p(m_1|\theta)$ cases. These distributions can be equidistant to the $SB1+p(\varpi)$ and $SB1+p(m_1|\theta)$ cases, if both sources of information are equally likely according to the model and the available data, or can be near to one of them, if one source of information is more likely than the other. The similarity of the posterior distribution observed in the $SB1+p(\varpi)+p(m_1|\theta)$ scenario to one of the simpler cases, $SB1+p(\varpi)$ or $SB1+p(m_1|\theta)$, allows to determine the most reliable prior according to the model and the observations. For example, if the posterior distributions (in the trio of parameters $\varpi,m_1,q$) of the mixed priors case are nearest to the distribution of the $SB1+p(m_1|\theta)$ than to the $SB1+p(\varpi)$ case, then one might conclude that the constrain imposed by the mass of the primary object is more reliable than the trigonometric parallax constrain, since the last one is less relevant in the inference process in the mixed scenario. This is particularly important in the context of the already noted differences between the orbital and trigonometric parallaxes for HIP 111170 and HIP 117186. We should recall that it is a well known fact that Hipparcos´ trigonometric parallaxes were indeed biased due to the orbital motion of the binary (i.e., the parallax and orbit signal are blended), as shown by \cite{Soder1999} (see, in particular his Section~3.1, and Table~2), and it is likely that Gaia will suffer from a similar problem\footnote{For example, according to Tokovinin's multiple star catalogue, HIP 64421 contains a binary with a 27~yr orbit. Its Hipparcos parallax is 8.6~mas, its dynamical parallax is 8.44~mas, and its Gaia DR2 parallax is 3~mas. However, Gaia does give a consistent parallax for the C component at 1.9~arcsec: $9.7 \pm 0.3$~mas, see \url{http://www.ctio.noao.edu/~atokovin/stars/}. There are other examples like this in the cited catalog.}. Indeed During each observation Gaia is not expected to resolve systems closer than about 0.4~arcsec, though over the mission there will be a final resolution of 0.1~arcsec. This is shown graphically in Figure~1 from \citet{Ziegleret2018}, where the current resolution of the second Gaia data release is shown to be around 1~arcsec, being a function of the magnitude difference between primary and secondary. It is expected that, from the third Gaia data release and on ($\sim 2024$), the treatment of binary stars will be much improved, by incorporating orbital motion (and its impact on the photocenter position of unresolved pairs) into the overall astrometric solution, thus suppressing/alleviating the parallax bias significantly.

Since we are also adopting a spectral type (and a luminosity class) as a proxy for the mass of the primary, we should be just as careful as with the trigonometric parallaxes, since these assumptions could also bias our inference if the assumed spectral type is in error. Therefore, it is important to asses the reliability of our adopted spectral types. One very important source of comparison are compilations of spectral types, the most authoritative being the "Catalogue of Stellar Spectral Classifications" by \citet{SkiffCat}, which provides a compilation  of spectral classifications determined for stars from spectra only (i.e., no narrow-band photometry), collected from the literature, and which is updated regularly in VizieR (catalog B/mk/mktypes, currently containing more than 90.000 stars). One interesting case in question is HIP 111170, for which we have adopted to be an F8V from SIMBAD. However Skiff´s catalog gives the possible range F7V to G0V, and even F9IV sub-giant, from five different literature sources. If we consider its reported $V$-band mag in SIMBAD ($V=6.160$), and its trigonometric parallax in Table~\ref{tab:SB2/SB2_plx&m1}, this implies an $M_V=+4.126$ which is totally consistent with an F8-F9 spectral type of luminosity class V from \cite{Abuset2020}, whereas an F9IV should have an $M_V =+2.88$, completely off our value. This suggest that our adopted spectral type is indeed correct (unless, of course, the Gaia parallax is completely off).

The estimation from the join use of astrometry (positions) and RV observations, as well as the incorporation of priors, have a non-negligible impact on the estimated posterior distribution of some the identifiable orbital parameters as well. Indeed, we observe cases where the impact on the posterior uncertainty of including additional sources of information (priors) is of the same order of magnitude than the uncertainty reduction obtained from actual observations (measurements). For example, the binary system HIP 89000 exemplifies the impact of the priors on the posterior distribution of the identifiable orbital parameters $a$ and $i$. In contrast, other binary system, like  HIP 108917, exhibit null impact on the marginal distributions of the identifiable orbital parameters when adding priors.

The estimated posterior distribution and the MAP estimates on the observation space presents no appreciable differences between all the cases studied ($SB2$, $SB1+p(\varpi)$, $SB1+p(m_1|\theta)$ and $SB1+p(\varpi)+p(m_1|\theta)$). The only significant difference is observed in the orbit of the system HIP 89000, where the mixed case $SB1+p(\varpi)+p(m_1|\theta)$ presents the lowest uncertainty, even lower than the full-information case $SB2$, but at the cost of a slight worst fitting of some of the orbital observations.

Based on the posterior distribution for the mass ratio $q$, we can see from Table~(\ref{tab:SB2/q_tab}), that the case that offers the highest similarity with the full-information scenario $SB2$, according to the KLD measure, is the $SB1+p(\varpi)$ case, followed by the mixed case $SB1+p(\varpi)+p(m_1|\theta)$ and the $SB1+p(m_1|\theta)$ case. However, the lowest mean absolute error between the MAP estimates, as well as the high posterior density interval range, is obtained in the $SB1+p(\varpi)+p(m_1|\theta)$ scenario (4.92\%), followed by the $SB1+p(\varpi)$ (7.41\%) and $SB1+p(m_1|\theta)$ (7.44\%) cases. There is a rather complex interplay between orbital parameters and the final value of $q$: In the $SB1$ scenario, the mass of the primary is mostly constrained by the prior imposed by its spectral type, hence, in principle, the largest the value of $m_1$ (in comparison to the value of $m_1$ derived from the equivalent $SB2$), the smallest the value of the inferred $q$. However, $m_2$ itself is inferred from the mass sum of the orbital solution, i.e., $m_2 \sim a^3/\left( \varpi^3 \, P^2 \right) - m_1$. Therefore if the trigonometric $\varpi$ (imposed in the prior) is different from the orbital parallax, this will also have an impact on the estimated $m_2$. Our three described cases are a clear example of this complex behavior: For HIP 89000 the trigonometric and orbital parallaxes are almost the same while the mass of the primary derived form the $SB2$ is almost the same as that from its spectral type (see Figures~\ref{fig:SB2/YSC132AaAb_params} and~\ref{fig:SB2/YSC132AaAb_params+}). As a consequence the $q$ in both cases are very similar, 0.973 vs. 0.956 respectively (see first and last line for this object in Table~\ref{tab:SB2/SB1+}). HIP 111170 exhibits a different situation: Its trigonometric parallax is larger than its orbital parallax (thus implying a smaller mass sum), while its spectral mass is smaller than its orbital mass for the primary, they both seem to compensate each other, ending with a similar mass ratio of 0.541 vs. 0.591 for the $SB2$ and $SB1$+priors cases respectively. Finally, for HIP 117186 we have that its trigonometric parallax is smaller than its orbital parallax, and its spectral mass is smaller than its orbital for the primary. In this case, one would expect that the mass of the secondary is substantially larger than in the $SB2$ case (and thus a larger $q$ as well), however this is compensated by a smaller value of the semi-major axis (see Figure~\ref{fig:SB2/HIP117186_params+}), which decreases the estimated value of $m_2$. The next result in this cases is that the $q$ values are, again, similar, 0.816 vs. 0.896 for the $SB2$ and $SB1$+priors cases respectively. We can surely envision cases where this combination could be more damaging, in the sense of rendering erroneous $q$ values (not seen among our benchmark systems though, compare the first and last lines in Table~\ref{tab:SB2/SB1+}). We emphasize however that ours is a method to provide rough informed estimates (in a statistical sense) of the mass ratio, but definitive values are still provided by $SB2$s. 

In summary, the obtained results indicate that the closest estimation to the full-information scenario is obtained through the incorporation of a prior information on the parallax $\varpi$. However, the most robust point estimates is obtained by incorporating both sources of priors, allowing to correct the estimation through $p(m_1|\theta)$ when $p(\varpi)$ is biased. This is clearly illustrated by the system HIP 89000, at a cost of a slight increase on the average estimation error. The lowest performance according to all the comparison metrics is achieved by the $SB1+p(m_1|\theta)$ case. This is attributed to the fact that the additional information on $m_1$ comes from an approximate empirical rule that relates the mass with the spectral type of the object, while the additional information on $\varpi$ has a direct relationship to the system´s geometry.

\section{Application to unresolved $SB1$s with a visual orbit} \label{sec:application}

In this section we apply the methodology described in the previous sections to study twelve $SB1$s systems with a visual orbit using the $SB1$, $SB1+p(\varpi)$, $SB1+p(m_1|\theta)$ and $SB1+p(\varpi)+p(m_1|\theta)$ scenarios. We note that, for all these binaries, there is no published joint estimation of their orbital parameters using the available astrometric and spectroscopic data, so this is the first study of these binaries from this point of view as well, with the exception of HIP 7918 which has a combined solution by \citet{2015A&A...574A...6A}. They represent an heterogeneous group of binaries, with masses in the range between 0.85$M_\odot$ to slightly above 20$M_\odot$, located at distances between 12 to 150~pc, and with very different data quality and orbital phase coverage. The adopted parallax, primary object's spectral type and derived mass for the priors of each system are presented in Table~\ref{tab:SB1/SB1_plx&m1}. These values are also visualized as error bars ($\pm 2\sigma$) in their corresponding marginal posterior distribution plot. We emphasize that, through this exercise, we do not intend to carry out an in-depth analysis of the selected objects (which will be presented in a forthcoming paper), but rather as a proof-of-concept of the methodology introduced previously.

\small
\begin{table}[h]
\centering
\caption{Reported trigonometric parallax, primary object's spectral type, and derived primary mass of $SB1$ stellar systems.}
\begin{tabular}{ccccc}
\hline\hline
HIP \# & Discovery   & $\varpi$   & SpTyp & $m_1$  \\
       & Designation & [mas] &       & [M$_\odot]$ \\
\hline
171 & BU733AB &$79.0696\pm0.5621$ &  G5V &   $0.955\pm0.015$ \\
3504 & NOI3Aa,Ab & $4.705\pm0.431(a)$ &  B5III &   $5.623\pm1.190$ \\
6564 & BU1163  &$21.44\pm0.61$ & F4V &   $1.250\pm0.050$ \\
7918 &  MCY2 &$76.5204\pm0.2142$ & G1V &   $1.023\pm0.024$ \\
65982 & HDS1895 & $37.7582\pm0.6272$ &    G8V &   $0.912\pm0.020$ \\
69962 & HDS2016AB  & $44.3763\pm0.7670$ &  K5 &   $0.646\pm0.040$ \\
78401 & LAB3 &  $6.64\pm0.89$ & B0IV &  $21.878\pm6.524$ \\
79101 & NOI2 &  $14.8990\pm0.4039$ &  B9V &   $2.390\pm0.140$ \\
81023 & DSG7Aa,Ab & $24.0090\pm0.3582$ & K0V &   $0.850\pm0.040$ \\
99675 &  WRH33Aa,Ab &  $4.3432\pm0.3464$ & K3Ib &  $10.000\pm0.707$ \\
109951 & HDS3158 &  $15.1176\pm0.5342$ & G5V  &   $0.955\pm0.015$ \\
115126 & MCA74Aa,Ab & $44.8996\pm0.5572$ & G8IV &   $1.202\pm0.025$ \\
\hline\hline
(a): From Gaia DR2.
\end{tabular}

\label{tab:SB1/SB1_plx&m1}
\end{table}
\normalsize

Similarly to what we showed for our benchmark systems, the orbital parameter estimates and their uncertainties are compared in both the parameters space, through a visualization of the posterior marginal distributions, as well as in the observations space, through the projection of $1000$ randomly selected samples of the posterior distribution on the observation space, drawing trajectories from the time of the first observations $t_0$ to the first completion of the orbit $t_0+P$. The MAP estimators and their respective $95\%$ high posterior density interval estimates for these systems are summarized in Table~\ref{tab:SB1/SB1}. For comparison purposes, in this table we also report the values provided by the Orb6 and SB9 catalogues for each target. From this table we see, in general, good agreement between our orbital parameters, and those from orbital or spectral fitting (see, e.g., HIP 171, 7918, 65982, 69962\footnote{As an aside, for this object \citet{SkiffCat} report spectral types from K5V to M0V, while we adopt K5V. With a $V=9.112$ from SIMBAD, and the trigonometric parallax in Table~\ref{tab:SB1/SB1_plx&m1}, its implied $M_V = +7.35$ which indeed corresponds a K5V from \citet{Abuset2020}, while an M0V would have $M_V=+8.83$, completely off the measured value. This confirms the adequacy of our assumption on its adopted spectral type.}, 81023, 109951). We note that the argument of periapsis ($\omega$) is typically well determined by RV observations as long as the distinction between primary and secondary is unambiguous (difficult, e.g., for equal-mass binaries), and ambiguous from astrometric data alone. On the other hand, the longitude of the ascending node ($\Omega$) can be well determined from astrometric observations alone, but it suffers from the same ambiguity in the case of equal-brightness binaries. In general, purely astrometric solutions can exhibit an offset of  $180^{\circ}$ in both $\omega$ and $\Omega$ which would affect both angles simultaneously. Two textbook examples of this behavior in our sample are HIP 6564 and 78401 (adding $180^{\circ}$ to both $\omega$ and $\Omega$ lands in our results).  The few cases in which there are discrepancies of unclear origin with SB9 and/or Orb6 could be due, e.g., to quadrant ambiguities in the input astrometric position angle data, like in HIP 6564, 79101, 99675, and 115126, all of which indeed have rather large predicted values of $q$ (and could thus be more prone to quadrant ambiguity).

To avoid redundancy, the analysis is focused only on three of the twelve systems studied: HIP 3504, HIP 99675 and HIP 109951. However, we provide estimated marginal posterior distribution and MAP estimates of orbital parameters as well as orbit and RV curves for all our studied systems, similar to Figures~\ref{fig:SB1/HIP3504_params+} and~\ref{fig:SB1/HIP3504_obs+}, in our website \url{http://www.das.uchile.cl/~rmendez/B_Research/MV_RAM_SB1/SB1/}.

\subsection{HIP 3504}

This a very challenging system, because the available data consists in few and imprecise astrometric observations which only covers three distinct points on the orbit. Very few and imprecise RV observations of the primary object, covering less than one period, are also available. The observations and their errors are visualized in Figure~\ref{fig:SB1/HIP3504_obs+}.

\begin{figure}[!h]
    \centering
    \includegraphics[width=\textwidth]{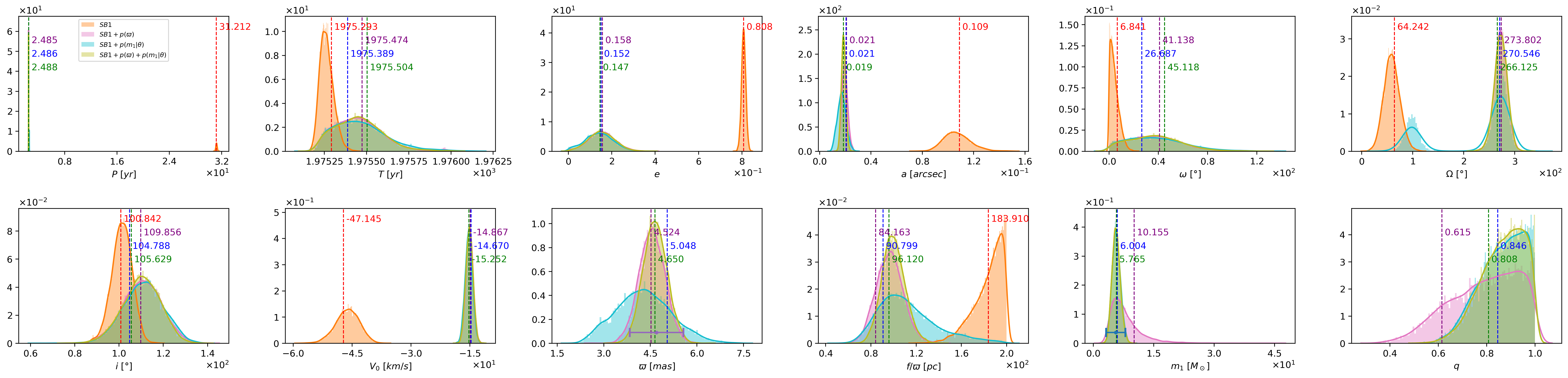}
    \caption{Marginal posterior distribution and MAP estimates of orbital parameters for the HIP 3504 binary system in the $SB1$, $SB1+p(\varpi)$, $SB1+p(m_1|\theta)$ and $SB1+p(\varpi)+p(m_1|\theta)$ cases.}
    \label{fig:SB1/HIP3504_params+}
\end{figure}

\begin{figure}[!h]
    \centering
    \includegraphics[width=\textwidth]{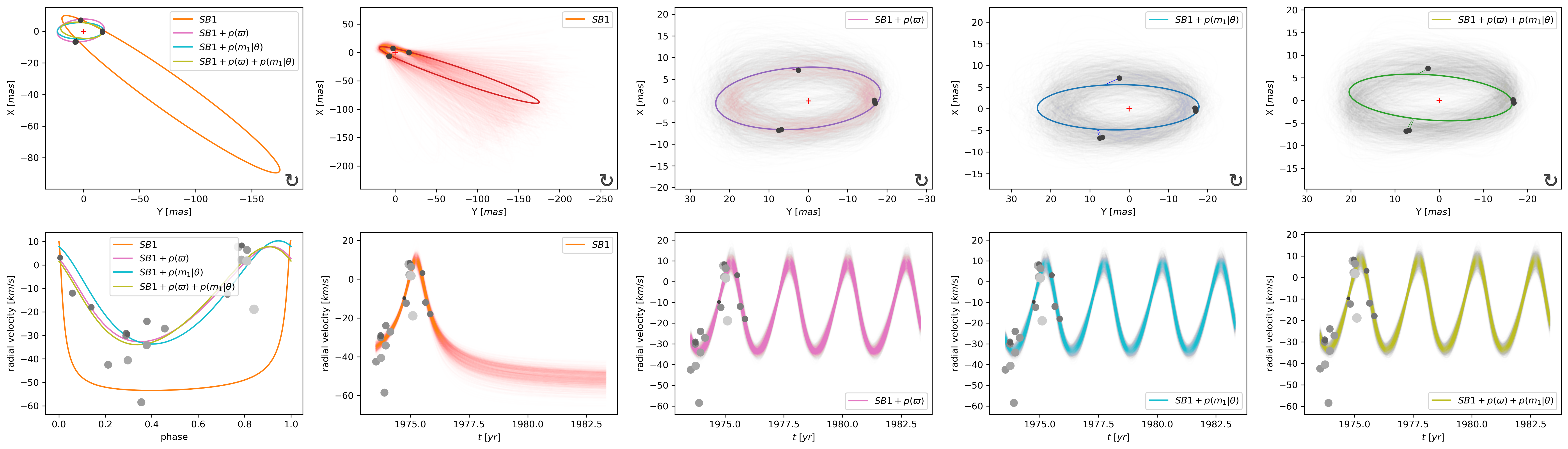}
    \caption{Estimated orbit and RV curves for the HIP 3504 binary system. First column: MAP point estimate projection of the posterior distribution for the $SB1$, $SB1+p(\varpi)$, $SB1+p(m_1|\theta)$ and $SB1+p(\varpi)+p(m_1|\theta)$ cases. Second to fifth columns: Projected posterior distribution of the $SB1$ $SB1+p(\varpi)$, $SB1+p(m_1|\theta)$, and $SB1+p(\varpi)+p(m_1|\theta)$ cases.}
    \label{fig:SB1/HIP3504_obs+}
\end{figure}

Figure~\ref{fig:SB1/HIP3504_params+} shows the posterior distributions of the $SB1$ cases with priors as well as the reference case without priors (i.e., a traditional $SB1$). These distributions are considerably different in their MAP values and dispersion compared to the reference $SB1$ case. The most significant difference is presented in the period $P$, where the posterior distribution of the $SB1$ case presents an extremely high uncertainty with a MAP of $31.15$ $[yr]$, while the posterior distribution of all the cases with priors presents a much less uncertainty with a MAP of only $2.48$ $[yr]$. The uncertainty of the orbital parameters $T,e,\omega,i$ is higher in the cases with priors compared to the $SB1$ case, while the orbital parameters $P,a,\Omega,V_0,f/\varpi$ is lower. The marginal posterior distribution of the mass ratio $q$ shows that the uncertainty of the $SB1+p(\varpi)$ case is the highest, while the uncertainty of $SB1+p(m_1|\theta)$ and $SB1+p(\varpi)+p(m_1|\theta)$ are almost identical.
 
The posterior distributions on the observation space are presented in Figure~\ref{fig:SB1/HIP3504_obs+}. The MAP estimate in the orbit space of the cases with priors are almost identical, but very different than the $SB1$ case. This is consistent with the important difference of the posterior distributions obtained for the period $P$. The orbital uncertainty of the $SB1$ case is very high in this analysis, projecting a dense cloud of feasible orbits. In contrast, the orbital uncertainty with priors $SB1+p(\varpi)$, $SB1+p(m_1|\theta)$ and $SB1+p(\varpi)+p(m_1|\theta)$ are confined to a much more limited ring of possible orbits. All those solutions are less uncertain than that of the $SB1$. The projected uncertainty in the orbit space of the $SB1+p(\varpi)$ and $SB1+p(m_1|\theta)$ cases are very similar, while the mixed case $SB1+p(\varpi)+p(m_1|\theta)$ presents a slightly lower uncertainty. Similarly, the uncertainty in the RV curve on the $SB1$ case is reduced around the epoch of observations, but very high at future epochs. In contrast, the RV uncertainty of the cases with priors presents no variation between the times with and without observations, preserving the uncertainty of the $SB1$ case at the epochs with observations. No significant differences in the RV curve's uncertainty are observed between the cases with priors. Another relevant difference between the cases with and without priors is that the period of the RV curve in the $SB1+p(\varpi)$, $SB1+p(m_1|\theta)$, $SB1+p(\varpi)+p(m_1|\theta)$ cases presents a visible lower period than the $SB1$ case. The period of the first one is visible in the figure's time window, while the period of the last case is much larger and is not visible in the figure's time window. This behavior coincides with the dramatic differences between the marginal posterior distributions of the period in the cases with and without priors presented in Figure~\ref{fig:SB1/HIP3504_params+}. Evidently, RV and/or astrometric observations over a short time-scale will quickly resolve if this is indeed a short period system.

The tremendous differences observed between the cases with and without priors show the crucial role that prior information can play, helping to pin-down some orbital parameters, the orbit, and the RV curves of binary systems when not enough observations are available.
 
\subsection{HIP 99675}
The data for this object consists in few and highly imprecise astrometric observations in two extreme zones of the orbit, and abundant and highly precise RV observations of the primary object. The observations and their errors are visualized in Figure~\ref{fig:SB1/HIP99675_obs+}.

\begin{figure}[!h]
    \centering
    \includegraphics[width=\textwidth]{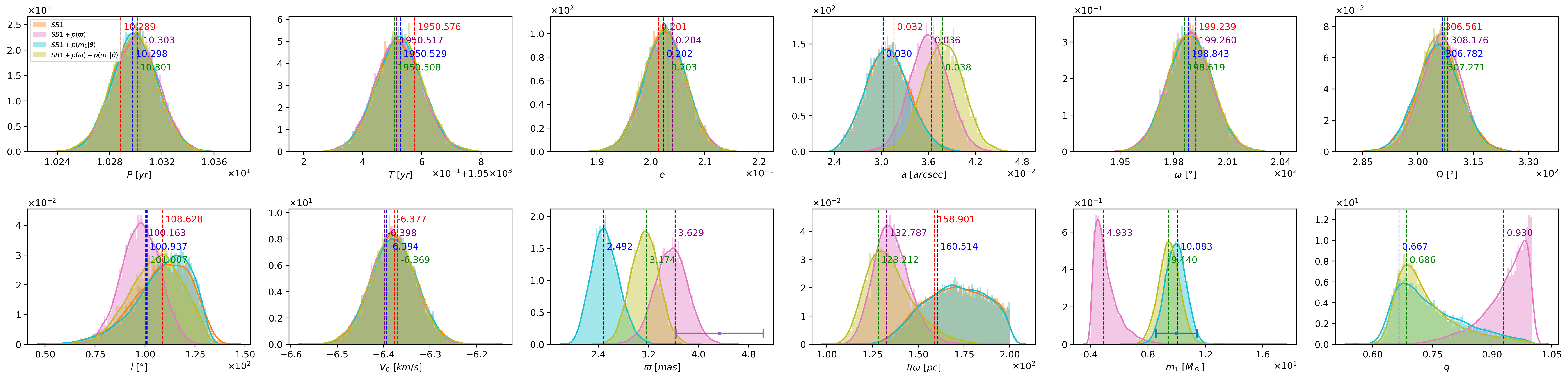}
    \caption{Same as Figure~\ref{fig:SB1/HIP3504_params+} but for the HIP 99675 binary system.}
    \label{fig:SB1/HIP99675_params+}
\end{figure}

\begin{figure}[!h]
    \centering
    \includegraphics[width=\textwidth]{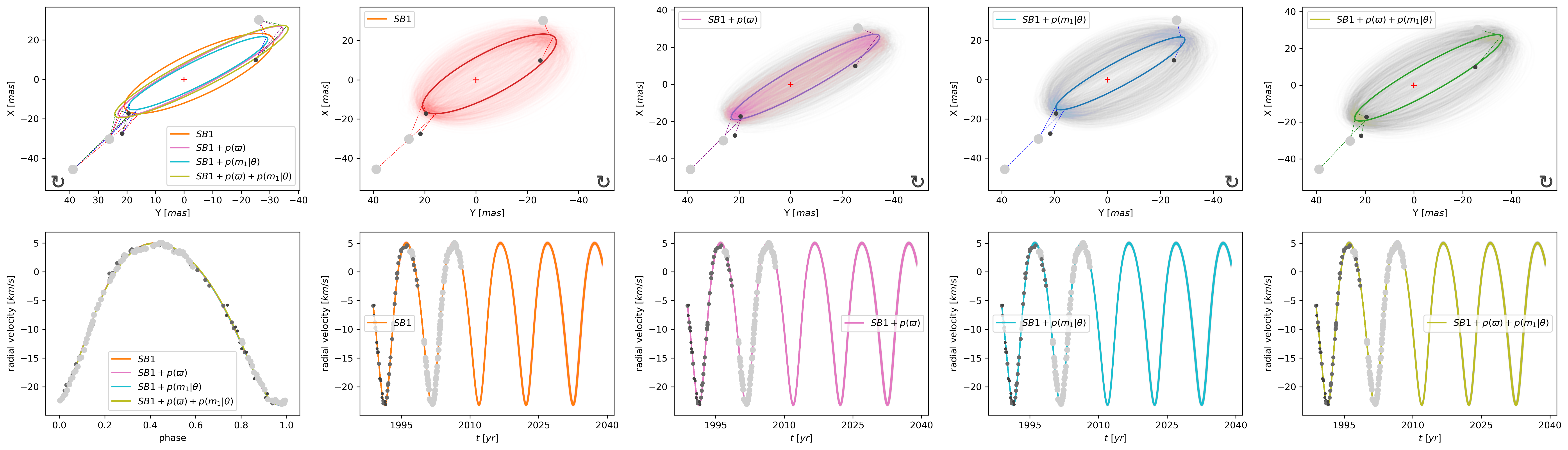}
    \caption{Same as Figure~\ref{fig:SB1/HIP3504_obs+} but for the HIP 99675 binary system.}
    \label{fig:SB1/HIP99675_obs+}
\end{figure}

Figure~\ref{fig:SB1/HIP99675_params+} shows that the posterior distributions of all the cases are identical for almost all the orbital parameters, with the exception of $a,i,\varpi,f/\varpi,m_1,q$, where the most significant differences are observed on the parameters $\varpi,m_1,q$. The marginal posterior distribution $\varpi$ of the mixed case $SB1+p(\varpi)+p(m_1|\theta)$ is in between the $SB1+p(\varpi)$ and $SB1+p(m_1|\theta)$ cases presenting the lowest uncertainty, where its MAP estimation is almost equidistant to the other cases. The marginal posterior distribution $m_1$ of the mixed case is almost equal to the obtained with the $SB1+p(m_1|\theta)$ case, but the MAP estimate is very different than the one obtained with the $SB1+p(\varpi)$ case, which has the lowest uncertainty. The posterior distributions of the mass ratio $q$ of the $SB1+p(\varpi)+p(m_1|\theta)$ and $SB1+p(m_1|\theta)$ are very similar, but quite different to the $SB1+p(\varpi)$ case. Overall, the $SB1+p(\varpi)$ offers the posterior distribution with the least uncertainty, followed by the $SB1+p(\varpi)+p(m_1|\theta)$ and $SB1+p(m_1|\theta)$ cases. Unlike all the previous systems studied, the estimated marginal posterior distribution of $\varpi$ is not equal to the prior $p(\varpi)$ (represented with the purple error bar) in the case $SB1+p(\varpi)$, presenting an appreciable bias even considering that the parameter $\varpi$ is soft-identifiable (i.e., identifiable through $p(\varpi)$). However, this phenomena turns explainable noting that the estimated posterior distribution of $q$ is saturated (to it upper bound $1$) in this case, not allowing the estimated posterior distribution of $\varpi$ to fit the imposed prior $p(\varpi)$. 
 
The obtained posterior distributions in the observation space are presented in Figure~\ref{fig:SB1/HIP99675_obs+}. The MAP estimator in the orbit space of all the cases are significantly different and none of them fit the positional data particularly well. The orbit uncertainty obtained from those distributions is extremely high in all the cases, that is expressed in a dense cloud of possible orbits. The $SB1$ case presents the highest uncertainty in the orbit space, followed by the $SB1+p(m_1|\theta)$ and $SB1+p(\varpi)+p(m_1|\theta)$ cases, the $SB1+p(\varpi)$ case being the less uncertain. In contrast, the MAP estimates in RV space of all the cases are identical, which offer a highly precise estimation of the RV measurements not directly observed. In this scenario, we could conclude that future measurements in the orbit space could really contribute to improve the estimation of parameters while no further evidence is needed from RV observations. 
 
\subsection{HIP 109951}
In this case, the available data consists in abundant and precise astrometric observations which covers less than a half of the orbit with only one less precise observation in the other half. There are abundant and highly imprecise RV observations of the primary object, that are mostly concentrated in a small segment of the phase, near periastron. The observations and their errors are visualized in Figure~\ref{fig:SB1/HIP109951_obs+}.

\begin{figure}[!h]
    \centering
    \includegraphics[width=\textwidth]{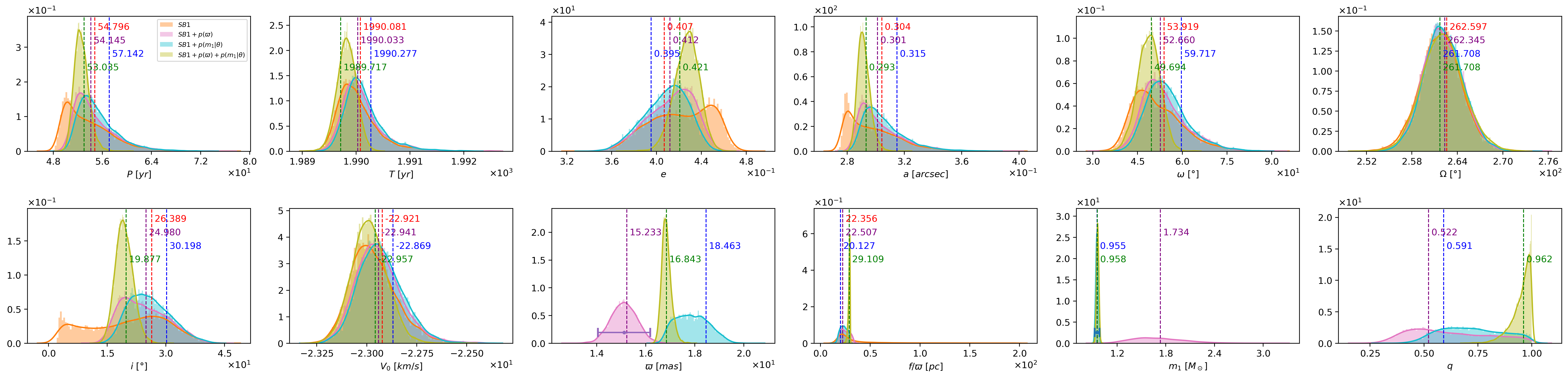}
    \caption{Same as Figure~\ref{fig:SB1/HIP3504_params+} but for the HIP 109951 binary system.}
    \label{fig:SB1/HIP109951_params+}
\end{figure}

\begin{figure}[!h]
    \centering
    \includegraphics[width=\textwidth]{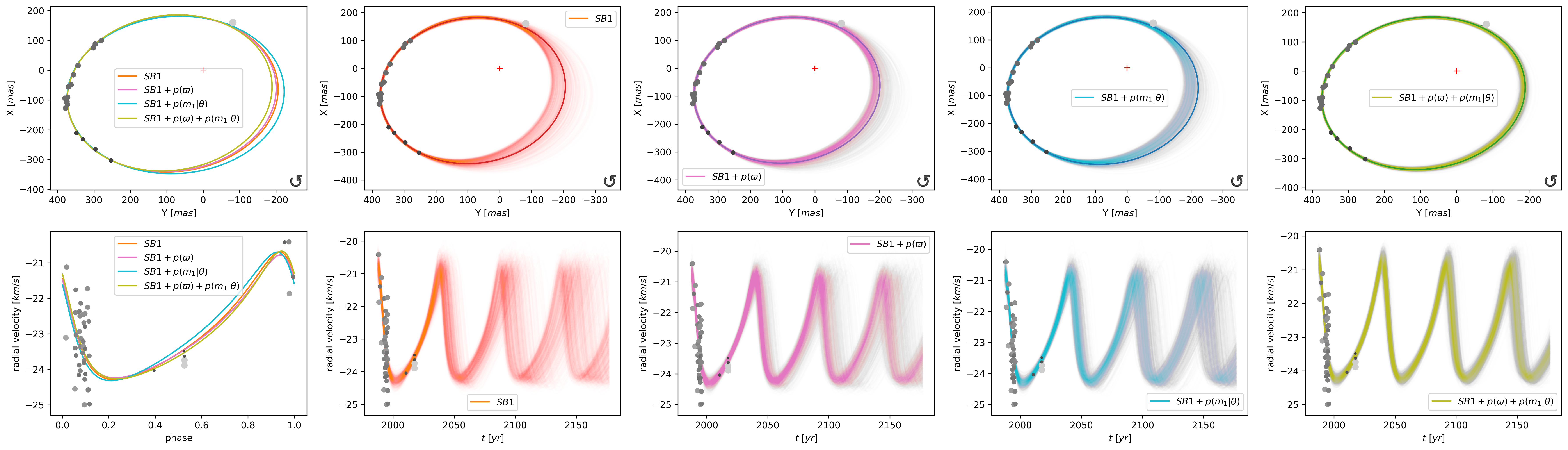}
    \caption{Same as Figure~\ref{fig:SB1/HIP3504_obs+} but for the 
  HIP 109951 binary system.}
    \label{fig:SB1/HIP109951_obs+}
\end{figure}

Figure~\ref{fig:SB1/HIP109951_params+} shows the posterior (marginal) distributions of all the cases and parameters. They are similar in almost all the orbital parameters, with the exception of the parameters $i,\varpi,f/\varpi,m_1,q$. The most significant differences are observed in the parameters $\varpi,m_1,q$. The marginal posterior distribution $\varpi$ of the mixed case $SB1+p(\varpi)+p(m_1|\theta)$ is in between the $SB1+p(\varpi)$ and $SB1+p(m_1|\theta)$ cases, exhibiting the lowest uncertainty, where its MAP estimation is almost equidistant to the other cases. The marginal posterior distribution $m_1$ of the mixed case is almost equal to the $SB1+p(m_1|\theta)$ case, but very different in its MAP estimates to the $SB1+p(\varpi)$ case, which has the lowest uncertainty. The posterior distribution of the mass ratio $q$ of the $SB1+p(\varpi)$ and $SB1+p(m_1|\theta)$ cases are very similar, but they are quite different to the $SB1+p(\varpi)+p(m_1|\theta)$ case, where the mixed case $SB1+p(\varpi)+p(m_1|\theta)$ presents the lowest uncertainty, distantly followed by the $SB1+p(m_1|\theta)$ and $SB1+p(\varpi)$ scenario. In all the cases, the uncertainty of the period $P$ is very high due to the poor orbital coverage in the astrometric and RV data. This is reflected in a wide dispersion of the corresponding marginal posterior distributions.
 
The obtained posterior distributions in the observation space are presented in Figure~\ref{fig:SB1/HIP109951_obs+}. The MAP estimates in the orbit and RV spaces shows slight but almost negligible differences between all the cases. The orbit uncertainty (obtained from the posterior distribution) is high in the segment of the orbit with no observations, and very small in the segment of the orbit with observations, which is an expected pattern. The $SB1$ case presents the highest uncertainty in the orbit space, followed by the $SB1+p(m_1|\theta)$ and $SB1+p(\varpi)$ cases, and concluding with the $SB1+p(\varpi)+p(m_1|\theta)$ case. This last case has considerable lower uncertainty when compared to all the other cases. The uncertainty in the RV is low in the zones with observations, but high in the zones with no observations. Unlike all the other studied systems, the uncertainty in the RV increases with time, which is a very interesting feature. This behavior is attributed to the high uncertainty of the orbital period $P$, since all the positional and RV observations are constrained to a small portion of the orbit. This fact does not allow an accurate estimation of the orbital period; The lack of a well- determined period causes the possible RV trajectories to get out of phase, increasing the uncertainty as time progresses. Similarly to the results obtained in the orbit space, the $SB1$ case presents the highest posterior uncertainty in RV space, followed by the results obtained in the $SB1+p(m_1|\theta)$ and $SB1+p(\varpi)$ cases. In this scenario, the mixed case $SB1+p(\varpi)+p(m_1|\theta)$ has the lowest uncertainty, being considerably lower compared to the uncertainties in the RV space of all the other cases.
 
\small
\begin{sidewaystable}[h]
\centering
\caption{MAP estimates and 95\% HDPIs derived from the marginal posterior distributions of the orbital parameters incorporating priors on the mass of the primary ($m_1$) and the trigonometric parallax ($\varpi$) of $SB1$ binary systems with astrometric data as derived from this work. In the first two lines of each target we report the values provided by the Orb6 and SB9 catalogues respectively, preserving the significant figures included in those catalogues as an indication of their precision (continued on next page).}
\resizebox{\textwidth}{!}{\begin{tabular}{cccccccccccccc}
\hline\hline
     HIP~\# &                     System &                 $P$ &                       $T$ &                     $e$ &              $a$ &                   $\omega$  &                $\Omega$ &                     $i$  &                $V_0$&              $\varpi$ &                $f/\varpi$ &         $m_1$  &                     $q$ \\
 & & [yr] & [yr] &  & [arcsec] &   [\textdegree] &  [\textdegree] & [\textdegree] & [km/s] & [mas] & [pc] &  [M$_\odot]$ & \\
\hline
171 & Orb6 & 26.28 & 1989.4 & 0.38 & 0.83 & 290. & 96. & 49. & --- & --- & --- & --- & --- \\
 & SB9 & 26.31 & 1989.57 & 0.372 & --- & 285.0 & --- & --- & -36.22 & --- & --- & --- & --- \\
         &                      $SB1$ & $26.606_{26.589}^{26.627}$ & $1882.987_{1882.887}^{1883.058}$ & $0.357_{0.355}^{0.360}$ & $0.817_{0.814}^{0.820}$ & $279.150_{278.804}^{279.427}$ & $109.239_{108.948}^{109.675}$ &    $49.758_{49.475}^{50.058}$ & $-36.292_{-36.325}^{-36.250}$ &                          — &       $5.650_{5.568}^{5.761}$ &                         — &                       — \\
          &               $SB1+p(\varpi)$ & $26.609_{26.590}^{26.628}$ & $1882.979_{1882.898}^{1883.065}$ & $0.357_{0.354}^{0.360}$ & $0.817_{0.814}^{0.820}$ & $279.137_{278.788}^{279.402}$ & $109.350_{108.965}^{109.681}$ &    $49.745_{49.481}^{50.064}$ & $-36.285_{-36.326}^{-36.251}$ & $78.971_{77.995}^{80.177}$ &       $5.662_{5.571}^{5.766}$ &   $0.863_{0.811}^{0.910}$ & $0.809_{0.780}^{0.844}$ \\
          &        $SB1+p(m_1|\theta)$ & $26.614_{26.589}^{26.628}$ & $1882.953_{1882.891}^{1883.061}$ & $0.357_{0.354}^{0.360}$ & $0.818_{0.814}^{0.820}$ & $279.111_{278.790}^{279.410}$ & $109.307_{108.956}^{109.677}$ &    $49.797_{49.485}^{50.059}$ & $-36.279_{-36.327}^{-36.251}$ & $77.075_{76.198}^{77.714}$ &       $5.660_{5.568}^{5.760}$ &   $0.950_{0.925}^{0.983}$ & $0.774_{0.753}^{0.794}$ \\
          & $SB1+p(\varpi)+p(m_1|\theta)$ & $26.603_{26.587}^{26.627}$ & $1882.997_{1882.897}^{1883.070}$ & $0.358_{0.355}^{0.360}$ & $0.819_{0.816}^{0.822}$ & $279.052_{278.767}^{279.371}$ & $109.314_{108.960}^{109.671}$ &    $49.912_{49.592}^{50.161}$ & $-36.296_{-36.329}^{-36.255}$ & $77.789_{77.016}^{78.314}$ &       $5.626_{5.531}^{5.714}$ &   $0.927_{0.907}^{0.959}$ & $0.778_{0.754}^{0.795}$ \\
\hline
  3504  & Orb6 & 2.835 & 2006.966 & 0.019 & 0.017 & 92.9 & 88.6 & 113.4. & --- & --- & --- & --- & --- \\
 & SB9 & 2.828 & 1972.95 & 0.11 & --- & 79. & --- & --- & -18.7 & --- & --- & --- & --- \\
 &                      $SB1$ & $31.212_{31.040}^{31.426}$ & $1975.293_{1975.190}^{1975.346}$ & $0.808_{0.790}^{0.828}$ & $0.109_{0.088}^{0.128}$ &      $6.841_{0.003}^{11.662}$ &    $64.242_{28.695}^{89.024}$ &  $100.842_{92.580}^{110.403}$ & $-47.145_{-52.121}^{-40.366}$ &                          — & $183.910_{157.710}^{199.999}$ &                         — &                       — \\
          &               $SB1+p(\varpi)$ &    $2.485_{2.473}^{2.499}$ & $1975.474_{1975.202}^{1975.705}$ & $0.158_{0.040}^{0.281}$ & $0.021_{0.016}^{0.024}$ &     $41.138_{0.002}^{74.341}$ & $273.802_{247.323}^{297.566}$ &  $109.856_{92.933}^{127.432}$ & $-14.867_{-16.746}^{-13.255}$ &    $4.524_{3.762}^{5.391}$ &   $84.163_{73.702}^{119.872}$ & $10.155_{2.958}^{15.745}$ & $0.615_{0.537}^{1.000}$ \\
          &        $SB1+p(m_1|\theta)$ &    $2.486_{2.472}^{2.591}$ & $1975.389_{1975.208}^{1975.772}$ & $0.152_{0.003}^{0.250}$ & $0.021_{0.011}^{0.024}$ &     $26.687_{0.059}^{80.380}$ &  $270.546_{80.951}^{296.478}$ &  $104.788_{94.145}^{129.009}$ & $-14.670_{-17.029}^{-13.440}$ &    $5.048_{2.638}^{6.008}$ &   $90.799_{69.584}^{166.468}$ &   $6.004_{3.830}^{7.825}$ & $0.846_{0.700}^{1.000}$ \\
          & $SB1+p(\varpi)+p(m_1|\theta)$ &    $2.488_{2.472}^{2.498}$ & $1975.504_{1975.211}^{1975.702}$ & $0.147_{0.031}^{0.264}$ & $0.019_{0.015}^{0.022}$ &     $45.118_{0.053}^{73.485}$ & $266.125_{247.068}^{297.564}$ &  $105.629_{93.597}^{126.030}$ & $-15.252_{-16.741}^{-13.262}$ &    $4.650_{3.925}^{5.436}$ &   $96.120_{80.178}^{120.024}$ &   $5.765_{3.812}^{7.726}$ & $0.808_{0.707}^{1.000}$ \\
\hline
  6564  & Orb6 & 16.110 & 1988.84 & 0.93 & 0.199 & 348.00 & 29. & 117. & --- & --- & --- & --- & --- \\
 & SB9 & 16.140 & 1972.74 & 0.92 & --- & 0.0 & --- & --- & --- & --- & --- & --- & --- \\
 &                      $SB1$ & $16.085_{16.069}^{16.105}$ & $1924.558_{1924.497}^{1924.604}$ & $0.933_{0.926}^{0.939}$ & $0.197_{0.195}^{0.198}$ & $170.355_{167.258}^{172.614}$ & $209.454_{207.753}^{210.863}$ & $120.621_{118.557}^{122.872}$ &   $99.743_{92.647}^{105.991}$ &                          — &    $29.468_{24.258}^{34.375}$ &                         — &                       — \\
          &               $SB1+p(\varpi)$ & $16.078_{16.064}^{16.095}$ & $1924.575_{1924.525}^{1924.619}$ & $0.940_{0.936}^{0.945}$ & $0.196_{0.195}^{0.198}$ & $168.056_{165.407}^{170.139}$ & $207.952_{206.326}^{209.158}$ & $122.092_{120.661}^{124.688}$ &    $92.260_{89.137}^{94.517}$ & $20.870_{19.795}^{22.108}$ &    $23.193_{21.506}^{24.880}$ &   $1.663_{1.375}^{1.969}$ & $0.938_{0.848}^{1.000}$ \\
          &        $SB1+p(m_1|\theta)$ & $16.075_{16.063}^{16.093}$ & $1924.582_{1924.530}^{1924.621}$ & $0.942_{0.938}^{0.947}$ & $0.197_{0.195}^{0.198}$ & $166.560_{164.749}^{169.507}$ & $207.002_{205.779}^{208.712}$ & $123.031_{121.150}^{125.147}$ &    $89.722_{87.754}^{91.881}$ & $22.854_{22.061}^{23.324}$ &    $21.365_{20.490}^{22.533}$ &   $1.261_{1.177}^{1.372}$ & $0.954_{0.905}^{1.000}$ \\
          & $SB1+p(\varpi)+p(m_1|\theta)$ & $16.080_{16.065}^{16.095}$ & $1924.569_{1924.523}^{1924.614}$ & $0.942_{0.938}^{0.947}$ & $0.196_{0.195}^{0.197}$ & $167.417_{165.156}^{169.743}$ & $207.453_{206.090}^{208.844}$ & $122.719_{121.119}^{125.011}$ &    $90.702_{88.307}^{92.123}$ & $22.440_{21.938}^{22.976}$ &    $21.923_{21.103}^{22.750}$ &   $1.317_{1.225}^{1.396}$ & $0.968_{0.926}^{1.000}$ \\
\hline
  7918 & Orb6 & 18.12 & 1993.630 & 0.434 & 0.631 & 144.3 & 205.0 & 98.8. & --- & --- & --- & --- & --- \\
 & SB9 & 19.73 & 1996.999 & 0.43 & --- & 203.4 & --- & --- & 3.31 & --- & --- & --- & --- \\
&                      $SB1$ & $19.731_{19.701}^{19.761}$ & $1996.969_{1996.914}^{1997.007}$ & $0.439_{0.434}^{0.444}$ & $0.608_{0.603}^{0.613}$ & $201.731_{200.528}^{202.536}$ & $208.513_{208.049}^{208.814}$ &    $97.549_{97.045}^{97.998}$ &       $3.323_{3.305}^{3.331}$ &                          — &       $2.928_{2.893}^{2.964}$ &                         — &                       — \\
          &               $SB1+p(\varpi)$ & $19.727_{19.702}^{19.762}$ & $1996.955_{1996.913}^{1997.007}$ & $0.440_{0.434}^{0.444}$ & $0.607_{0.602}^{0.613}$ & $201.486_{200.488}^{202.522}$ & $208.430_{208.038}^{208.798}$ &    $97.653_{97.008}^{97.981}$ &       $3.319_{3.305}^{3.331}$ & $76.577_{76.120}^{76.951}$ &       $2.932_{2.893}^{2.965}$ &   $0.991_{0.968}^{1.032}$ & $0.290_{0.284}^{0.294}$ \\
          &        $SB1+p(m_1|\theta)$ & $19.729_{19.701}^{19.761}$ & $1996.962_{1996.913}^{1997.008}$ & $0.439_{0.434}^{0.444}$ & $0.609_{0.603}^{0.613}$ & $201.562_{200.508}^{202.548}$ & $208.476_{208.037}^{208.798}$ &    $97.511_{97.038}^{98.000}$ &       $3.320_{3.305}^{3.331}$ & $76.054_{74.771}^{77.192}$ &       $2.925_{2.894}^{2.966}$ &   $1.026_{0.976}^{1.069}$ & $0.286_{0.281}^{0.292}$ \\
          & $SB1+p(\varpi)+p(m_1|\theta)$ & $19.730_{19.703}^{19.763}$ & $1996.959_{1996.913}^{1997.006}$ & $0.438_{0.434}^{0.443}$ & $0.610_{0.604}^{0.613}$ & $201.562_{200.557}^{202.583}$ & $208.344_{208.033}^{208.803}$ &    $97.405_{97.023}^{97.980}$ &       $3.320_{3.306}^{3.332}$ & $76.555_{76.067}^{76.856}$ &       $2.915_{2.892}^{2.958}$ &   $1.008_{0.981}^{1.035}$ & $0.287_{0.284}^{0.292}$ \\
\hline
 65982 & Orb6 & 3.2448 & 1994.324 & 0.551 & 0.0938 & 352.3 & 315.2 & 10. & --- & --- & --- & --- & --- \\
 & SB9 & 3.2526 & 1994.332 & 0.641 & --- & 355.8 & --- & --- & -51.148 & --- & --- & --- & --- \\
 &                      $SB1$ &    $3.232_{3.229}^{3.236}$ & $1984.702_{1984.663}^{1984.732}$ & $0.517_{0.503}^{0.537}$ & $0.099_{0.096}^{0.101}$ &      $5.378_{0.023}^{12.820}$ & $303.150_{297.656}^{308.949}$ &    $25.078_{16.692}^{29.869}$ & $-51.128_{-51.295}^{-51.020}$ &                          — &       $5.755_{4.465}^{8.515}$ &                         — &                       — \\
          &               $SB1+p(\varpi)$ &    $3.232_{3.229}^{3.237}$ & $1984.702_{1984.668}^{1984.734}$ & $0.517_{0.505}^{0.539}$ & $0.099_{0.096}^{0.101}$ &      $5.834_{0.007}^{12.737}$ & $303.021_{297.249}^{308.552}$ &    $23.468_{14.228}^{29.192}$ & $-51.184_{-51.297}^{-51.021}$ & $37.852_{36.578}^{39.030}$ &       $6.017_{4.444}^{9.592}$ &   $1.310_{0.955}^{1.549}$ & $0.295_{0.201}^{0.567}$ \\
          &        $SB1+p(m_1|\theta)$ &    $3.232_{3.229}^{3.237}$ & $1984.706_{1984.666}^{1984.733}$ & $0.518_{0.504}^{0.538}$ & $0.099_{0.096}^{0.101}$ &      $7.146_{0.014}^{13.025}$ & $301.829_{297.396}^{308.687}$ &    $23.828_{15.341}^{29.627}$ & $-51.161_{-51.299}^{-51.026}$ & $42.138_{39.069}^{44.251}$ &       $6.162_{4.463}^{9.126}$ &   $0.909_{0.873}^{0.951}$ & $0.351_{0.238}^{0.551}$ \\
          & $SB1+p(\varpi)+p(m_1|\theta)$ &    $3.233_{3.229}^{3.237}$ & $1984.701_{1984.671}^{1984.735}$ & $0.537_{0.523}^{0.546}$ & $0.096_{0.095}^{0.097}$ &      $5.802_{0.291}^{12.429}$ & $302.935_{296.521}^{308.101}$ &    $14.492_{11.058}^{17.042}$ & $-51.202_{-51.291}^{-51.021}$ & $38.245_{36.710}^{39.264}$ &     $10.120_{8.577}^{12.578}$ &   $0.920_{0.878}^{0.955}$ & $0.631_{0.502}^{0.865}$ \\
\hline
 69962  & Orb6 & 18.67 & 2003.74 & 0.302 & 0.348 & 71.0 & 285.0 & 107.9 & --- & --- & --- & --- & --- \\
 & SB9 & 18.67 & 1984.79 & 0.306 & --- & 67.1 & --- & --- & 11.55 & --- & --- & --- & --- \\
 &                      $SB1$ & $18.802_{18.706}^{18.924}$ & $1984.933_{1984.798}^{1985.069}$ & $0.301_{0.290}^{0.313}$ & $0.349_{0.345}^{0.352}$ &    $71.741_{70.489}^{73.544}$ & $285.358_{284.712}^{285.900}$ & $107.950_{107.294}^{108.443}$ &    $11.548_{11.391}^{11.752}$ &                          — &       $7.163_{6.571}^{7.656}$ &                         — &                       — \\
          &               $SB1+p(\varpi)$ & $18.815_{18.703}^{18.921}$ & $1984.916_{1984.790}^{1985.066}$ & $0.301_{0.290}^{0.312}$ & $0.349_{0.345}^{0.352}$ &    $71.962_{70.453}^{73.483}$ & $285.430_{284.693}^{285.877}$ & $107.836_{107.323}^{108.467}$ &    $11.503_{11.389}^{11.744}$ & $44.504_{42.880}^{45.843}$ &       $7.167_{6.599}^{7.669}$ &   $0.928_{0.827}^{1.059}$ & $0.468_{0.412}^{0.525}$ \\
          &        $SB1+p(m_1|\theta)$ & $18.838_{18.707}^{18.928}$ & $1984.886_{1984.791}^{1985.071}$ & $0.301_{0.290}^{0.313}$ & $0.348_{0.345}^{0.352}$ &    $71.996_{70.372}^{73.510}$ & $285.375_{284.693}^{285.841}$ & $107.840_{107.308}^{108.453}$ &    $11.564_{11.393}^{11.747}$ & $48.766_{47.657}^{51.324}$ &       $7.243_{6.597}^{7.680}$ &   $0.664_{0.566}^{0.721}$ & $0.546_{0.488}^{0.609}$ \\
          & $SB1+p(\varpi)+p(m_1|\theta)$ & $18.818_{18.716}^{18.933}$ & $1984.900_{1984.759}^{1985.033}$ & $0.298_{0.287}^{0.310}$ & $0.347_{0.344}^{0.351}$ &    $71.621_{70.232}^{73.357}$ & $285.200_{284.737}^{285.920}$ & $108.155_{107.461}^{108.585}$ &    $11.556_{11.404}^{11.757}$ & $46.793_{45.742}^{47.815}$ &       $7.428_{6.945}^{7.994}$ &   $0.752_{0.700}^{0.814}$ & $0.533_{0.481}^{0.593}$ \\
\hline
\hline
\end{tabular}
}
\label{tab:SB1/SB1}
\end{sidewaystable}
\normalsize

\small
\begin{sidewaystable}[h]
\centering
\caption{Table 6 (Contd.). MAP estimates and 95\% HDPIs derived from the marginal posterior distributions of the orbital parameters incorporating priors on the mass of the primary ($m_1$) and the trigonometric parallax ($\varpi$) of $SB1$ binary systems with astrometric data as derived from this work. In the first two lines of each target we report the values provided by the Orb6 and SB9 catalogues respectively, preserving the significant figures included in those catalogues as an indication of their precision.}
\resizebox{\textwidth}{!}{\begin{tabular}{cccccccccccccc}
\hline\hline
     HIP~\# &                     System &                 $P$ &                       $T$ &                     $e$ &              $a$ &                   $\omega$  &                $\Omega$ &                     $i$  &                $V_0$&              $\varpi$ &                $f/\varpi$ &         $m_1$  &                     $q$ \\
 & & [yr] & [yr] &  & [arcsec] &   [\textdegree] &  [\textdegree] & [\textdegree] & [km/s] & [mas] & [pc] &  [M$_\odot]$ & \\
\hline
 78401 & Orb6 & 10.802 & 2011.501 & 0.9373 & 0.09894 & 359.5 & 175.0 & 34.12 & --- & --- & --- & --- & --- \\
 & SB9 & 10.58 & 2000.69 & 0.94 & --- & 179. & --- & --- & -6.0 & --- & --- & --- & --- \\
 &                      $SB1$ & $10.808_{10.807}^{10.810}$ & $1979.072_{1979.069}^{1979.076}$ & $0.935_{0.934}^{0.936}$ & $0.099_{0.099}^{0.099}$ & $173.337_{173.243}^{173.422}$ &       $0.022_{0.000}^{0.074}$ &    $35.170_{34.400}^{35.937}$ &    $-5.756_{-6.690}^{-5.131}$ &                          — &    $53.311_{51.447}^{54.849}$ &                         — &                       — \\
          &               $SB1+p(\varpi)$ & $10.808_{10.807}^{10.809}$ & $1979.079_{1979.075}^{1979.082}$ & $0.939_{0.938}^{0.940}$ & $0.099_{0.099}^{0.099}$ & $180.473_{179.422}^{181.696}$ & $354.172_{353.085}^{355.012}$ &    $32.224_{31.268}^{33.017}$ &    $-6.417_{-7.127}^{-5.664}$ &    $6.345_{5.120}^{8.381}$ &    $56.352_{54.467}^{58.265}$ & $20.917_{5.995}^{38.771}$ & $0.557_{0.381}^{0.868}$ \\
          &        $SB1+p(m_1|\theta)$ & $10.808_{10.807}^{10.809}$ & $1979.078_{1979.075}^{1979.082}$ & $0.939_{0.938}^{0.940}$ & $0.099_{0.099}^{0.099}$ & $180.786_{179.448}^{181.696}$ & $353.898_{353.137}^{355.054}$ &    $32.222_{31.313}^{33.062}$ &    $-6.308_{-7.157}^{-5.657}$ &    $6.360_{5.513}^{8.378}$ &    $56.308_{54.391}^{58.241}$ & $20.763_{5.792}^{29.902}$ & $0.558_{0.439}^{0.885}$ \\
          & $SB1+p(\varpi)+p(m_1|\theta)$ & $10.808_{10.807}^{10.809}$ & $1979.079_{1979.075}^{1979.082}$ & $0.939_{0.938}^{0.940}$ & $0.099_{0.099}^{0.099}$ & $180.822_{179.417}^{181.757}$ & $353.867_{353.075}^{355.062}$ &    $32.074_{31.248}^{33.037}$ &    $-6.571_{-7.170}^{-5.665}$ &    $6.611_{5.549}^{7.687}$ &    $56.231_{54.481}^{58.350}$ & $18.090_{8.753}^{29.606}$ & $0.592_{0.453}^{0.770}$ \\
\hline
 79101 &  Orb6 & 1.546 & 1996.105 & 0.522 & 0.0321 & 351.9 & 9.1 & 12.1 & --- & --- & --- & --- & --- \\
 & SB9 & 1.535 & 1969.829 & 0.47 & --- & 357. & --- & --- & -16.79 & --- & --- & --- & --- \\
 &                      $SB1$ &    $1.545_{1.544}^{1.547}$ & $1966.730_{1966.719}^{1966.748}$ & $0.469_{0.446}^{0.487}$ & $0.034_{0.031}^{0.035}$ & $351.266_{348.282}^{357.003}$ & $186.628_{180.994}^{192.459}$ &     $26.920_{0.929}^{34.953}$ & $-16.684_{-16.769}^{-16.649}$ &                          — &     $7.265_{4.199}^{143.186}$ &                         — &                       — \\
          &               $SB1+p(\varpi)$ &    $1.546_{1.544}^{1.546}$ & $1966.730_{1966.719}^{1966.748}$ & $0.468_{0.446}^{0.488}$ & $0.033_{0.031}^{0.035}$ & $352.102_{348.254}^{357.043}$ & $187.885_{180.877}^{192.396}$ &     $26.389_{5.418}^{31.660}$ & $-16.691_{-16.767}^{-16.648}$ & $14.886_{14.101}^{15.649}$ &      $7.480_{5.522}^{32.833}$ &   $4.136_{2.035}^{4.975}$ & $0.125_{0.080}^{0.933}$ \\
          &        $SB1+p(m_1|\theta)$ &    $1.544_{1.543}^{1.546}$ & $1966.758_{1966.751}^{1966.765}$ & $0.455_{0.439}^{0.480}$ & $0.033_{0.031}^{0.035}$ &       $0.492_{0.000}^{1.777}$ & $179.048_{175.432}^{183.776}$ &     $20.520_{5.355}^{31.150}$ & $-16.770_{-16.813}^{-16.712}$ & $17.446_{14.306}^{18.710}$ &      $9.437_{5.134}^{32.671}$ &   $2.381_{2.104}^{2.657}$ & $0.197_{0.104}^{0.937}$ \\
          & $SB1+p(\varpi)+p(m_1|\theta)$ &    $1.546_{1.544}^{1.547}$ & $1966.728_{1966.719}^{1966.748}$ & $0.475_{0.450}^{0.490}$ & $0.033_{0.031}^{0.034}$ & $351.371_{348.320}^{357.075}$ & $188.084_{181.365}^{192.976}$ &       $7.186_{5.485}^{8.725}$ & $-16.700_{-16.762}^{-16.644}$ & $15.258_{14.369}^{15.716}$ &    $27.491_{22.763}^{34.582}$ &   $2.350_{2.147}^{2.661}$ & $0.723_{0.544}^{1.000}$ \\
\hline
 81023  &  Orb6 & 0.61867 & 1988.4316 & 0.3067 & 0.016 & 20.46 & 162.7 & 117.9 & --- & --- & --- & --- & --- \\
 & SB9 & 0.6190 & 1988.431 & 0.3114 & --- & 20.4 & --- & --- & -51.33 & --- & --- & --- & --- \\
  &                      $SB1$ &    $0.619_{0.618}^{0.619}$ & $1987.813_{1987.806}^{1987.820}$ & $0.310_{0.294}^{0.328}$ & $0.016_{0.014}^{0.018}$ &    $20.195_{16.889}^{23.999}$ & $161.069_{153.844}^{170.946}$ & $119.061_{112.704}^{130.369}$ & $-51.260_{-51.621}^{-50.997}$ &                          — &    $33.102_{28.415}^{43.433}$ &                         — &                       — \\
          &               $SB1+p(\varpi)$ &    $0.619_{0.618}^{0.619}$ & $1987.819_{1987.812}^{1987.824}$ & $0.290_{0.279}^{0.311}$ & $0.023_{0.022}^{0.024}$ &    $24.221_{20.134}^{27.072}$ & $159.663_{152.259}^{165.390}$ & $107.091_{103.961}^{110.598}$ & $-51.259_{-51.590}^{-50.999}$ & $23.314_{22.638}^{23.984}$ &    $21.298_{20.627}^{21.966}$ &   $1.232_{1.117}^{1.342}$ & $0.986_{0.957}^{1.000}$ \\
          &        $SB1+p(m_1|\theta)$ &    $0.619_{0.618}^{0.619}$ & $1987.816_{1987.811}^{1987.823}$ & $0.321_{0.304}^{0.339}$ & $0.017_{0.015}^{0.019}$ &    $21.412_{18.731}^{25.428}$ & $153.044_{145.578}^{160.873}$ & $104.075_{101.142}^{107.873}$ & $-51.527_{-51.858}^{-51.277}$ & $18.017_{16.349}^{20.571}$ &    $27.688_{24.016}^{30.199}$ &   $1.042_{0.987}^{1.095}$ & $0.996_{0.988}^{1.000}$ \\
          & $SB1+p(\varpi)+p(m_1|\theta)$ &    $0.619_{0.618}^{0.619}$ & $1987.820_{1987.814}^{1987.826}$ & $0.311_{0.291}^{0.324}$ & $0.022_{0.021}^{0.022}$ &    $23.834_{20.695}^{27.220}$ & $156.601_{148.529}^{161.970}$ & $102.925_{100.258}^{106.190}$ & $-51.549_{-51.785}^{-51.210}$ & $23.645_{22.849}^{24.166}$ &    $21.083_{20.645}^{21.844}$ &   $1.019_{0.972}^{1.076}$ & $0.994_{0.989}^{1.000}$ \\
\hline
 99675  &  Orb6 & 10.040 & 1963.012 & 0.118 & 0.043 & 129.8 & 144.9 & 104.7 & --- & --- & --- & --- & --- \\
 & SB9 & 10.358 & 2002.19 & 0.2084 & --- & 204.5 & --- & --- & -6.421 & --- & --- & --- & --- \\
 &                      $SB1$ & $10.289_{10.266}^{10.334}$ & $1950.576_{1950.371}^{1950.673}$ & $0.201_{0.195}^{0.210}$ & $0.032_{0.026}^{0.036}$ & $199.239_{196.551}^{201.265}$ & $306.561_{294.070}^{317.579}$ &  $108.628_{81.075}^{134.172}$ &    $-6.377_{-6.469}^{-6.281}$ &                          — & $158.901_{142.079}^{199.955}$ &                         — &                       — \\
          &               $SB1+p(\varpi)$ & $10.303_{10.266}^{10.333}$ & $1950.517_{1950.374}^{1950.673}$ & $0.204_{0.195}^{0.210}$ & $0.036_{0.032}^{0.041}$ & $199.260_{196.552}^{201.314}$ & $308.176_{295.824}^{317.199}$ &  $100.163_{78.550}^{115.655}$ &    $-6.398_{-6.476}^{-6.292}$ &    $3.629_{3.065}^{4.082}$ & $132.787_{117.708}^{155.199}$ &   $4.933_{4.031}^{7.012}$ & $0.930_{0.808}^{1.000}$ \\
          &        $SB1+p(m_1|\theta)$ & $10.298_{10.266}^{10.332}$ & $1950.529_{1950.367}^{1950.668}$ & $0.202_{0.195}^{0.210}$ & $0.030_{0.026}^{0.036}$ & $198.843_{196.568}^{201.305}$ & $306.782_{294.111}^{316.830}$ &  $100.937_{82.350}^{131.660}$ &    $-6.394_{-6.476}^{-6.287}$ &    $2.492_{2.094}^{2.943}$ & $160.514_{142.528}^{199.997}$ & $10.083_{8.536}^{11.302}$ & $0.667_{0.623}^{0.924}$ \\
          & $SB1+p(\varpi)+p(m_1|\theta)$ & $10.301_{10.264}^{10.331}$ & $1950.508_{1950.379}^{1950.679}$ & $0.203_{0.195}^{0.210}$ & $0.038_{0.033}^{0.043}$ & $198.619_{196.656}^{201.390}$ & $307.271_{296.018}^{317.073}$ &  $101.007_{82.384}^{129.779}$ &    $-6.369_{-6.474}^{-6.286}$ &    $3.174_{2.716}^{3.570}$ & $128.212_{113.424}^{162.508}$ &  $9.440_{8.048}^{10.949}$ & $0.686_{0.631}^{0.910}$ \\
\hline
109951  &  Orb6 & 50.49 & 1989.88 & 0.451 & 0.2834 & 45.2 & 206.0 & 17.7 & --- & --- & --- & --- & --- \\
 & SB9 & 50.49 & 1989.88 & 0.452 & --- & 45.2 & --- & --- & -23.04 & --- & --- & --- & --- \\
&                      $SB1$ & $54.796_{48.371}^{61.195}$ & $1990.081_{1989.393}^{1990.829}$ & $0.407_{0.373}^{0.465}$ & $0.304_{0.275}^{0.334}$ &    $53.919_{38.455}^{69.076}$ & $262.597_{256.406}^{267.213}$ &     $26.389_{2.533}^{34.976}$ & $-22.921_{-23.167}^{-22.726}$ &                          — &   $22.356_{13.693}^{139.427}$ &                         — &                       — \\
          &               $SB1+p(\varpi)$ & $54.145_{49.892}^{61.842}$ & $1990.033_{1989.503}^{1990.923}$ & $0.412_{0.371}^{0.448}$ & $0.301_{0.282}^{0.336}$ &    $52.660_{41.197}^{70.223}$ & $262.345_{256.813}^{267.118}$ &    $24.980_{15.139}^{36.338}$ & $-22.941_{-23.142}^{-22.705}$ & $15.233_{14.064}^{16.111}$ &    $22.507_{17.239}^{33.170}$ &   $1.734_{1.174}^{2.337}$ & $0.522_{0.348}^{0.974}$ \\
          &        $SB1+p(m_1|\theta)$ & $57.142_{50.329}^{61.911}$ & $1990.277_{1989.567}^{1990.943}$ & $0.395_{0.369}^{0.442}$ & $0.315_{0.285}^{0.336}$ &    $59.717_{42.404}^{70.114}$ & $261.708_{257.161}^{267.265}$ &    $30.198_{16.674}^{36.155}$ & $-22.869_{-23.137}^{-22.710}$ & $18.463_{16.757}^{19.099}$ &    $20.127_{17.627}^{29.597}$ &   $0.955_{0.926}^{0.985}$ & $0.591_{0.506}^{0.992}$ \\
          & $SB1+p(\varpi)+p(m_1|\theta)$ & $53.035_{50.393}^{54.951}$ & $1989.717_{1989.475}^{1990.167}$ & $0.421_{0.405}^{0.448}$ & $0.293_{0.284}^{0.301}$ &    $49.694_{41.413}^{56.121}$ & $261.708_{256.664}^{267.486}$ &    $19.877_{15.208}^{23.874}$ & $-22.957_{-23.153}^{-22.814}$ & $16.843_{16.488}^{17.234}$ &    $29.109_{26.896}^{30.299}$ &   $0.958_{0.934}^{0.990}$ & $0.962_{0.859}^{1.000}$ \\
\hline
115126  &  Orb6 & 6.321 & 2012.301 & 0.173 & 0.189 & 28.3 & 314.9 & 44.5 & --- & --- & --- & --- & --- \\
 & SB9 & 6.292 & 2006.032 & 0.1620 & --- & 212.13 & --- & --- & 9.9 & --- & --- & --- & --- \\
&                      $SB1$ &    $6.325_{6.321}^{6.328}$ & $1980.738_{1980.724}^{1980.755}$ & $0.160_{0.158}^{0.163}$ & $0.191_{0.189}^{0.193}$ & $212.995_{212.552}^{213.564}$ & $341.293_{340.775}^{341.807}$ &    $49.712_{48.803}^{50.741}$ &    $-1.613_{-1.622}^{-1.594}$ &                          — &       $8.695_{8.503}^{8.856}$ &                         — &                       — \\
          &               $SB1+p(\varpi)$ &    $6.325_{6.322}^{6.328}$ & $1980.737_{1980.724}^{1980.755}$ & $0.161_{0.158}^{0.163}$ & $0.191_{0.189}^{0.193}$ & $213.001_{212.543}^{213.586}$ & $341.154_{340.807}^{341.829}$ &    $49.900_{48.771}^{50.711}$ &    $-1.606_{-1.622}^{-1.595}$ & $44.749_{43.846}^{46.018}$ &       $8.658_{8.512}^{8.859}$ &   $1.188_{1.055}^{1.282}$ & $0.633_{0.609}^{0.676}$ \\
          &        $SB1+p(m_1|\theta)$ &    $6.325_{6.322}^{6.328}$ & $1980.738_{1980.724}^{1980.754}$ & $0.160_{0.158}^{0.163}$ & $0.190_{0.189}^{0.192}$ & $213.114_{212.569}^{213.577}$ & $341.212_{340.773}^{341.790}$ &    $49.573_{48.802}^{50.741}$ &    $-1.610_{-1.623}^{-1.595}$ & $44.299_{43.910}^{45.266}$ &       $8.726_{8.503}^{8.853}$ &   $1.217_{1.154}^{1.253}$ & $0.630_{0.615}^{0.649}$ \\
          & $SB1+p(\varpi)+p(m_1|\theta)$ &    $6.326_{6.322}^{6.328}$ & $1980.739_{1980.724}^{1980.755}$ & $0.160_{0.158}^{0.163}$ & $0.191_{0.189}^{0.193}$ & $213.227_{212.546}^{213.570}$ & $341.296_{340.764}^{341.813}$ &    $49.887_{48.835}^{50.741}$ &    $-1.611_{-1.623}^{-1.596}$ & $44.758_{44.095}^{45.270}$ &       $8.647_{8.507}^{8.837}$ &   $1.195_{1.151}^{1.241}$ & $0.631_{0.615}^{0.648}$ \\
\hline
\hline
\end{tabular}
}
\end{sidewaystable}
\normalsize

\subsection{Concluding remarks}
The experimental results presented in this section show the adequacy and usefulness of the proposed Bayesian inference methodology to characterize and visualize the posterior uncertainty of $SB1$ visual-spectroscopic binary systems. 
Interesting results are obtained in some of the binary systems evaluated. For example, HIP 3504 shows an extremely high uncertainty on its orbit and RV curve, due to the few available observations. However, the incorporation the priors allows us to radically reduce the (posterior) uncertainty of the estimates, reaching completely different solutions to the ones obtained without the prior information. This is reflected in a tighter orbit and in a RV curve with a lower period in Fig.\ref{fig:SB1/HIP3504_obs+}. The drastic change in period, reflects also in a large difference in the predicted systemic velocity $V_0$ (see Table~\ref{tab:SB1/SB1}).

The results obtained for the system HIP 99675 show that the effect of the prior incorporation is mostly concentrated in the trio of soft-identifiable parameters $\varpi$, $m_1$ and $q$, affecting only slightly the other orbital parameters. This behavior express the robustness of the estimation when abundant and precise observations are available. On the soft-identifiable parameters, we observe that the posterior distributions of the $SB1+p(m_1|\theta)$ and the $SB1+p(\varpi)+p(m_1|\theta)$ cases are similar, but usually quite different to the case $SB1+p(\varpi)$. This behavior can be interpreted as the prior information of the primary object mass $p(m_1|\theta)$ being more reliable than the prior on the parallax $p(\varpi)$ for this system. Finally, the results obtained for the system HIP 109951 show the usefulness of providing a good characterization of the uncertainty. In this case, we observe that the uncertainty in the half portion of the orbit with no observations is considerably higher than the other half, but most importantly, the uncertainty of the RV curve visibly increases with time. This behavior is attributed to the large uncertainty of the period of the system, making the trajectories of the RV get out of phase with time. 

On the prior information incorporated, we observe that the posterior distribution of the cases $SB1+p(\varpi)$ and $p(m_1|\theta)$ are very similar to the one obtained in the $SB1$ case. However the mixed case $SB1+p(\varpi)+p(m_1|\theta)$ shows a visible uncertainty reduction in all the orbital parameters with a very similar MAP value. Regarding the posterior distribution of the mass ratio $q$ for HIP 109951, we observe that the cases $SB1+p(\varpi)$ and $p(m_1|\theta)$ provide a poor estimation, reflected in a very high uncertainty. However, the mixed case $SB1+p(\varpi)+p(m_1|\theta)$ exhibits a much more constrained posterior distribution than the other cases, considerably differing in their MAP estimates. It is important to note that, as the posterior distribution on the mass ratio $q$ of the $SB1+p(\varpi)$ and $p(m_1|\theta)$ cases present a very high uncertainty (with the shape of an almost uniform distribution), its corresponding MAP estimates are not very reliable, but the mixed case $SB1+p(\varpi)+p(m_1|\theta)$ does. This last result shows the relevance of incorporating both sources of prior information to obtain more robust estimates, and it explains the large variation in the predicted $q$ in the three cases with priors. Indeed, a similar behavior was already seen in the case of the benchmark HIP 89000 (see the bottom right panel on Figure~\ref{fig:SB2/YSC132AaAb_params+}).

By looking at Table~\ref{tab:SB1/SB1} we can see that, besides HIP 109951 discussed above, two other cases have large variation in the predicted $q$ depending on the prior used, namely HIP 65982 and 79101, and we discuss them in turn. For HIP 65982 the mixed case induces a large bias in some orbital parameters and a considerable increase in the uncertainty for $q$ (see the corresponding PDFs in our web site). For these reasons, in this case, one should probably favor the $SB1+p(\varpi)$ solution (which is similar to the $SB1++p(m_1|\theta)$ case), and a rather small value of $q\sim 0.3$

For HIP 79101, the most notable "feature" is a bi-modal distribution of the PDF (see our web site) for the auxiliary parameter $f/\pi$ in the $SB1+p(\pi)$ and $SB1+p(m1|\theta)$ scenarios (the MAP value is the left peak) which is "resolved" by the mixed priors case (uni-modal), in which the MAP value coincides now with the right peak (which shows also that the value is not biased), all of which is also reflected in the distribution of $q$ values. Therefore, in this case it would seem that the mixed case is preferred since it has less uncertainty in $q$ (for the other scenarios the distribution of $q$ is very broad), and it is not biased. A final note regarding this system: Skiff´s catalogue of spectral types \citep{SkiffCat} gives a broad range of possibilities between B8V to A0V, B9IV, B9III and even A0II, from fifteen different sources, our adopted value being a B9V. Considering its reported $V=4.27$ on SIMBAD, and its adopted trigonometric parallax in Table~\ref{tab:SB1/SB1_plx&m1}, the predicted $M_V = +0.14$, this corresponds indeed to a B8V-B9V from \cite{Abuset2020}. On the other hand, a B9IV should have $M_V=-0.05$, a B9II should have $M_V=-0.50$, and an A0II should have $M_V=-3.4$, all of them far from our observed value. We thus conclude that our adopted spectral type is indeed reasonable.

\section{Conclusions and final comments} \label{sec:conclusions}

The Bayesian methodology for the inference of orbital parameters in $SB1$ binary systems with a visual orbit proposed in this paper allows us to provide a computationally-efficient, robust, and precise estimation of the corresponding joint posterior distributions of these parameters. This inference is implemented through the No-U-Turn sampler MCMC algorithm, which allows the incorporation of prior information of the stellar system to constrain the inference in scenarios with imprecise or missing data. The flexibility of this sampling scheme to add prior information is very useful for an estimation of individual component masses in $SB1$ visual-spectroscopic binaries.

An exhaustive experimental analysis has been carried out for the validation of the proposed methodology. We study the quality of the inference by comparing the estimated posterior distribution of well-studied $SB2$ visual-spectroscopic binaries with their $SB1$ visual spectroscopic counterparts by omitting the RV observations of the companion object. Our results show a negligible difference between the estimated posterior distributions of the orbital parameters (and their uncertainties) when compared to the benchmark (full observation) case, in which the RV of both components are considered. This is a very promising result, showing that partial observations ($SB1$ case) offers a good estimation performance.

Our empirical results indicate that the incorporation of prior information of the system (through the trigonometric parallax and the mass of the primary object) allows an estimation of the mass ratio of the system (and hence the individual component masses) with good precision. The incorporation of the prior distributions makes those parameters identifiable, where the derived estimations — position, dispersion, and shape of the marginal posterior distribution — strongly depend on the prior chosen. This prior knowledge has also an influence on the estimated posterior distribution of the other orbital parameters. The impact and relevance of incorporating priors on the inference of previous-identifiable orbital parameters (that are already identifiable without the incorporation of the priors) depends on the abundance, precision, and orbital coverage of the available observations. In particular, it is observed that if the system is precisely determined, the impact of the prior on the estimation of those set of parameters is negligible, as expected.

Our numerical results show that the lowest estimation error (from the optimal MAP estimator of all the systems analyzed) on the system's mass ratio, with respect to the full-information scenario (with both RV observations), was achieved by the mixed case that incorporates prior information of trigonometric parallax and a mass for the primary object simultaneously ($4.92 \%$), while the highest error was obtained by incorporating a prior only on the mass of the primary alone ($7.44\%$), achieving a percentage error lower than $8\%$. It is shown that the closest marginal posterior distribution to the full-information scenario —in the KLD sense— was achieved by the incorporation of a prior on the system parallax alone. The lowest similitude was obtained by the incorporation of a prior on the primary object mass alone, which is attributed to the fact that the parallax prior information is probably more reliable and precise than the prior on the mass of the primary star, at least for our benchmark systems. Overall, the incorporation of both priors was the most beneficial to the accuracy of the MAP estimates, where, when more information is provided, better estimation can be obtained. Large differences in the posterior distribution of parameters depending on the prior imposed could signal that one of them is biased, and should thus be taken with caution.

Taking advantage of the flexibility and richness of our sample-based methodology, the differences between the estimated posterior distributions in all the studied cases was also analyzed in the corresponding observations space. This novel analysis provides a better understanding of the effect of the different sources of information on the shape and uncertainty of the orbit and RV curves of the stellar systems. Finally, we applied the proposed Bayesian framework to twelve previously unstudied $SB1$s with astrometric data, providing a complete analysis of the obtained results.

The present work addresses the classical orbital parameters estimation of binary stellar systems through a Bayesian perspective, emphasizing the importance of providing not only an estimation, but also a complete characterization of the posterior distribution of the orbital parameters. This approach allows us to provide an uncertainty quantification of the inference process (as many classical optimization-fitting methods roughly provide), but it also allows us to visualize the uncertainty of the orbit itself in the observation space. This last dimension in our analysis is fundamental to decide what (and when) future measurements are required to improve the precision of the estimation. In this way we show how an in-depth statistical analysis can provide important insights from an observational planing perspective. 

\begin{acknowledgments}

We are very grateful to the referee for his/her careful reading of our manuscript and, in particular, for pointing out the relevance of verifying the reliability of the adopted spectral types for the priors.

MV and RAM acknowledge support from FONDECYT/ANID grant No. 1190038. MV and JS acknowledge support from FONDECYT/ANID grant No. 1210315. MO \& JS acknowledge support from the Advanced Center for Electrical and Electronic Engineering, AC3E, Basal Project FB0008, ANID. We are very grateful for the continuous support of the Chilean National Time Allocation Committee under programs CN2018A-1, CN2019A-2, CN2019B-13, CN2020A-19, CN2020B-10 and CN2021B-17. 

This research has made use of the Washington Double Star Catalog maintained at the U.S. Naval Observatory and of the SIMBAD database, operated at CDS, Strasbourg, France. This work has made use of data from the European Space Agency (ESA) mission Gaia (\url{https://www.cosmos.esa.int/gaia}), processed by the Gaia Data Processing and Analysis Consortium (DPAC, \url{https://www.cosmos.esa.int/web/gaia/dpac/consortium}). Funding for the DPAC has been provided by national institutions, in particular the institutions participating in the Gaia Multilateral Agreement.
\end{acknowledgments}

\vspace{5mm}
\facilities{AURA: SOAR, Gemini.}

\software{Stan \citep{carpenter2017stan},  
          NumPy \citep{harris2020array}, 
          Matplotlib \citep{hunter2007matplotlib},
          Seaborn \citep{Waskom2021seaborn},
          ArviZ \citep{arviz_2019}.
          }

\appendix

\section{Orbital model}
\label{Sec1:binary}

\subsection{Visual binary systems}
\label{Sec1:VB}
A visual binary (or visual double system) corresponds to a gravitationally bound binary system in which the relative position of both components are observable. The positional or astrometric observations in binary systems measures the relative position of the fainter or companion object with respect to the brighter or primary object. Depending on the technique used to obtain the astrometric observations, these can be classified roughly into micrometric, photographic, or interferometric positional measurements.

The solution of the differential equation of the motion is described by Kepler´s laws (Equation~(\ref{eq:mean_anomaly})), assuming both objects behave as point masses (detached binaries), where in the case of binary stars, it corresponds to an elliptical orbit where the primary star is in the focus and the area swept by the radius vector is constant per unit time. This elliptical orbit, denoted as real (relative) orbit, is characterized by four orbital parameters: 
\begin{itemize}
    \item \textbf{Period ($P$):} The revolution period in years.
    \item \textbf{Time of periastron passage ($T$):} One epoch of passage through periastron (minimum true distance between the components) in years and fraction of a year.
    \item \textbf{Semi-major axis ($a$):} The major semiaxis of the elliptical true orbit in seconds of arc.
    \item \textbf{Eccentricity ($e$):} The numerical eccentricity.
\end{itemize}

The astrometric observations are position measurements of the projection of the real orbit in the plane of the sky relative to the observer (plane of reference), denoted as apparent orbit. Three additional parameters are necessary to project the real orbit into the apparent orbit: 
\begin{itemize}
    \item \textbf{Longitude of the ascending node ($\Omega$):} The position angle from a reference direction to the ascending node\footnote{Point where the real orbit of the object passes through the plane of reference.} in the plane of reference (ranging from 0° to 360°).
    \item \textbf{Argument of periapsis ($\omega$):} The angle from the node to the periastron in the real orbit, following the direction of motion (ranging from 0° to 360°).
    \item \textbf{Inclination ($i$):} The angle between the plane of projection and that of the true orbit (ranging from 0° to 180°).
\end{itemize}

It is worth pointing out that two values for the longitude of the ascending node ($\Omega$ and $\Omega+\varpi$) results in identical apparent orbits. Therefore, the ascending node cannot be identified by positional observations. By convention in astronomy, if the ascending node is undetermined, the value of $\Omega$ is placed in the first two quadrants, i.e., from 0$^o$ to 180$^o$.

On the specifics, the position on the apparent orbit $(\rho,\theta)$ at a certain time $t$ (the ephemerides formulae) involves the determination of the position in the real orbit and its projection to the apparent orbit. The position on the real orbit involves the determination of the three orbital anomalies: the true anomaly $\nu(t)$, the eccentric anomaly $E(t)$ and the mean anomaly $M(t)$, in terms of the orbital parameters $P$, $T$, $a$ and $e$.

The true anomaly $\nu(t)$ is defined as the angle between the periapsis and the current position of the companion object in the orbit, as seen from its main focus (the position of the primary object). This angular parameter can be determined by the following geometrical identity:
\begin{equation}
    \tan{\frac{\nu(t)}{2}}=\sqrt{\frac{1+e}{1-e}}\tan{\frac{E(t)}{2}}.
    \label{eq:nu}
\end{equation}
where $E(t)$ is the eccentric anomaly. The eccentric anomaly is defined as the angle between the periapsis and the intersection of a perpendicular line to the semi-major axis of the orbit and the position of the companion object in the orbit, as seen from its central point. This angular parameter can be determined by the numerical resolution of Kepler´s equation\footnote{The Kepler equation is commonly resolved using the Newton-Raphson method \cite{ypma1995historical}.}:
\begin{equation}
    M(t)=\frac{2\pi(t-T)}{P}=E(t)-e\sin E(t),
    \label{eq:mean_anomaly}
\end{equation}
where $M(t)$ is the mean anomaly of the orbit. The mean anomaly represents the angular movement of the companion object in the orbit (similar to the true anomaly) but at a uniform rate. The uniform rate of movement is represent by a circle circumscribed to the orbit. Therefore, the mean anomaly is defined as the angle between the periapsis and the point in the circle circumscribed to the orbit, as seen from its central point, that covers the same area per unit time as the true anomaly. This angular parameter corresponds to the revolution period of the companion object in the orbit, i.e., $M(t)=2\pi(t-T)/P$.
 
Finally, the position on the real orbit is projected to the apparent orbit through the angular parameters $\omega$, $\Omega$ and $i$:
\begin{equation}
\begin{split}
    \tan(\theta(t)-\Omega)&=\tan(\nu(t)+\omega)\cos (i), \\
    \rho(t)&=r(t)\cos(\nu(t)+\omega)\sec(\theta(t)+\Omega),
\end{split}
\label{eq:pos_polar}
\end{equation}
with $r(t)=a(1-e^2)/(1+e\cos(\nu(t)))$ is the radius vector.

The procedure to compute the position on the apparent orbit in rectangular coordinates ($X,Y$) at a certain time $t$ involves determining the normalized rectangular coordinates in the true orbit $x,y$:
\begin{equation}
\begin{split}
    x(t)&=\cos{E(t)}-e, \\
    y(t)&=\sqrt{1-e^2}\sin{E(t)}, 
\end{split}
\label{eq:pos_rectangular}
\end{equation}
with $E(t)$ the eccentric anomaly determined in (\ref{eq:mean_anomaly}). Therefore, the position on the true orbit is computed by a ponderation of the normalized coordinates:
\begin{equation}
    \begin{split}
        X(t)&=Ax(t)+Fy(t),\\
        Y(t)&=Bx(t)+Gy(t),
    \end{split}
    \label{eq:pos2}
\end{equation}
with $A,B,F$ and $G$ are the so-called the Thiele-Innes elements, defined as:
\begin{equation}
\begin{split}
  A&=a(\cos{\omega}\cos{\Omega}-\sin{\omega}\sin{\Omega}\cos{i}),\\
  B&=a(\cos{\omega}\sin{\Omega}+\sin{\omega}\cos{\Omega}\cos{i}),\\
  F&=a(-\sin{\omega}\cos{\Omega}-\cos{\omega}\sin{\Omega}\cos{i}),\\
  G&=a(-\sin{\omega}\sin{\Omega}+\cos{\omega}\cos{\Omega}\cos{i}).
\end{split}
\end{equation}
The terms $(A/a,B/a)$ and $(F/a,G/a)$ are interpreted as the direction cosines of the major and minor axis, respectively, of the orbit in the rectangular coordinate system formed by the tangential plane and the North direction (more specifically, $X$ (or $x$) and $Y$ (or $y$) point in the North and East directions respectively). The Thiele-Innes elements form a one-to-one correspondence with the elements $a,\Omega,\omega,i$.

\subsection{Spectroscopic binary systems}
\label{Sec1:SB}

A spectroscopic binary system corresponds to a binary system in which the spectral lines of the light emitted by its components are observable. The movement of the stars in the orbit produces variations on the spectral lines observed as a consequence of the Doppler's effect: Blue or red-shifted lines are measured when the stars moves towards or away from the observer. The Doppler shift of the components' spectral lines measured through a spectrometer results in RV observations of the objects.

The RV $V$ at a certain time $t$ can be calculated as the sum of the RV of the system´s center of mass $V_{CoM}$ (a constant since the system is assumed to be free from external forces), and the radial part of the orbital velocity of the observed component relative to the center of mass of the system, $\dot{z}=dz/dt$:
\begin{equation}
    V(t)=V_0+\dot{z}(t).
\end{equation}
Following the resolution of the two-body problem%presented in the Appendix \ref{appA}
, the radial component of the system can be expressed as:
\begin{equation}
    z(t)=r\sin(\nu(t)+\omega)\sin (i),
\end{equation}
and therefore, by taking the first temporal derivative, we have that:
\begin{equation}
    \dot{z}(t)=\frac{2\pi a\sin (i)}{P\sqrt{1-e^2}}[e\cos \omega + \cos(\nu(t) + \omega)].
\end{equation}

Denoting $K=2\pi a\sin i / (P\sqrt{1-e^2})$ as the semi-amplitude of the RV curve, the above expression becomes: 
\begin{equation}
    V(t)=V_0+K[e\cos \omega + \cos(\nu(t) + \omega)].
    \label{eq:RV_general}
\end{equation}
Therefore, the RV $V(t)$ is characterized by six orbital parameters:
\begin{itemize}
    \item \textbf{Period ($P$):} The revolution period in days.
    \item \textbf{Time of periastron passage ($T$):} One epoch of passage through periastron, typically expressed in Julian Date (J.D.).
    \item \textbf{Eccentricity ($e$):} The numerical eccentricity.
    \item \textbf{Argument of periapsis ($\omega$):} The periastron longitude, counted from the maximum of the RV curve.
    \item \textbf{Semi-amplitude ($K$):} The semi-amplitude of the RV curve in km/s.
    \item \textbf{Velocity of the center of mass, ($V_0$):} The RV of the center of mas of the system in km/s (sometimes referred to as the systemic velocity as well).
\end{itemize}
Although the parameters $P$ and $T$ do not appear directly in the expression of the RV Equation~(\ref{eq:RV_general}), they are implicit in the determination of the true anomaly $\nu$ (\ref{eq:nu}). According to the convention for the units of the orbital parameters involved, the semi-amplitude $K$ is measured in km/s, whereby, $a\sin(i)$ must be measured in km and $P$ must be converted to seconds through $P[s]=86400\cdot P[days]$.

Expression (\ref{eq:RV_general}) is valid for the relative orbit $\vec{r}=\vec{r}_2-\vec{r}_1$, however, the RV observations are relative to the center of mass of the system. To correct this discrepancy, the semi-major axis $a$ of the relative orbit must be replaced by their counterparts $a_1$ and $a_2$ of the components relative to the center of mass of the system.

By the definition of the center of mass of a system composed by two particles of mass $m_1$ and $m_2$, the following relation is directly obtained:
\begin{equation}
\begin{split}
    m_1a_1=m_2a_2,
\end{split}
\label{eq:a1/a2}
\end{equation}
and noticing that $a=a_1+a_2$, the following expressions for $a_1$ and $a_2$ are obtained:
\begin{equation}
    \begin{split}
    a_1&=a\cdot \frac{m_2}{m_1+m_2}=a\cdot \frac{q}{1+q}, \\ 
    a_2&=a\cdot \frac{m_1}{m_1+m_2}=a\cdot
    \frac{1}{1+q},
    \end{split}
    \label{eq:a_1,2}
\end{equation}
where $q=m_2/m_1$ is defined as the mass ratio. Consequently, the semi-amplitude of each component becomes:
\begin{equation}
    \begin{split}
    K_{1}&=\frac{1}{86400}\frac{2\pi a_1\sin (i)}{P\sqrt{(1-e^2)}}, \\
    K_{2}&=\frac{1}{86400}\frac{2\pi a_2\sin (i)}{P\sqrt{(1-e^2)}}.
    \end{split}
    \label{eq:K_1,2}
\end{equation}

Finally, noting that the argument of periapsis of the primary $\omega_1$ and companion $\omega_2$ components of the system differs by 180°, we have that $\omega=\omega_1=\omega_2+\pi$, the expression for the RV of each component of the binary system becomes:
\begin{equation}
    \begin{split}
        V_1(t)&=V_0+K_1[e\cos (\omega) + \cos(\nu(t) + \omega)], \\
        V_2(t)&=V_0-K_2[e\cos (\omega) + \cos(\nu(t) + \omega)].
    \end{split}
    \label{eq:RV}
\end{equation}
It is important to note that the terms $a_1\sin(i)$ and $a_2\sin(i)$ in the definition of the semi-amplitudes (\ref{eq:K_1,2}) can not be separated through RV observations.

When the spectra of both components are distinguishable, i.e., the RV of the primary and the companion object are observable, the system is denoted as \textbf{$SB2$} and is characterized by the set of orbital parameters $\vartheta_{SB2}=\{P,T,e,K_1,K_2,V_0\}$. However, this is an infrequent case since the most of the spectroscopic binary systems observations are only from the primary (brighter) object ($\sim 80\%$). When the primary object spectra is the only observable, the system is denoted as \textbf{$SB1$} and is characterized by the set of orbital parameters $\vartheta_{SB2}=\{P,T,e,K_1,V_0\}$, since the parameter $K_2$ is undetermined.

\subsection{Visual-spectroscopic binary systems}
\label{Sec1:Combined}
Visual-spectroscopic binary systems corresponds to binary systems in which the relative position and the RV of its components are observable. Since four orbital parameters $P,T,e,\omega$ are common in the visual and spectroscopic binary systems, the dynamical equations are coupled and a joint modeling that describes both sources of information (positional and RV observations) allows to lift the ambiguities and indeterminacy of each individual set of equations. Furthermore, the joint modeling allows to determine the individual masses of the system, and its parallax, called orbital parallax.

RV observations allows us to solve the indeterminacy of the longitude of the ascending node $\Omega$ in the visual binary case, since the maximum/minimum of the RV curve of each component is reached in the ascending/descending node. Conversely, the positional observations allow us to decouple the term $a\sin(i)$ in the spectroscopic binary case, since we can determine  the inclination $i$ of the orbit.

The individual masses of a binary system can be computed through the determination of the total mass of the system $m_1+m_2$ and the mass ratio $q=m_2/m_1$. According to the Third Law of Kepler, the total mass of the system is  obtained through:
\begin{equation}
    m_1+m_2=\frac{a[AU]^3}{P^2},
    \label{eq:kep_3_law}
\end{equation}
where $a[\text{AU}]$ is the relative semi-major axis of the system (in astronomical units), $P$ the period of the system (in seconds) and $m_1,m_2$ the mass of the primary and the companion object (in solar masses), respectively. Since positional observations in the plane of the sky only allow us to determine the semi-major axis $a$ in angular units (seconds of arc), the conversion to linear distance units (AU) is determined by the following expression: 
\begin{equation}
    a[AU]=\frac{a['']}{\varpi},
    \label{eq:a_conver}
\end{equation}
where $\varpi$ is the system parallax (in seconds of arc), which becomes then an additional orbital parameter required to determine the individual masses.

For the computation of the mass ratio $q$, the combination of positional and RV observations are required. Considering that the RV observations can determine the terms $a_1\sin(i)$ and $a_2\sin(i)$ and the positional observations can determine the inclination $i$, both sources of observations allow to determine the individual semi-major axis of each component ($a_1$ and $a_2$) and the mass ratio $q=a_1/a_2$ (\ref{eq:a1/a2}).

Considering Equation~(\ref{eq:a_conver}), the RV expression in (\ref{eq:RV}) becomes\footnote{The units conversion between the common orbital parameters of Section \ref{Sec1:VB} and Section \ref{Sec1:SB} is omitted for simplicity.}:
\begin{equation}
\begin{split}
    V_{1}(t)&=V_{0}+\dfrac{2\pi a_1 \sin{i}}{P\sqrt{1-e^2}}[\cos(\omega + \nu(t))+e\cos{\omega}],\\
    V_{2}(t)&=V_{0}-\dfrac{2\pi a_2 \sin{i}}{P\sqrt{1-e^2}}[\cos(\omega + \nu(t))+e\cos{\omega}],
    \label{eq:V_comb}
\end{split}
\end{equation}
with $a_1=a''/\varpi\cdot q/(1+q)$, $a_2=a''/\varpi\cdot 1/(1+q)$ and $a''$ the semi-major axis in arcseconds.

If the RV observations of each component ($V_1(t)$ and $V_2(t)$) are available ($SB2$ case), the combined model that describes the positional and RV observations is characterized by the set of orbital parameters $\vartheta_{VB+SB2}=\{P,T,e,a,\omega,\Omega,i,V_0,\varpi,q\}$. However, if the RV observations of only one component are available ($SB1$ case), the parameters $q$ and $\varpi$ cannot be simultaneously determined.

\section{The Hamiltonian Monte Carlo and the No-U-Turn sampler}
\label{app:hmc&nuts}

\subsection{Hamiltonian Monte Carlo}
\label{app:hmc}
The Hamiltonian Monte Carlo \citep{neal2011mcmc,betancourt2017conceptual}, also known as Hybrid Monte Carlo, is an instance of the Metropolis-Hastings algorithm that makes use of the geometry of the target probability distribution to guide the transitions of the Markov chain. This allows to perform the sampling very efficiently, avoiding a random walk behavior of the solution, and overly sensitivity to correlated parameters. These features facilitate convergence on high-dimensional target distributions much more quickly than with other simpler methods, such as the random walk Metropolis-Hastings \citep{metropolis1953equation,hastings1970monte} or the Gibbs sampler \citep{geman1984stochastic}.

The core of the Hamiltonian Monte Carlo method is to sample from the zone of the parameter space that contributes highly to the computation of the expectation of a target distribution $P(x)$ given a parametrization $f(x)$ of the parameter space $\mathcal{X}$:
\begin{equation}
    \mathbb{E}_P(f)=\int_\mathcal{X} f(x)P(x)dx,
    \label{eq:typical_set}
\end{equation}
or in other words, to sample from the area of the parameter space of highest mass $P(x)dx$ given a parametrization $f(x)$. This zone is denoted as the \textit{typical set}. 

For that purpose, the transitions of the Markov chain must be guided by a vector field in the direction of the typical set by exploiting the differential structure of the target distribution. Hence, the vector field is generated by using the gradient of the target distribution together with auxiliary parameters of momentum that compensates the attractive force of the gradient to the target distribution mode, preserving a dynamical equilibrium that allows to align the generated vector field with the typical set.

A conservative dynamic in physical systems requires that any compression or expansion in the position space must be compensated with a respective expansion or compression in the momentum space, preserving the volume in the joint space of position and momentum. To ensure this conservative dynamic behavior, the transition probabilities of the chain follows the Hamiltonian dynamics. The Hamiltonian dynamical system is described by a function over the position $x$ and momentum $p$ variables, known as the Hamiltonian function $H(x,p)$.

Let $x_n\in \mathbb{R}^d$ be the vector of parameters of the space state and $P(x)$ the target distribution, each dimension of the space state is complemented by a fictitious momentum variable $x$:
\begin{equation}
    x_n\rightarrow (x_n,p_n),
\end{equation}
where $p_n\in\mathbb{R}^d$. The combined space of the parameters $(x_n,p_n)\in\mathbb{R}^{2d}$ is denoted as the phase space and the respective induced distribution $P(x,p)$ is denoted as the canonical distribution. 

To mimic the conservative dynamic behaviour of the space variables and the momentum variables, the canonical distribution is written in terms of the Hamiltonian function\footnote{It follows the Boltzmann canonical distribution $P(x)=z^{-1}e^{-E(x)/t}$, with $z$ a normalization constant and $t$ the temperature variable fixed to one.}:
\begin{equation}
    P(x,p)=e^{-H(x,p)},
\end{equation}
which implies that:
\begin{equation}
    H(x,p)=-\log{P(x,p)},
\end{equation}
Hence, the Hamiltonian function captures the probabilistic structure of the phase space, and consequently, the geometry of its typical set.

A marginalization of the canonical distribution $P(x,p)$ in terms of the state variable $p$ induces the following decomposition of the Hamiltonian function $H(x,p)$:
\begin{equation}
\begin{split}
    H(x,p)&=-\log P(p|x)-\log P(x)\\
          &\equiv K(p,x)+V(x).
\end{split}
\label{eq:hamiltonian_descomp}
\end{equation}
The decomposition can be interpreted in terms of a kinetic energy $K(p,x)$ function, dependant upon both the spatial and momentum variables, and a potential energy function $V(x)$, dependant upon the momentum variables only. The potential function is simply the negative logarithm of the target distribution, while the kinetic energy is usually expressed as a quadratic term on $p$:
  \begin{equation}
  K(p)=\frac{1}{2}p^T\cdot M^{-1}\cdot p,
  \label{eq:kinetic_energy}
\end{equation}
where $M$ is a symmetric, positive-definite matrix denoted as mass-matrix. The mass matrix is typically a scalar multiple of the identity matrix, but can explicitly depends on $x$ as in (\ref{eq:hamiltonian_descomp}).

With these elements, the vector field oriented in the direction of the typical set can be defined through the Hamiltonian equations:
\begin{equation}
\begin{split}
\frac{dx}{dt}&=+\frac{\partial H}{\partial p}=\frac{\partial K}{\partial p}\\
 \frac{dp}{dt}&=-\frac{\partial H}{\partial x}=-\frac{\partial K}{\partial x}-\frac{\partial V}{\partial x}.
\end{split}
\label{eq:hamiltonian_equations}
\end{equation}

Following the vector field (determined by the Hamiltonian equations for a time $t$) we can generate trajectories $\phi_t(x,p)$ that move along the typical set. To compute these trajectories, the solution of (\ref{eq:hamiltonian_equations}) is obtained by numerical methods. In particular, the trajectory $\phi_T(x,p)$ of a variable $i$ can be approximated by the \textit{leap-frog integration} method iterating the following expressions:
\begin{equation}
    \begin{split}
        p_i(t+\epsilon/2)&= p_i(t) - \frac{\epsilon}{2}\frac{\partial V}{\partial x_i}(x_i(t))\\
    x_i(t+\epsilon) &= x_i(t) + \frac{\epsilon}{m_i} p_i(t+\epsilon/2) \\
    p_i(t+\epsilon) &= p_i(t+\epsilon/2) - \frac{\epsilon}{2}\frac{\partial V}{\partial x_i}(x_i(t+\epsilon)),\,\,\,\,\forall i\in{1,...,L}.
    \end{split}
\end{equation}
$L$ is the number of steps, $\epsilon\in\mathbb{R}$ is the step size and $T=\lfloor L/\epsilon\rfloor$ the integration time. The adequate selection of the algorithm hyper-parameter $\epsilon$ and $L$ is crucial for a good sampling performance.

In summary, the Markov chain that samples from the target distribution $P(x)$ will follow the Metropolis-Hastings algorithm defined on the phase space $(x,p)$ with transition probabilities $T_{(x_0,p_0),(x_L,p_L)}$ determined by the solution of the Hamiltonian equations following the leap-frog integration method for a fixed number of steps $L$ and step size $\epsilon$. The momentum variables are sampled from a proposal marginal distribution $P(p|x)$ and the final samples of $P(x)$ are obtained by projecting the samples of the phase space on the state space $(x,p)\rightarrow x$, i.e., ignoring the momentum variables. 

The transition probabilities of the current Markov Chain must be modified since the transition ratio in the Metropolis-Hastings acceptance probability $T_{(x_L,p_L),(x_0,p_0)}/T_{(x_0,p_0),(x_L,p_L)}=0/1=0$, because the leap-frog integration does not allow reverse trajectories. Thus, the transition probabilities are modified to be reversible by augmenting the numerical integration with a negation step that flips the sign of momentum $(x,p)\rightarrow(x,-p)$. Thereby, the Metropolis-Hastings acceptance rate becomes:
\begin{equation}
    \begin{split}
        \mathcal{A}((x_L,-p_L),(x_0,p_p))&=\min\left\{1,\frac{T_{(x_L,-p_L),(x_0,p_0)}P(x_L,-p_L)}{T_{(x_0,p_0),(x_L,-p_L)}P(x_0,p_0)}\right\}\\
        &=\min\left\{1,\frac{\delta(x_L-x_L)\delta(-p_l+p_L)P(x_L,-p_L)}{\delta(x_0-x_0)\delta(p_0-p_0)P(x_0,p_0)}\right\}\\
        &=\min\left\{1,\frac{P(x_L,-p_L)}{P(x_0,p_0)}\right\}\\
        &=\min\left\{1,e^{-H(x_L,-p_L)+H(x_0,p_0)}\right\}.
    \end{split}
\end{equation}
The complete Hamiltonian Monte Carlo sampling procedure is described in Algorithm \ref{alg:hmc}.
 
\iffalse
\begin{algorithm}[H]
  \SetKwInOut{Parameter}{Parameters}
  \Parameter{$N, \epsilon, T, M$}
  Initialise $x^{(0)}$ \\
  \For{$0 \leq i < N$}{
    Sample $u \sim \mathcal{U}([0,1]), p \sim \mathcal{N}(0,M)$ \\
    $x_{0} \gets x^{(i)}, p_{0} \gets p, L \gets \lfloor \frac{T}{\epsilon}\rfloor$ \\
    \For{$0 \leq l < L$}{
    $p_{l + \frac{1}{2}} \gets p_{l} - \frac{\epsilon}{2}\frac{\partial V}{\partial x}(x_{l})$ \\
    $x_{l+1} \gets x_{l} + \epsilon p_{l+\frac{1}{2}}$ \\
    $p_{l+1} \gets p_{l+\frac{1}{2}} - \frac{\epsilon}{2}\frac{\partial V}{\partial x}(x_{l})$
    }
    \eIf{$u < \min\left(1,e^{-H(x_L,-p_L)+H(x_0,p_0)}\right)$}{
    $x^{(i+1)} \gets x_{L}$}{
    $x^{(i+1)} \gets x^{(i)}$}
  }
  \caption{Hamiltonian Monte Carlo.}
  \label{alg:hmc}
\end{algorithm}
\fi

\subsection{No-U-Turn sampler}
\label{Sec2:NUTS}
The No-U-Turn Sampler algorithm \citep{hoffman2014no} is an extension of the Hamiltonian Monte Carlo algorithm that adaptively sets the number of steps $L$ of the trajectories, facilitating the use of the sampling tool by avoiding a low performance selection of the Hamiltonian Monte Carlo user-defined hyper-parameters.
The selection of the number of steps $L$ follows the computation of forward and backward exploration trajectories of the Hamiltonian Monte Carlo algorithm until an end condition is met, where the new sample is obtained by a random selection of the generated trajectories.

To generate the exploration trajectories, a binary tree is constructed iteratively. Let $(x_n(0),p_n(0))$ be an initial particle composed by a position and momentum of the $n$-th iteration of the Markov chain, $(x_n^+,p_n^+)$ be a forward in time particle and $(x_n^-,p_n^-)$ be a backward in time particle. In each iteration $j$, the binary tree selects at random uniformly to move the $(j-1)$-particle forwards or the backward particle backwards in time, with $2^j$ number of leap-frog integration steps.

The iterative procedure continues until the following condition (namely the U-Turn condition) is met:
\begin{equation}
    (x_n^+-x_n^-)\cdot p_n^- < 0 \,\,\vee\,\, (x_n^+-x_n^-)\cdot p_n^+ < 0,
\end{equation}
or when the Hamiltonian trajectory generated by the leap-frog integration becomes imprecise in the sense that:
 \begin{equation}
    \begin{gathered}
    e^{-H(x_n^+,p_n^+)+\Delta_{max}}<U_n \,\,\vee\,\, e^{-H(x_n^-,p_n^-)+\Delta_{max}}<U_n,
    \end{gathered}
\end{equation}
where $U_n\sim \mathcal{U}\left(0,e^{-H(q_n(0),p_n(0))}\right)$ is a slice random variable sample and $\Delta_{max}$ is a maximum energy hyper-parameter.
The idea of the No-U-Turn condition is avoiding the generation of redundant trajectories by stopping the exploration when the trajectory begins to turn back to previous explored zones.

Finally, the new sample $(x_{n+1},p_{n+1})$ is selected by a uniform sampling of the generated trajectory that satisfies the precision condition $U_n<e^{-H(x_{n+1},p_{n+1})}$.
The complete No-U-Turn sampler sampling procedure is described in Algorithm \ref{alg:nuts}.

\subsection{Implementation considerations in the binary stars context}
\label{appB}
The implementation of the proposed Bayesian inference methodology based on the No-U-Turn sampler algorithm requires to compute the gradient of the posterior function, which can be analytically derived from the Keplerian model formulae presented in Section \ref{Sec1:binary} by taking the partial derivatives with respect to each orbital parameter that characterizes the binary stellar system. However, a special consideration must be taken with the partial derivatives of the eccentric anomaly $E$, since it is not analytically calculated, but rather numerically approximated. The non-zero partial derivatives of the eccentric anomaly can be expressed as a function of the variable itself as follows:
\begin{equation}
    \begin{split}
    \frac{\partial E}{\partial e}&=\frac{\sin{E}}{1-e\cos{E}},\\
    \frac{\partial E}{\partial T}&=-\frac{2\pi}{P}\cdot\frac{1}{1-e\cos{E}},\\
    \frac{\partial E}{\partial P}&=-\frac{2\pi(t-T)}{P^2}\cdot\frac{1}{1-e\cos{E}},
    \end{split}
\end{equation}
with the value of $E$ previously approximated by any numerical method (e.g., the Newton-Raphson method \citep{ypma1995historical}).

\bibliography{sample631}{}

\begin{thebibliography}{}
\expandafter\ifx\csname natexlab\endcsname\relax\def\natexlab#1{#1}\fi
\providecommand{\url}[1]{\href{#1}{#1}}
\providecommand{\dodoi}[1]{doi:~\href{http://doi.org/#1}{\nolinkurl{#1}}}
\providecommand{\doeprint}[1]{\href{http://ascl.net/#1}{\nolinkurl{http://ascl.net/#1}}}
\providecommand{\doarXiv}[1]{\href{https://arxiv.org/abs/#1}{\nolinkurl{https://arxiv.org/abs/#1}}}

\bibitem[{{Abushattal} {et~al.}(2020){Abushattal}, {Docobo}, \&
  {Campo}}]{Abuset2020}
{Abushattal}, A.~A., {Docobo}, J.~A., \& {Campo}, P.~P. 2020, \aj, 159, 28,
  \dodoi{10.3847/1538-3881/ab580a}

\bibitem[{{Agati} {et~al.}(2015){Agati}, {Bonneau}, {Jorissen}, {Souli{\'e}},
  {Udry}, {Verhas}, \& {Dommanget}}]{2015A&A...574A...6A}
{Agati}, J.~L., {Bonneau}, D., {Jorissen}, A., {et~al.} 2015, \aap, 574, A6,
  \dodoi{10.1051/0004-6361/201323056}

\bibitem[{Aller {et~al.}(1996)Aller, Appenzeller, Baschek, Butler, De~Loore,
  Duerbeck, El~Eid, Fink, Herczeg, Richtler, {et~al.}}]{aller1996landolt}
Aller, L., Appenzeller, I., Baschek, B., {et~al.} 1996, lbor, 3

\bibitem[{Betancourt(2017)}]{betancourt2017conceptual}
Betancourt, M. 2017, arXiv preprint arXiv:1701.02434

\bibitem[{Bouffanais \& Porter(2019)}]{bouffanais2019bayesian}
Bouffanais, Y., \& Porter, E.~K. 2019, Physical Review D, 100, 104023

\bibitem[{Brown {et~al.}(2018)Brown, Vallenari, Prusti, De~Bruijne, Babusiaux,
  Bailer-Jones, Biermann, Evans, Eyer, Jansen, {et~al.}}]{brown2018gaia}
Brown, A., Vallenari, A., Prusti, T., {et~al.} 2018, Astronomy \& astrophysics,
  616, A1

\bibitem[{Burgasser {et~al.}(2012)Burgasser, Luk, Dhital, Gagliuffi, Nicholls,
  Prato, West, \& L{\'e}pine}]{burgasser2012discovery}
Burgasser, A.~J., Luk, C., Dhital, S., {et~al.} 2012, The Astrophysical
  Journal, 757, 110

\bibitem[{Carpenter {et~al.}(2017)Carpenter, Gelman, Hoffman, Lee, Goodrich,
  Betancourt, Brubaker, Guo, Li, \& Riddell}]{carpenter2017stan}
Carpenter, B., Gelman, A., Hoffman, M.~D., {et~al.} 2017, Journal of
  statistical software, 76, 1

\bibitem[{{Carroll} \& {Ostlie}(2006)}]{2006ima..book.....C}
{Carroll}, B.~W., \& {Ostlie}, D.~A. 2006, {An introduction to modern
  astrophysics and cosmology}

\bibitem[{Cid~Palacios(1958)}]{cid1958necessary}
Cid~Palacios, R. 1958, AJ, 63, 395

\bibitem[{{Claveria} {et~al.}(2019){Claveria}, {Mendez}, {Silva}, \&
  {Orchard}}]{Clavet2019}
{Claveria}, R.~M., {Mendez}, R.~A., {Silva}, J.~F., \& {Orchard}, M.~E. 2019,
  \pasp, 131, 084502, \dodoi{10.1088/1538-3873/ab22e2}

\bibitem[{Docobo(1985)}]{docobo1985analytic}
Docobo, J. 1985, Celestial Mechanics, 36, 143

\bibitem[{Docobo {et~al.}(1992)Docobo, Ling, \& Prieto}]{docobo1992adaptation}
Docobo, J., Ling, J., \& Prieto, C. 1992, in International Astronomical Union
  Colloquium, Vol. 135, Cambridge University Press, 220--222

\bibitem[{Docobo {et~al.}(2018)Docobo, Tamazian, Campo, \&
  Piccotti}]{docobo2018visual}
Docobo, J.~A., Tamazian, V.~S., Campo, P.~P., \& Piccotti, L. 2018, The
  Astronomical Journal, 156, 85

\bibitem[{Ford(2005)}]{ford2005quantifying}
Ford, E.~B. 2005, The Astronomical Journal, 129, 1706

\bibitem[{Geman \& Geman(1984)}]{geman1984stochastic}
Geman, S., \& Geman, D. 1984, IEEE Transactions on pattern analysis and machine
  intelligence, 721

\bibitem[{{Gray}(2008)}]{2008oasp.book.....G}
{Gray}, D.~F. 2008, {The Observation and Analysis of Stellar Photospheres}

\bibitem[{Gregory(2005)}]{gregory2005bayesian}
Gregory, P. 2005, The Astrophysical Journal, 631, 1198

\bibitem[{Gregory(2011)}]{gregory2011bayesian}
Gregory, P.~C. 2011, Monthly Notices of the Royal Astronomical Society, 410, 94

\bibitem[{Habets \& Heintze(1981)}]{habets1981empirical}
Habets, G., \& Heintze, J. 1981, Astronomy and Astrophysics Supplement Series,
  46, 193

\bibitem[{Hajian(2007)}]{hajian2007efficient}
Hajian, A. 2007, Physical Review D, 75, 083525

\bibitem[{Halbwachs {et~al.}(2016)Halbwachs, Boffin, Le~Bouquin, Kiefer,
  Famaey, Salomon, Arenou, Pourbaix, Anthonioz, Grellmann,
  {et~al.}}]{halbwachs2016masses}
Halbwachs, J.-L., Boffin, H., Le~Bouquin, J.-B., {et~al.} 2016, Monthly Notices
  of the Royal Astronomical Society, 455, 3303

\bibitem[{Harris {et~al.}(2020)Harris, Millman, van~der Walt, Gommers,
  Virtanen, Cournapeau, Wieser, Taylor, Berg, Smith,
  {et~al.}}]{harris2020array}
Harris, C.~R., Millman, K.~J., van~der Walt, S.~J., {et~al.} 2020, Nature, 585,
  357

\bibitem[{Hastings(1970)}]{hastings1970monte}
Hastings, W.~K. 1970

\bibitem[{Hoffman {et~al.}(2014)Hoffman, Gelman, {et~al.}}]{hoffman2014no}
Hoffman, M.~D., Gelman, A., {et~al.} 2014, J. Mach. Learn. Res., 15, 1593

\bibitem[{Hou {et~al.}(2012)Hou, Goodman, Hogg, Weare, \&
  Schwab}]{hou2012affine}
Hou, F., Goodman, J., Hogg, D.~W., Weare, J., \& Schwab, C. 2012, The
  Astrophysical Journal, 745, 198

\bibitem[{Hummel {et~al.}(1994)Hummel, Armstrong, Quirrenbach, Buscher,
  Mozurkewich, Elias, \& Wilson}]{hummel1994very}
Hummel, C., Armstrong, J., Quirrenbach, A., {et~al.} 1994, The Astronomical
  Journal, 107, 1859

\bibitem[{Hunter(2007)}]{hunter2007matplotlib}
Hunter, J.~D. 2007, Computing in science \& engineering, 9, 90

\bibitem[{Ji {et~al.}(2017)Ji, Banks, Budding, \& Rhodes}]{ji2017investigation}
Ji, Y., Banks, T., Budding, E., \& Rhodes, M. 2017, Astrophysics and Space
  Science, 362, 1

\bibitem[{Kumar {et~al.}(2019)Kumar, Carroll, Hartikainen, \&
  Martin}]{arviz_2019}
Kumar, R., Carroll, C., Hartikainen, A., \& Martin, O. 2019, Journal of Open
  Source Software, 4, 1143, \dodoi{10.21105/joss.01143}

\bibitem[{Liu \& Nocedal(1989)}]{liu1989limited}
Liu, D.~C., \& Nocedal, J. 1989, Mathematical programming, 45, 503

\bibitem[{Lucy(2014)}]{lucy2014mass}
Lucy, L. 2014, Astronomy \& Astrophysics, 563, A126

\bibitem[{Lucy(2018)}]{lucy2018binary}
---. 2018, Astronomy \& Astrophysics, 618, A100

\bibitem[{MacKnight \& Horch(2004)}]{macknight2004calculating}
MacKnight, M., \& Horch, E. 2004, AAS, 204, 07

\bibitem[{{Mason}(2015)}]{Mas2015}
{Mason}, B.~D. 2015, in IAU General Assembly, Vol.~29, 2300709

\bibitem[{{Mason} {et~al.}(2001){Mason}, {Wycoff}, {Hartkopf}, {Douglass}, \&
  {Worley}}]{WDSCat2001}
{Mason}, B.~D., {Wycoff}, G.~L., {Hartkopf}, W.~I., {Douglass}, G.~G., \&
  {Worley}, C.~E. 2001, \aj, 122, 3466, \dodoi{10.1086/323920}

\bibitem[{{Mendez} {et~al.}(2021){Mendez}, {Claver{\'\i}a}, \&
  {Costa}}]{Mendet2021}
{Mendez}, R.~A., {Claver{\'\i}a}, R.~M., \& {Costa}, E. 2021, \aj, 161, 155,
  \dodoi{10.3847/1538-3881/abdb28}

\bibitem[{Mendez {et~al.}(2017)Mendez, Claveria, Orchard, \&
  Silva}]{mendez2017orbits}
Mendez, R.~A., Claveria, R.~M., Orchard, M.~E., \& Silva, J.~F. 2017, The
  Astronomical Journal, 154, 187

\bibitem[{Metropolis {et~al.}(1953)Metropolis, Rosenbluth, Rosenbluth, Teller,
  \& Teller}]{metropolis1953equation}
Metropolis, N., Rosenbluth, A.~W., Rosenbluth, M.~N., Teller, A.~H., \& Teller,
  E. 1953, The journal of chemical physics, 21, 1087

\bibitem[{Morbey(1975)}]{morbey1975synthesis}
Morbey, C. 1975, Publications of the Astronomical Society of the Pacific, 87,
  689

\bibitem[{Mor{\'e}(1978)}]{more1978levenberg}
Mor{\'e}, J.~J. 1978, in Numerical analysis (Springer), 105--116

\bibitem[{Muterspaugh {et~al.}(2010)Muterspaugh, Hartkopf, Lane, O’Connell,
  Williamson, Kulkarni, Konacki, Burke, Colavita, Shao,
  {et~al.}}]{muterspaugh2010phases}
Muterspaugh, M.~W., Hartkopf, W.~I., Lane, B.~F., {et~al.} 2010, The
  Astronomical Journal, 140, 1623

\bibitem[{Neal {et~al.}(2011)}]{neal2011mcmc}
Neal, R.~M., {et~al.} 2011, Handbook of markov chain monte carlo, 2, 2

\bibitem[{Nelson {et~al.}(2013)Nelson, Ford, \& Payne}]{nelson2013run}
Nelson, B., Ford, E.~B., \& Payne, M.~J. 2013, The Astrophysical Journal
  Supplement Series, 210, 11

\bibitem[{Pourbaix(1994)}]{pourbaix1994trial}
Pourbaix, D. 1994, Astronomy and Astrophysics, 290, 682

\bibitem[{Pourbaix(1998)}]{pourbaix1998simultaneous}
---. 1998, Astronomy and Astrophysics Supplement Series, 131, 377

\bibitem[{{Pourbaix}(2000)}]{Pour2000}
{Pourbaix}, D. 2000, \aaps, 145, 215, \dodoi{10.1051/aas:2000237}

\bibitem[{{Pourbaix} {et~al.}(2004){Pourbaix}, {Tokovinin}, {Batten}, {Fekel},
  {Hartkopf}, {Levato}, {Morrell}, {Torres}, \& {Udry}}]{Pouret2004}
{Pourbaix}, D., {Tokovinin}, A.~A., {Batten}, A.~H., {et~al.} 2004, \aap, 424,
  727, \dodoi{10.1051/0004-6361:20041213}

\bibitem[{Prusti {et~al.}(2016)Prusti, De~Bruijne, Brown, Vallenari, Babusiaux,
  Bailer-Jones, Bastian, Biermann, Evans, Eyer, {et~al.}}]{prusti2016gaia}
Prusti, T., De~Bruijne, J., Brown, A.~G., {et~al.} 2016, Astronomy \&
  Astrophysics, 595, A1

\bibitem[{Sahlmann {et~al.}(2013)Sahlmann, Lazorenko, S{\'e}gransan,
  Mart{\'\i}n, Queloz, Mayor, \& Udry}]{sahlmann2013astrometric}
Sahlmann, J., Lazorenko, P., S{\'e}gransan, D., {et~al.} 2013, Astronomy \&
  Astrophysics, 556, A133

\bibitem[{Schmidt-Kaler {et~al.}(1982)Schmidt-Kaler, Schaifers, \&
  Voigt}]{schmidt1982landolt}
Schmidt-Kaler, T., Schaifers, K., \& Voigt, H. 1982, Stars and Star Clusters,
  vol

\bibitem[{Shabram {et~al.}(2020)Shabram, Batalha, Thompson, Hsu, Ford,
  Christiansen, Huber, Berger, Catanzarite, Nelson,
  {et~al.}}]{shabram2020sensitivity}
Shabram, M.~I., Batalha, N., Thompson, S.~E., {et~al.} 2020, The Astronomical
  Journal, 160, 16

\bibitem[{{Skiff}(2014)}]{SkiffCat}
{Skiff}, B.~A. 2014, VizieR Online Data Catalog, B/mk

\bibitem[{{S{\"o}derhjelm}(1999)}]{Soder1999}
{S{\"o}derhjelm}, S. 1999, \aap, 341, 121

\bibitem[{{Straizys} \& {Kuriliene}(1981)}]{1981Ap&SS..80..353S}
{Straizys}, V., \& {Kuriliene}, G. 1981, \apss, 80, 353,
  \dodoi{10.1007/BF00652936}

\bibitem[{Thiele(1883)}]{thiele1883neue}
Thiele, T.~N. 1883, Astronomische Nachrichten, 104, 245

\bibitem[{Tokovinin(1992)}]{tokovinin1992frequency}
Tokovinin, A. 1992, Astronomy and Astrophysics, 256, 121

\bibitem[{Wang {et~al.}(2009)Wang, Kulkarni, \& Verd{\'u}}]{wang2009divergence}
Wang, Q., Kulkarni, S.~R., \& Verd{\'u}, S. 2009, IEEE Transactions on
  Information Theory, 55, 2392

\bibitem[{Waskom(2021)}]{Waskom2021seaborn}
Waskom, M.~L. 2021, Journal of Open Source Software, 6, 3021,
  \dodoi{10.21105/joss.03021}

\bibitem[{{Wenger} {et~al.}(2000{\natexlab{a}}){Wenger}, {Ochsenbein}, {Egret},
  {Dubois}, {Bonnarel}, {Borde}, {Genova}, {Jasniewicz}, {Lalo{\"e}},
  {Lesteven}, \& {Monier}}]{2000A&AS..143....9W}
{Wenger}, M., {Ochsenbein}, F., {Egret}, D., {et~al.} 2000{\natexlab{a}},
  \aaps, 143, 9, \dodoi{10.1051/aas:2000332}

\bibitem[{{Wenger} {et~al.}(2000{\natexlab{b}}){Wenger}, {Ochsenbein}, {Egret},
  {Dubois}, {Bonnarel}, {Borde}, {Genova}, {Jasniewicz}, {Lalo{\"e}},
  {Lesteven}, \& {Monier}}]{SIMBAD}
---. 2000{\natexlab{b}}, \aaps, 143, 9, \dodoi{10.1051/aas:2000332}

\bibitem[{Ypma(1995)}]{ypma1995historical}
Ypma, T.~J. 1995, SIAM review, 37, 531

\bibitem[{{Ziegler} {et~al.}(2018){Ziegler}, {Law}, {Baranec}, {Morton},
  {Riddle}, {De Lee}, {Huber}, {Mahadevan}, \& {Pepper}}]{Ziegleret2018}
{Ziegler}, C., {Law}, N.~M., {Baranec}, C., {et~al.} 2018, \aj, 156, 259,
  \dodoi{10.3847/1538-3881/aad80a}

\end{thebibliography}
\bibliographystyle{aasjournal}



\end{document}